\documentclass[11pt]{article}
\newcommand{\Comment}[1]{{}}
\usepackage{amsfonts,amsthm,amsmath,amssymb,slashed}
\usepackage[textwidth = 430 pt, textheight = 630 pt]{geometry}
\usepackage{color}
\usepackage[section]{placeins}
\usepackage{comment}

\Comment{\usepackage{color}
\definecolor{MyDarkBlue}{rgb}{0.15,0.15,0.45}
\usepackage[linktocpage=true]{hyperref}
\hypersetup{
colorlinks=true,
citecolor=MyDarkBlue,
linkcolor=MyDarkBlue,
urlcolor=MyDarkBlue,
pdfauthor={Horatiu Nastase and Jacob Sonnenschein},
pdftitle={The title},
pdfsubject={hep-th}
}

\usepackage[numbers,sort&compress]{natbib}
\usepackage{hypernat}}
\usepackage{graphicx}
\usepackage{cite}
\usepackage{url}

\newcommand\ignore[1]{}
\def\one{{\,\hbox{1\kern-.8mm l}}}

\def\bra#1{\left\langle #1\right|}
\def\ket#1{\left| #1\right\rangle}
\newcommand{\braket}[2]{\langle #1 | #2 \rangle}

\def\Tr{{\rm Tr\, }}

\def\sgn{\textrm{sgn}}

\def\a{\alpha}\def\b{\beta}

\def\d{\partial}
\def\dag{\dagger}

\def\Tr{\mathop{\rm Tr}\nolimits}

\newcommand{\Cset}{{\,\,{{{^{_{\pmb{\mid}}}}\kern-.45em{\mathrm C}}}}}

\newcommand{\be}{\begin{equation}}
\newcommand{\bea}{\begin{eqnarray}}

\newcommand{\ee}{\end{equation}}
\newcommand{\eea}{\end{eqnarray}}

\providecommand{\lsim}{\lesssim}
\providecommand{\gsim}{\gtrsim}

\begin{document}

\renewcommand{\thefootnote}{\fnsymbol{footnote}}

\makeatletter
\@addtoreset{equation}{section}
\makeatother
\renewcommand{\theequation}{\thesection.\arabic{equation}}

\rightline{}
\rightline{}
%   \vspace{1.8truecm}

%\begin{flushright}
% preprint nrs.
%\end{flushright}

%\vspace{10pt}

%\begin{document}
\begin{center}
{\LARGE \bf{\sc Krylov complexity, path integrals, and instantons}} 
\end{center} 
 \vspace{1truecm}
\thispagestyle{empty} \centerline{
{\large \bf {\sc Cameron Beetar${}^{a},$}}\footnote{E-mail address: \Comment{\href{mailto:btrcam001@myuct.ac.za}}
{\tt btrcam001@myuct.ac.za}}
{\large \bf {\sc Eric L Graef${}^{b},$}}\footnote{E-mail address: \Comment{\href{mailto:eric.graef@unesp.br}}
{\tt eric.graef@unesp.br}}
{\large \bf {\sc Jeff Murugan${}^{a},$}}\footnote{E-mail address: \Comment{\href{mailto:jeff.murugan@uct.ac.za}}
{\tt jeff.murugan@uct.ac.za}} }
\centerline{
{\large \bf {\sc Horatiu Nastase${}^{b}$}}\footnote{E-mail address: \Comment{\href{mailto:horatiu.nastase@unesp.br}}
{\tt horatiu.nastase@unesp.br}}
{\bf{\sc and}}
{\large \bf {\sc Hendrik J R  Van Zyl${}^{a}$}}\footnote{E-mail address: \Comment{\href{mailto:hjrvanzyl@gmail.com}}{\tt hjrvanzyl@gmail.com}}
                                                        }

\vspace{.5cm}

%\vspace{.3cm}

\centerline{{\it ${}^a$The Laboratory for Quantum Gravity and Strings,}}
\centerline{{\it Department of Mathematics and Applied Mathematics, }} 
\centerline{{\it University of Cape Town, Cape Town, South Africa}}
\vspace{.3cm}
\centerline{{\it ${}^b$Instituto de F\'{i}sica Te\'{o}rica, UNESP-Universidade Estadual Paulista}} 
\centerline{{\it R. Dr. Bento T. Ferraz 271, Bl. II, Sao Paulo 01140-070, SP, Brazil}}

\vspace{1truecm}

%%%%%%%%%%%%%%%%%
\thispagestyle{empty}

\centerline{\sc Abstract}

\vspace{.4truecm}

\begin{center}
\begin{minipage}[c]{380pt}
{\noindent 

Krylov complexity has emerged as an important tool in the description of quantum information and, in particular,
quantum chaos. Here we formulate Krylov complexity $K(t)$ for quantum mechanical systems
as a path integral, and argue that at large times, 
for classical chaotic systems with at least two minima of the potential, 
that have a plateau for $K(t)$, the value of the plateau is described 
by quantum mechanical instantons, as is the case for standard transition amplitudes. We explain and test 
these ideas in a simple toy model. 
}
\end{minipage}
\end{center}

\vspace{.5cm}

\setcounter{page}{0}
\setcounter{tocdepth}{2}

\newpage

\tableofcontents
\renewcommand{\thefootnote}{\arabic{footnote}}
\setcounter{footnote}{0}

\linespread{1.1}
\parskip 4pt

%{}~
%{}~

%---------------------------------------------------------

%%%%%%%%%%%%%%%%%%%%%%%%%%%%%%%%%%%%%%%%%%%%%%%%%%%%%%%%%%%%%%%%%%%%%%%%%%%%%%%%%%%%%%%%

\section{Introduction}

\noindent
Understanding how information spreads and reorganizes itself in time has proved to be a powerful way of 
classifying quantum dynamics across fields as diverse as condensed–matter theory, quantum chaos and 
holography.  Among the available diagnostics, \emph{Krylov complexity} \cite{Parker:2018yvk, Nandy:2024evd, Rabinovici:2025otw,Baiguera:2025dkc} has come to occupy 
a singular position.  
Introduced by Parker \textit{et al.}~\cite{Parker:2018yvk}, it measures the mean position 
of an evolving operator 
in the Lanczos–Krylov chain and is entirely algorithmic, basis–independent and free 
of the geometric ambiguities 
that beset Nielsen–style circuit complexity~\cite{Nielsen:2005mkt,Jefferson:2017sdb}
(even in the gravity dual formulation, we can have complexity equals volume \cite{Stanford:2014jda}, 
complexity equals action \cite{Brown:2015bva}, or a varying notion, 
``complexity = anything" \cite{Belin:2021bga}).\\

\noindent
Numerical results over the 
past several years have revealed a striking phenomenology: finite integrable systems display bounded 
oscillations; unbounded potentials lead to an ever–increasing Krylov complexity; and quantum–chaotic models 
rise rapidly before settling to a long-lived plateau whose height correlates with more traditional chaos indicators 
such as out-of-time-order correlators and random–matrix 
statistics~\cite{Balasubramanian:2021mxo, Rabinovici:2022beu, Erdmenger:2023wjg, Hashimoto:2023swv}.  Despite this remarkable progress, analytic 
control, especially over the late-time plateau, has remained elusive.\\

\noindent
The present work closes part of that gap.  We first recast the Lanczos recursion \cite{Lanczos1950AnIM} in a continuous path integral 
language: the amplitudes $|\psi_n(t)|^{2}$ of the Lanczos chain can be written as a Feynman path integral for a 
world-line particle hopping on an auxiliary semi-infinite lattice whose  (classical) potential $V(x)$
is fixed by the Hamiltonian in the Krylov basis.  
In this description the physical time acts as an Euclidean proper time, immediately inviting 
semiclassical analysis.  We show that, for any Hamiltonian whose Krylov complexity saturates, 
generally meaning for Hamiltonians displaying classically chaotic behaviour in $V(x)$ 
and allowing for instanton solutions, 
the late-time plateau for the Krylov complexity $K(t)$ (as well as the form of $|\psi_n(t)|^2$) are 
governed by a single Euclidean instanton that tunnels between the minima of the classical
potential.  The 
result is a compact formula,
\begin{equation}
K_\infty \;=\;\mathcal{N}\,e^{-\frac{2}{\hbar}S_{\text{inst}}},\qquad 
\mathcal{N}= \sum_{n} n \, 
\bigl|K^{*}_{n}({x}_1)\psi({x}_2)+K^{*}_{n}({x}_2)\psi({x}_1)\bigr|^{2},
\label{eq:intro-plateau}
\end{equation}
which expresses a late-time dynamical observable in terms of the instanton action $S_{\text{inst}}$ and the 
values of the Krylov wave functions at the (distinct) turning points ${x}_{i,j}$.\footnote{For some recent works
in which path integral methods were used to calculate the time evolution of Krylov complexity in the 
DSSYK model, but not at late times, see 
\cite{Aguilar-Gutierrez:2025pqp,Aguilar-Gutierrez:2025mxf,Aguilar-Gutierrez:2025hty}.}\\

%\noindent
%The present work closes part of that gap. We argue that, generally, for Hamiltonians displaying classically 
%chaotic behavior and allowing for instanton solutions, late-time transition amplitudes are governed by these 
%instanton solutions in the context of the semiclassical approximation for the path integral. In particular, we can 
%recast the Lanczos recursion in a continuous path integral language, such that, at late times, the amplitudes $|
%\psi_n(t)|^{2}$ and the Krylov complexity are governed by instantons. The 
%result is a compact formula,
%\begin{equation}
%    K(t\rightarrow\infty)
 %  \simeq \sum_n n \left| \sum_{\{x_1\},\{x_2\}
%   ={\rm minima}}K_n^*(\{x_2\}) \psi(\{x_1\}) e^{-\frac{1}{\hbar}
%   S_E({\rm inst}; x_1,x_2)}\right|^2\,
%\end{equation}
%which expresses a late-time dynamical observable in terms of the instanton action $S_{\text{inst}}$ and the 
%values of the Krylov wave functions at the turning points ${x}_{1,2}$.\\

\noindent
To bring this formalism under quantitative control we develop two auxiliary tools.  First, a \emph{return 
amplitude Lanczos algorithm} that eliminates ill-behaved Lanczos coefficients by working directly with the 
survival amplitude; the method, detailed in Appendix A, regularises bosonic models where the naive recursion 
diverges.  Second, a \emph{position-space implementation} of the Lanczos chain that iterates basis vectors 
directly in the continuous variables $(x_{1},x_{2})$ without imposing a Fock cut-off (Appendix B).  Both 
algorithms are independent of the particular model studied here and may be of wider use.\\

%Armed with these tools we study a solvable test-bed consisting of two harmonic oscillators coupled through a 
%weak exponential interaction.  The model is simple enough to admit: (i) an analytic early-time expansion 
%obtained from the moment method; (ii) an explicit semiclassical evaluation of $S_{\text{inst}}$; and (iii) a fully 
%numerical determination of $K(t)$ up to times $\omega_{1}t\simeq120$.  Throughout the parameter window 
%$0.05\lesssim g\lesssim0.20$ and $|\alpha|m^{2}\gtrsim12$ the instanton formula~\eqref{eq:intro-plateau} 
%reproduces the numerically extracted plateau height to within a few per-cent, and the plateau always exceeds 
%the envelope of the early-time oscillations—as expected when classical contributions cancel chaotically. Taken 
%together, these results turn Krylov complexity into a quantity that can be attacked with the full toolbox of 
%semiclassical field theory, opening a new analytic window on the late-time dynamics of quantum chaos.\\

\noindent
Armed with these tools, we study a solvable test-bed consisting of two harmonic oscillators coupled through a 
weak exponential interaction.  The model is simple enough to admit: (i) a numerical determination of the 
Lanczos coefficients and of the wave functions of the Krylov vectors at small $n$; (ii) a numerical early-time 
determination of $K(t)$ up to $t\cdot g\lsim 30$;  (iii) an analytic finite-time expansion
in perturbation theory, showing that classical contributions cancel chaotically;
(iv) an explicit semiclassical evaluation of $S_{\text{inst}}$ and of the formula 
(\ref{eq:intro-plateau}) up to $n\le 30$, and find no inconsistencies;. We explore different sets of 
parameters and initial states.\\

\noindent
Taken 
together, our proposals turn the Krylov complexity plateau into a quantity that can be attacked
 with the full toolbox of 
semiclassical field theory, opening a new analytic window on the late-time dynamics of quantum chaos.\\

\noindent
The remainder of the paper is organized as follows.  Section~\ref{sec:KC-review} summarizes the standard 
Lanczos construction and reviews the three canonical late-time regimes of Krylov complexity.  
Section~\ref{sec:path integral} develops the path integral representation and derives the instanton formula for 
late-time complexity.  Section~\ref{sec:toy} presents the coupled-oscillator model, combining analytic estimates 
with numerical data.  We conclude in Section~\ref{sec:conclusions} with a discussion of implications, open 
problems and possible experimental signatures.

\section{Krylov complexity - standard  formulation}\label{sec:KC-review}

\subsection{Review of Krylov complexity}

Given an hermitian Hamiltonian $H$ and starting state $|K_0\rangle$, it is a well-appreciated fact that the 
Lanczos algorithm \cite{Lanczos1950AnIM} generates an orthonormal basis $\{|K_n\rangle\}$ in which the 
Hamiltonian acquires a tridiagonal form,
\begin{eqnarray}
    H|K_n\rangle & = & a_n |K_n\rangle + b_n |K_{n-1}\rangle + b_{n+1}  |K_{n+1}\rangle \, ,
\end{eqnarray}
with $b_0 = 0$.  The coefficients $a_n, b_n$ are called the Lanczos coefficients and the orthonormal basis is 
called the Krylov basis.   Using this, the spread complexity \cite{Balasubramanian:2022tpr} of a time-evolved refernce state, $|\psi(t)\rangle$ 
can be defined as
\be
K(t)=\langle \psi(t)|\left(\sum_n n|K_n\rangle\langle K_n|\right)|\psi(t)\rangle=
\langle\psi(t)|\hat K|\psi(t)\rangle=\sum_n n |\psi_n(t)|^2\;,\label{Kpsin}
\ee
where $\hat{K}$ denotes the Krylov complexity operator defined in the Krylov basis.  The reference state at 
$t=0$ serves as the starting state for the Lanczos algorithm i.e. $|K_0\rangle = |\psi(0)\rangle$.  
%A natural generalization of this notion is the {\it spread complexity} introduced in \cite{Balasubramanian:2022tpr}, in which one replaces the weights $n$ with arbitrary coefficients $c_n$,
%\be
 %  C(t)=\langle \psi(t)|\left(\sum_n c_n |B_n\rangle\langle B_n|\right)|\psi(t)\rangle=\langle \psi(t)|\hat B|\psi(t)\rangle
%   =\sum_n c_n |\psi_n(t)|^2\,,
%\ee
%with the corresponding spread complexity operator $\hat{B}$. 
We adopt the notation of \cite{Haque:2022ncl,Chattopadhyay:2023fob} here.\footnote{Note that the 
Krylov complexity is reference state dependent, as shown explicitly for 2d CFTs in \cite{Kundu:2023hbk}.}  \\

\noindent
Alternatively, the Krylov basis can be defined, by its relation to the moments of the survival amplitude \cite{Balasubramanian:2022tpr}. 
Indeed, from the decomposition of the time-evolved reference state in the Krylov basis,
\be
|\psi(t)\rangle=\sum_n \psi_n(t)|K_n\rangle\,,
\ee
the {\em survival amplitude} may be defined as,
\be
S(t)\equiv \langle \psi(t)|\psi(0)\rangle=\langle \psi(0)|e^{\frac{i}{\hbar}Ht}|\psi(0)\rangle\equiv 
\langle K_0|e^{\frac{i}{\hbar}Ht}|K_0\rangle\,.   \label{SurvivalAmplitude}
\ee
It is then clear that the moments of this survival amplitude are related to the Krylov basis, since
\be
\mu_n=\left.\frac{d^n}{dt^n}S(t)\right|_{t=0}=\langle K_0|(\frac{i}{\hbar} H)^n|K_0\rangle\,.
\ee
Explicitly, the expansion coefficients may be obtained \cite{Beetar:2023mfn} in terms of the moments and Lanczos coefficients as
\begin{eqnarray}
\psi_n(t) & = & \sum_{m=0}^n k_{m,n} \  \overline{\mu_m}\;,    \nonumber \\
k_{m,n+1} & = & \frac{i k_{m-1,n} - a_n k_{m,n} - b_n k_{m,n-1}}{b_{n+1}}\;,  \nonumber \\
k_{0,0} & = & 1\;.
\end{eqnarray}
In appendix \ref{Lanczos2} we show how this may be reformulated in a way where the Lanczos coefficients are 
not needed\footnote{See also \cite{Beetar:2025mdz} for another approach that only requires the survival amplitude evaluated at discrete points in time}.\\

\noindent
A similar construction applies to operators rather than states 
\cite{Parker:2018yvk, Caputa:2021sib, Muck:2022xfc}. Associating the operator $\mathcal{O}$ with a state via
$|\mathcal{O}) \equiv \mathcal{O} |\psi\rangle$,
one defines the inner product
\begin{eqnarray}
    (\mathcal{O}_n | \mathcal{O}_m) = \langle \psi | \mathcal{O}_n^\dagger \mathcal{O}_m | \psi \rangle\,,
\end{eqnarray}
and constructs the Krylov basis $\{ |\mathcal{O}_n) \}$ via the Lanczos algorithm (there's more to be said about possible choices of inner product, but we won't get into that here). The time-evolved operator 
can then be expanded as
\be
  {\cal O}(t)\equiv \sum_n i^n \psi_n(t){\cal O}_n\;,\;\;
  \psi_n(t)=i^{-n}({\cal O}_n|{\cal O}(t))\,,
\ee
where at time $t$,
\be
   |{\cal O}(t))\equiv {\cal O}(t)|\psi\rangle.
\ee
At finite temperature, the inner product generalizes to mixed states as
\be
   ({\cal O}_n|{\cal O}_m)_\b\equiv \frac{1}{Z}\Tr\left[e^{-\b H}{\cal O}^\dagger_n {\cal O}_m\right]\;,
\ee
leading to
\be
   \psi_n(t)=i^{-n}({\cal O}_n|{\cal O}(t))_\b\,.
\ee
Alternatively, one may use the 2-point function,
\be
   C(t,\b)\equiv ({\cal O}|{\cal O}(-t))_\b\equiv \sum_n M_n\frac{(-it)^n}{n!}\,,
\ee
with moments
\be
   M_n= \frac{1}{(-i)^n}\left.\frac{d^n C(t,\b)}{dt^n}\right|_{t=0}
   = \int_{-\infty}^{+\infty}\frac{d\omega}{2\pi}\omega^n f(\omega)\,,
\ee
where $f(\omega)$ is the Fourier transform of $C(t,\b)$,
\be
   f(\omega)= \int_{-\infty}^{+\infty}dt \;e^{i\omega t}\,C(t,\b)\,.
\ee

\noindent
Yet another viewpoint employs the thermofield double (TFD) purification of the thermal density matrix at 
temperature $T$ (with $\beta = 1/(k_B T)$), 
\be
|\psi\rangle\rightarrow |0,\b\rangle\equiv  \frac{\sum_{n=\tilde n}
e^ {-\frac{\b E_n}{2}}|n\rangle\otimes |\tilde n\rangle}{
\left(\sum_m e^ {-\b E_m}\right)^ {1/2}}\;,
\ee
as a reference state to define Krylov complexity \cite{Balasubramanian:2022tpr}. The TFD state 
$|0,\beta\rangle$ or $|\psi_\beta\rangle$ can be time evolved with the Hamiltonian acting separately on the left 
and right Hilbert spaces,
\be
   |\psi_\b(t)\rangle=e^{-\frac{i}{\hbar}Ht}|\psi_\b\rangle=|\psi_{\b+2it}\rangle\,.
\ee
Then the analytically continued partition function, 
\be
   Z_{\b-it}=\sum_n e^{-(\b-\frac{i}{\hbar}t)E_n}\,,
\ee
defines the {\em spectral form factor (SFF)},
\be
   {\rm SFF}_{\b-it}\equiv\frac{|Z_{\b-it}|^2}{|Z_\b|^2}\,.
\ee
Correspondingly, the survival amplitude of the time-evolved TFD state is
\be
   S_\b(t)=\langle \psi_{\b+2it}|\psi_\b\rangle=\frac{Z_{\b-it}}{Z_\b}\Rightarrow  {\rm SFF}_{\b-it}=|S_\b(t)|^2\,,
\ee
and its moments again connect back to the Krylov basis.

\subsection{Some simple models and a test of quantum chaos}

To better understand the possible behaviour of $K(t)$, let us consider first some solvable models following 
closely the treatment in\cite{Balasubramanian:2022tpr}.

\begin{enumerate}
    
\item {\em A particle on the $Sl(2,\mathbb{R})$ group manifold:} The relevant algebra is
\be
   [L_0,L_\pm]=\mp L_\pm\;,\;\; [L_+,L_-]=2L_0\;,
\ee
with a general Hamiltonian of the form
\be
\frac{H}{\hbar}=\a(L_++L_-)+\gamma L_0+\delta \one.
\ee
The representation space is labeled by the scaling dimension $h$ through the states $|h,n\rangle$, satisfying
\bea
   L_0|h,n\rangle&=& (h+n)|h,n\rangle\;,\cr
   L_+|h,n\rangle&=&\sqrt{(n+1)(2h+n)}|h,n+1\rangle\;,\cr
   L_-|h,n\rangle&=& \sqrt{n(2h+n-1)}|h,n-1\rangle\;.
\eea
If the evolution begins from the highest weight state, it is natural to identify \hbox{$|h,n\rangle=|K_n\rangle$} 
as the Krylov basis. A useful point of contact is provided by the harmonic oscillator, whose partition function 
and survival amplitude are
\be
   Z(\b)= \frac{e^{\b\hbar\omega}}{e^{\b\hbar\omega}-1}\Rightarrow S_\b(t)
   =\frac{1-e^{-\b\hbar\omega}}{1-e^{-(\b-\frac{i}{\hbar}t)\hbar \omega}}\,.
\ee
One can match this to the motion on the $SL(2,\mathbb{R})$ group manifold by choosing the parameters
\be
   h=\frac{1}{2}\;,\;\;\; \gamma=\frac{\omega}{\tanh (\b\hbar\omega/2)}\;,\;\;
   \delta=-\frac{\omega}{2}\;,\;\;\;
   \a=\frac{\omega}{2\sinh(\b\hbar\omega/2)}\,.\label{harmoscmatch}
\ee
Accordingly, the initial thermofield double (TFD) state is simply
\be
   |\psi_\b\rangle=|K_0\rangle=|h=1/2\rangle\;,
\ee
and the associated Krylov complexity at temperature $T$ is found to be
\be
   K(t)=\frac{\sin^2(\omega t/2)}{\sinh^2(\b\hbar\omega/2)}\;,
\ee
which exhibits oscillatory behavior expected of the harmonic oscillator.

If instead one considers the inverted harmonic oscillator, obtained by analytically continuing 
$\omega \to -i \omega_i$, the complexity becomes 
\be
   K(t)=\frac{\sinh^2(\omega_i t/2)}{\sin^2(\b\hbar\omega_i/2)}\;,
\ee
which grows without bound, reflecting the dynamical instability of the inverted system.

\item {\em A particle moving on the $SU(2)$ group manifold}: The associated Lie algebra is given by
\be
   [L_0,L_\pm]=\pm L_\pm\;,\;\; [L_+,L_-]=2L_0\;,
\ee
with a natural Krylov basis for spin $j$ states chosen as
$|K_n\rangle = |j, -j + n\rangle$,
so that the initial reference state at zero temperature is
$|K_0\rangle = |j, -j\rangle$. Under the time evolution,
\be
   |\psi(t)\rangle=e^{-\frac{i}{\hbar}Ht}|j,-j\rangle\,,
\ee
the zero temperature Krylov complexity evaluates to 
\be
   K(t)=\frac{2j}{1+\frac{\gamma^2}{4\a^2}}\sin^2\left(\a t\sqrt{1+\frac{\gamma^2}{4\a^2}}\right)\,,
\ee
which is manifestly oscillatory. This behavior reflects both the integrability and the finite-dimensional 
nature of the $SU(2)$ Hilbert space.

\item {\em A particle moving on the Heisenberg-Weyl group:} In this setting, the generators are identified as
\begin{eqnarray}
    L_+ = a^\dagger,
    \quad
    L_- = a,
    \quad
    L_0 = N = a^\dagger a\,,
\end{eqnarray}
with parameters relabeled according to $\alpha \rightarrow \lambda$, and $\gamma \rightarrow \omega$, 
so that the Hamiltonian takes the form 
\be
   \frac{H}{\hbar}=\lambda(a^\dagger +a)+\omega N+\delta \one\;,
\ee
Choosing the vacuum $|\psi\rangle=|K_0\rangle=|0\rangle$, as the reference state, the time-evolved state is
\be
   |\psi(t)\rangle=e^{-\frac{i}{\hbar}Ht}|0\rangle\,.
\ee
The Krylov complexity at zero temperature is then found to be
\be
   K(t)=\frac{4\lambda^2}{\omega^2}\sin^2\left(\frac{\omega t}{2}\right)\,,
\ee
which is again oscillatory, reflecting the integrable and bounded structure of the Heisenberg–Weyl group.
\end{enumerate}

\noindent
Finally, for quantum chaotic systems — in particular the {\it Schwarzian theory, random matrix models,} 
and the {\it SYK model} — one observes that the Krylov complexity $K(t)$ approaches a plateau at late 
times \cite{Balasubramanian:2022tpr}. In these cases, the behavior has typically been studied through 
numerical simulations. However, it has been argued more recently that such a plateau can also appear in 
certain classical or even integrable systems. As a result, a more robust diagnostic of chaotic dynamics may be 
the peak preceding the plateau, as well as the spectrum of fluctuations around the plateau itself 
\cite{Balasubramanian:2023kwd, Espanol:2022cqr, Camargo:2024deu, Baggioli:2024wbz}.

\section{Krylov complexity - path integral formulation}\label{sec:path integral}

\noindent
Having reviewed the standard formulation of Krylov complexity, we now give an alternative, 
path integral formulation that will allow us to probe the late time behavior of $K(t)$ better. 
In particular, we show how the Euclidean path integral can be used to give a value for the complexity plateau.

\subsection{Path integral}

As noted earlier, the Krylov complexity can be expressed in terms of $\psi_n(t)$ via the relation 
\eqref{Kpsin}. Recall also that
\be
   \psi_n(t)=\langle K_n|\psi(t)\rangle=\langle K_n|e^{-\frac{i}{\hbar} Ht}|\psi\rangle\,.
\ee
Assuming the quantum system has $N$ degrees of freedom, we can insert a resolution of the identity 
in position space using the complete basis of $N$ positions $\{x_i\}_{i=1,\dots,N}$,
\be
\left(\prod_{i=1}^N\int dx_i\right) |\{x_i\}\rangle\langle\{x_i\}|=\one\,.
\ee
Inserting two such completeness relations on either side of the propagator yields
\be
   \psi_n(t)=\left(\prod_{i}\prod_{j}\int dx_{i}\int dx'_{j}\right)
   \; K_n^*(\{x_{i}\})\psi(\{x'_{j}\})\langle \{x_{i}\}|e^{-\frac{i}{\hbar}Ht}|\{x'_{j}\}\rangle\;.
\ee
The rightmost matrix element, interpreted as a transition amplitude, can be expressed as a path 
integral via the standard representation,
\be
   \langle \{x_i(t)\}|\{x'_j(0)\}\rangle=\langle\{x_i\}|e^{-\frac{i}{\hbar}Ht}|\{x'_j\}\rangle=
   \int _{\{x_i(t)\};\{x'_j(0)\}}{\mathcal D} \{x_m(t')\}e^{\frac{i}{\hbar}S[\{x_m(t')\}]}\,.
\ee
Consequently, the Krylov complexity takes the form
\bea
   K(t)&=&\sum_n n\left| \left(\prod_{i=1}^N\prod_{j=1}^N\int dx_{i}\int dx'_{j}\right)
   \; K_n^*(\{x_{i}\})\psi(\{x'_{j}\})\right.\times\cr
   &&\left.\times\int _{\{x_i(t)\};\{x'_j(0)\}}{\cal D} \{x_m(t')\}e^{\frac{i}{\hbar}S[\{x_m(t')\}]}\right|^2\,,
\eea
highlighting the path integral structure embedded in $\psi_n(t)$. 

\subsection{Instantonic plateau}

Our aim is to understand the late-time behavior of the Krylov complexity $K(t)$ for a (quantum) chaotic 
system\footnote{Since, as noted earlier, a plateau can also appear in certain non-chaotic situations, the 
reasoning, and the resulting plateau formula we obtain, may in fact apply more generally.}. Toward this 
end, consider first a simple propagator $\langle x(t) | x’(0) \rangle$ without integration over the endpoints. In 
that case, if a classical path connects the initial and final points, the path integral is dominated by the classical 
action, giving $\langle x(t)|x'(0)\rangle \simeq e^{\frac{i}{\hbar}S_{\rm cl}[x(t);x'(0)]}$.\\

\noindent
If however no classical path exists, the 
dominant contribution comes from an {\it instanton solution} — a saddle point of the Euclidean (Wick-rotated) 
action $S_E = -i S$, with trajectories in imaginary time $t_E$. For a single particle with potential $V(x)$, the 
instanton equation of motion is
\be
   \frac{d^2 x}{dt_E^2}-\frac{dV}{dx}=0\Rightarrow V_E(x)=-V(x)\,.
\ee
In other words, the Euclidean dynamics corresponds to motion in the inverted potential $V_E$. An instanton 
trajectory connecting two minima of $V$ (equivalently, two maxima of $V_E$) — say, $x^{(1)}$ and $x^{(2)}$ 
— with vanishing initial velocity, behaves asymptotically as
$t_E \to -\infty \quad \text{at } x^{(1)},
\quad
t_E \to +\infty \quad \text{at } x^{(2)}$,
meaning the total real time satisfies $i t = T_E = 2 t_E \to \infty$. The trajectory passes through an intermediate 
saddle corresponding to a maximum of $V$ (or a minimum of $V_E$).\\

\noindent
Summarizing, if a classical real-time trajectory exists, it dominates the path integral. If no classical trajectory is 
possible, a complex, imaginary-time instanton controls the amplitude. For arbitrary finite initial and 
final positions (with a finite initial velocity), as discussed in Landau \& Lifshitz \cite{LandauLifshitzvol3} (see 
also \cite{Nastase:2022}), the instanton is defined by an intermediate saddle associated with a pole of $V(x)$ 
in the complex plane. This generalizes the real maximum of $V$: the path proceeds from $x_i$ to a complex 
saddle $x_0$, then on to $x_f$. The projection of this saddle on the real line is the point of maximal $V$, or 
minimal $V_E$.\\

\noindent
However, in our case we are interested in: {\it (i)} the $t\rightarrow\infty$ limit; {\it (ii)}  integrating over both 
$x(t)$ and $x'(0)$; and 
{\it (iii)} a chaotic quantum system. In classical chaos, small changes in $x’(0)$ lead to exponentially divergent differences in $x(t)$. In quantum 
terms, this translates into a rapidly varying action $S$, so the phase factor $e^{\frac{i}{\hbar}S}$ becomes wildly oscillatory. 
When we integrate over initial and final positions, these wildly varying phases should average out to (almost) 
zero for contributions of real-time classical paths — especially for $t \to \infty$, covering essentially the entire 
chaotic phase space.\\

\noindent
This leads us to conjecture that {\it in quantum chaotic systems with integration over endpoints, the sum over classical 
real-time paths is effectively negligible at late times, provided the instanton action is not too large (so that $e^{-\frac{1}{\hbar}
S_E}$ is not suppressed with respect to the remainder of the sum over classical paths). 
As a result, the path integral is dominated by imaginary-time 
(instanton) trajectories describing genuine quantum transitions.}\\

\noindent
What kinds of potentials $V(x)$ are relevant? They must be bounded from below to avoid instabilities — 
otherwise, as shown in earlier examples, $K(t)$ would grow unbounded. A single minimum would allow the 
particle to simply sit there at long times, leading to a classical solution dominating the path integral. For chaotic 
systems, however, classical solutions should not dominate as their phase contributions should average to 
nearly zero.
Consequently, the relevant quantum chaotic systems should have at least two (local) minima $x_a$, with 
$a = 1,2,\dots$. This in turn allows for instanton solutions connecting them. At finite time and general positions, 
the instanton path will not exactly minimize the Euclidean action, but once we integrate over all initial and final 
positions, the path integral will be sharply peaked on the configurations with minimal $S_E$ — essentially a 
saddle-point approximation. Concretely,
\be
   \int dx_i \int dx_f e^{-\frac{1}{\hbar}S_E}\simeq \sum_{x_1,x_2={\rm minima}}e^{-\frac{1}{\hbar}S_E(x_i=x_1,x_f=x_2)}\,.
\ee
Hence the long-time wavefunction components are given by
\bea
   \label{eq:conjecture psi}
   \psi_n(t\rightarrow\infty) &\simeq& \sum_{\{x_1\},\{x_2\}={\rm minima}}K_n^*(\{x_2\}) \psi(\{x_1\})
   \langle \{x_2(t\rightarrow\infty)\}|\{x_1(0)\}\rangle\cr
   &\simeq& \sum_{\{x_1\},\{x_2\}={\rm minima}}K_n^*(\{x_2\}) \psi(\{x_1\}) e^{-\frac{1}{\hbar}S_E({\rm instanton}; x_1,x_2)}\,,
\eea
and the Krylov complexity reaches a plateau,
\bea
   K(t\rightarrow\infty)
   &=&\sum_n n|\psi_n(t\rightarrow\infty)|^2\cr
   &\simeq& \sum_n n \left| \sum_{\{x_1\},\{x_2\}
   ={\rm minima}}K_n^*(\{x_2\}) \psi(\{x_1\}) e^{-\frac{1}{\hbar}
   S_E({\rm instanton}; x_1,x_2)}\right|^2\,,    \label{instantonApprox}
\eea
which is a purely {\it topological} quantity, i.e., dependent only on the minima of the potential. In the simplest 
case with just two minima and a single instanton, one obtains
\be
   K(t\rightarrow\infty)
   \simeq e^{-\frac{2}{\hbar}S_E({\rm instanton})}\sum_n n \left|K_n^*(\{x_2\})\psi(\{x_1\})+K_n^*(\{x_1\})
   \psi(\{x_2\})\right|^2\,,\label{compplateau}
\ee
providing a remarkably simple formula for the plateau, requiring only knowledge of the wavefunction of the 
reference state $\psi$, the Krylov basis $K_n$ evaluated at the minima, and the instanton action!

\subsection{Previous formulae for the plateau value}

In the interest of completeness, we should now comment on previous formulae for the plateau value of Krylov 
complexity. For example, in \cite{Balasubramanian:2024lqk}, the authors studied the behavior of the plateau in 
the limit where the Hilbert space dimension $L$ goes to infinity, with $n \to \infty$ and $L \to \infty$ while 
keeping
$x = \frac{n}{L} \in [0,1)$
fixed. In this limit, the Krylov complexity simplifies to the scaled average of $x$ in the state $|\psi(t)\rangle$,
\be
   \frac{C(t)}{L}=\sum_n \frac{n}{L}|\psi_n(t)|^2\rightarrow \langle x\rangle\,.
\ee
The Lanczos coefficients also admit a continuum description as functions $a(x)$ and $b(x)$, 
from which we define 
\be
   E_L=a(x)-2b(x)\;,\;\; E_H=a(x)+2b(x)\,.
\ee
Then, as shown in \cite{Balasubramanian:2023kwd}, at finite temperature the long-time stationary distribution 
of the ThermoField Double (TFD) state in the Krylov basis becomes
\be
   |\psi_n(t\rightarrow \infty, L\rightarrow \infty)|^2\rightarrow 
   \omega(x)=\frac{L}{Z_\b}\int dE\frac{e^{-\b E}}{\pi\sqrt{4(b(x))^2-(E-a(x))^2}}\,,
\ee
leading to a plateau value $C_{\rm max}=L\int_0^1 dx\; x\;\omega(x)$, which, at low temperatures, can be 
approximated as
\bea
   \frac{C_{\rm max}}{L}=
   \langle x_{\rm max}\rangle&\simeq &\int dx\, \frac{Lx e^{-\b E_L(x)}}{Z_\b\sqrt{\b \pi (E_H(0)-E_L(0))}}\cr
   &=&\frac{L}{Z_\b\sqrt{\b \pi (E_H(0)-E_L(0))}}\int dE_L\; x(E_L)\frac{1}{E'_L(x)}e^{-\b E_L}\,.
\eea

\noindent
There are a few observations to be made at this point. First, this formula explicitly assumes finite temperature. 
Second, it relies on the $L \to \infty$ limit, neither of which are required in our framework. Third, their 
expression is formulated in terms of the continuum Lanczos coefficients $a(x)$ and $b(x)$, which is not the 
perspective we are taking. Moreover, as argued in \cite{Balasubramanian:2023kwd}, this form of the plateau is 
not uniquely tied to chaotic dynamics, since it can also appear in certain integrable systems. This might be due 
to the role of finite temperature, although our own formula also does not explicitly enforce chaos; rather, chaos 
enters implicitly, through the requirement that the average over classical real-time paths vanishes, leaving only 
the instanton contributions. The existence of a plateau formula independent of chaoticity does not necessarily 
mean that it applies equally to both chaotic and non-chaotic systems.\\

\noindent
Further, in \cite{Espanol:2022cqr, Camargo:2024deu} it was demonstrated that the plateau value — and 
whether its presence distinguishes integrability from chaos — can depend on the choice of initial state, the so-
called “seed”. For this reason, it was argued in \cite{Craps:2024suj} that one should instead consider a 
multi-seed Krylov complexity, averaging over the different seeds, to obtain a robust test of chaos. In our 
framework, this issue is connected to whether the initial state and the Krylov basis have a significant overlap with the minima 
involved in the instanton transitions. In other words, the values of the relevant wavefunctions appearing in our formula are all dependent on the initial state.

\subsection{Consistency with spin chain model dualities}

A quantum-mechanical instanton (that is, an instanton in $(0+1)$-dimensions) can be interpreted as a time-independent kink solution in a $(1+1)$-dimensional quantum field theory. By making the substitutions $t 
\rightarrow x$ and $X \rightarrow \Phi$, so that $X(t) \mapsto \Phi(x)$, one obtains a field $\Phi(x)$ that 
interpolates between two minima of the potential $V(\Phi)$ through a maximum — precisely the static kink 
solution. A kink, which is a classical solution in one spatial dimension, itself acts as the boundary value of a 
vortex (the static solution in two spatial dimensions). In this sense, the vortex plays the role of an “instanton” 
for the kink: if its second spatial dimension is interpreted\footnote{Pushing this pattern one dimension higher, a 
monopole (static solution in 3D) serves as an instanton interpolating between vacua with different vortex 
numbers, and a true four-dimensional Euclidean instanton interpolates between vacua with different monopole 
numbers.} as Euclidean time, then the vortex describes a transition between kink number zero at $t = -\infty$ 
and kink number one at $t = +\infty$.\\

\noindent
In this sequence of topological solutions, each interpolates between vacuum configurations that break some 
symmetry. This is most transparent in the kink case, which connects two minima related by a $\mathbb{Z}_2$ 
symmetry of the action ($\Phi \rightarrow -\Phi$ or equivalently $X \rightarrow -X$), given that the potential is 
even, $V(\Phi) = V(\Phi^2)$.\\

\noindent
Consider then the duality between the Ising spin chain in a magnetic field and the fermionic Kitaev chain, 
related via the Jordan–Wigner transformation, as studied in \cite{Murugan:2024ory}. The Jordan–Wigner map 
expresses the fermionic operators $c_i, c_i^\dagger$ in terms of nonlocal strings of spin operators:
\be
   c_j=\frac{1}{2}\left(\prod_{i<j}\sigma_i^z\right)\sigma_j^-\,,
\ee
with commutation relations
\be
   \{c_i,c_j^\dagger\}=\delta_{ij}\;,\;\;
  [\sigma_j^\mu,\sigma_k^\nu]=2i\epsilon^{\mu\nu\rho}\delta_{jk}\sigma^\rho_k\,.
\ee
However, this construction necessarily singles out the site $j=1$ on the spin chain.\\

\noindent
In \cite{Murugan:2024ory}, it was shown that in the open chain, where no nontrivial boundary terms appear, the 
Ising and Kitaev models remain identical, even in their Krylov complexity. In contrast, for the closed chain, the 
Jordan–Wigner transformed Ising Hamiltonian acquires nontrivial boundary terms because the map breaks 
translational invariance at site $j=1$, so that the Ising and Kitaev models no longer coincide. Physically, the 
spin chain can be understood as a discretized worldsheet — just like in the BMN string construction 
\cite{Berenstein:2002jq}. Each fermionic oscillator $c_i, c_i^\dagger$ corresponds to a coordinate $X_i(t)$, 
which in the continuum becomes $X(y,t)$ for the continuous spatial coordinate $y$, or equivalently, under the 
substitution $X \to \Phi$ and $t \to x$, $\Phi(y,x)$.\\

\noindent
An interesting observation in \cite{Murugan:2024ory} is that only operators which mix the fermionic parity,
\begin{eqnarray}
    P_F = (-1)^{ \sum_i c_i^\dagger c_i } = (-1)^{ \sum_i n_i }\,,
\end{eqnarray}
corresponding in the field language to
$X(y_i,t) \rightarrow -X(y_i,t)$ for all $y_i$,
or equivalently $\Phi(y,x) \rightarrow -\Phi(y,x)$, produce a significant boundary contribution to the Krylov 
complexity. This contribution to the plateau differs markedly between the two sides of the duality, and might 
seem surprising in the large-$N$ (continuum) limit. However, this is consistent with our instanton-based 
argument: generalizing from $(0+1)$-dimensional quantum mechanics for $X(t)$ to a $(1+1)$-dimensional 
worldsheet string for $X(y,t)$, the Krylov complexity should still be controlled by Euclidean instanton solutions. 
In this two-dimensional Euclidean picture (time and the discrete site $j$), these instantons correspond to 
“vortex instantons” interpolating between “no kink” and “kink” vacua at the spatial boundaries ($j=1$ and 
$j=L$). We have already seen that the kink vacua mix the $\mathbb{Z}_2$ (parity) sectors.
Therefore, the observed difference in the plateau of Krylov complexity, depending on whether the operator 
mixes parity or not, is consistent with our picture. The instanton transitions involve changes in the 
$\mathbb{Z}_2$ parity vacua, explaining why operators sensitive to these transitions acquire a boundary-dependent contribution to the plateau.

\subsection{Relation to general transition amplitudes and energy eigenfunctions}

The conjecture introduced in the previous subsections relates the long-time behavior of the amplitudes 
$\psi_n(t)$ to instanton contributions, and was primarily formulated in the context of Krylov complexity. 
However, the logic underlying the derivation is more general. Rather than computing
$\psi_n = \langle K_n | e^{-\frac{i}{\hbar} H t} | \psi \rangle$, we could instead start from a general transition 
amplitude
$\langle \phi_2 | e^{-\frac{i}{\hbar} H t} | \phi_1 \rangle$,
and proceed through analogous steps to arrive at a more general result valid in the large-$t$ limit,
\begin{equation}
    \bra{\phi_2} e^{-\frac{i}{\hbar}Ht}\ket{\phi_1}
    \simeq \sum_{x,x'} \phi_2^*(x') \phi_1(x) e^{-\frac{1}{\hbar}S_{E}(instanton;x,x')},\label{moregeneral}
\end{equation}
Here, $\phi_1$ and $\phi_2$ are arbitrary quantum states in position space, and the sum is over positions $x$ 
and $x'$ near the relevant minima of the potential. This suggests that, in systems where our assumptions 
hold, any long-time transition amplitude is instanton-dominated and can be approximated if the initial and final 
states are known.\\

\noindent
To explore one immediate consequence of this formula, consider the case of a trivial transition amplitude 
between identical energy eigenstates $|n\rangle$ of the Hamiltonian $H$,
\begin{equation}
    | \bra{n} e^{-\frac{i}{\hbar}Ht} \ket{n}|^2 = |e^{-\frac{i}{\hbar}E_nt}\langle n|n\rangle|^2=1.
\end{equation}
In this case, (\ref{moregeneral}) gives
\begin{equation}
    \left|\sum_{x,x'} n^*(x') n(x) e^{-\frac{1}{\hbar}S_{E}(instanton;x,x')}\right|^2 \approx 1\,,
\end{equation}
where $n(x)$ is the position-space wavefunction of the eigenstate $|n\rangle$. For the class of systems we are 
considering, it is generally not feasible to find analytic expressions for the eigenfunctions $n(x)$, but this 
prediction could be tested numerically for specific models.\\

\noindent
To gain further insight, assume there is a single 
dominant instanton and that the wavefunction $n(x)$ is real-valued, which is possible if the Hamiltonian is 
Hermitian and time-independent. Then the approximation simplifies to
\begin{align}
    &4|n(x')|^2 |n(x)|^2 e^{-\frac{2}{\hbar}S_{E}(instanton;x,x')} \simeq 1\Rightarrow\nonumber \\
    &4|n(x')|^2 |n(x)|^2 \simeq e^{\frac{2}{\hbar}S_{E}(instanton;x,x')}.
\end{align}
This relation implies that the eigenfunction $n(x)$ must have sharply peaked amplitude near the minima of the 
potential — in fact, exponentially large peaks if the instanton action is large. Because the wavefunction must 
be normalized, such tall peaks must also be correspondingly narrow. Moreover, increasing the separation 
between the minima typically increases the instanton action $S_E$,  making the peaks even higher and 
narrower.\\

\noindent
However, a subtle tension arises here. On one hand, the semiclassical approximation requires that 
$S_E$ be large so that the path integral is dominated by the saddle (instanton). On the other hand, our 
instanton-dominance conjecture — which assumes the suppression of classical real-time paths — holds only 
when $S_E$ is not too large, such that instanton contributions remain unsuppressed. This requires a delicate 
balance:
$S_E \gg \hbar$ for the semiclassical approximation, $S_E \lesssim \text{moderate}$ for instanton dominance. The overlap 
of these two conditions defines a narrow but nontrivial window where both approximations are simultaneously 
valid. Exploring this intermediate regime is crucial for validating the instanton-based estimates of long-time 
transition amplitudes and Krylov complexity plateaus.

\subsection{Classically chaotic potentials without instanton solutions}

We now comment on the implications of our proposal for classical chaotic systems that do not admit instanton 
solutions. A simple and well-studied example is the model analyzed in \cite{Akutagawa:2020qbj}, which 
consists of a two-dimensional harmonic oscillator with an additional quartic coupling of the form $g^2 x_1^2 
x_2^2$. This model arises as a dimensional reduction of the BFSS matrix model and serves as an interesting 
toy model for studying classical and quantum chaos.
It has been established that this system exhibits two distinct classical dynamical regimes: a low-energy regime 
with regular (integrable-like) motion, and a high-energy regime characterized by classical chaos. Moreover, 
quantum signatures of chaos — such as level repulsion in energy spectra and the exponential growth of out-of-time-ordered correlators (OTOCs) — have been observed in certain parameter regimes. However, these 
diagnostics are sensitive to the specific energy window, and care must be taken when interpreting them across 
the two regimes.\\

\noindent
From the standpoint of our proposal, this model falls outside the class of systems under consideration. As can 
be easily verified, the potential $V(x_1, x_2) = \tfrac{1}{2}(x_1^2 + x_2^2) + g^2 x_1^2 x_2^2$ has a single 
global minimum at $x_1 = x_2 = 0$. Consequently, there are no instanton solutions connecting distinct minima, 
and as we argued earlier, in such cases the path integral remains dominated by classical real-time trajectories, 
even when the classical system is chaotic. The presence of a regular low-energy regime violates one of our 
key assumptions — namely, the absence of integrable sectors — which further disqualifies this system from 
the domain of applicability of our framework. For this reason, we turn instead to a different toy model with a 
modified coupling structure between the two oscillators, which is better suited to exploring the implications of 
our instanton-based framework.

%One of the key points in the demonstration of our proposal is the large times cancellation of the contribution of classical trajectories in the semiclassical approximation for the path integral. One should expect this to happen regardless of the existence of instanton solutions. We could take this to mean that the plateau can only be reached in the presence of instantons. However, what would happen to complexity at late times in their absence if the system is classically chaotic? Alternatively, we could assume that maybe these systems can still reach a plateau, but then they sit outside the scope of our proposal and their behavior is explained by something unaccounted for.

\section{Toy model analysis}\label{sec:toy} 

In order to test these ideas, we would like to consider a simple toy model. 

\subsection{The model}

A minimal toy model compatible with our assumptions consists of two harmonic oscillators coupled via an 
exponential interaction term suppressed by a small coupling (although we will consider also 
larger couplings). The Hamiltonian is given by
\be
   H=\frac{p_1^2+p_2^2}{2m}+m\frac{\omega_1^2x_1^2}{2}+m\frac{\omega_2^2x_2^2}{2}
   +g\hbar(\omega_1+\omega_2)e^{\a m^2x_1x_2}\,,\label{coupHamexp}
\ee
with $g > 0$. The idea is as follows: for suitable choices of parameters (and initial states), the exponential 
interaction term is negligible at early times, and the system behaves approximately as two decoupled harmonic 
oscillators. Over time, however, the exponential coupling grows in significance, ultimately introducing strong 
nonlinearities and potentially chaotic dynamics — as the potential becomes highly sensitive to small variations 
in position.\\

\noindent However, for use in the standard Lanczos algorithm for the early times,
we could consider a slightly modified version of the model, in terms of $a_i,a_i^\dagger$,
$i=1,2$, namely
\be
H=\hbar\omega_1(a_1^\dagger a_1+1/2)+\hbar \omega_2(a_2^\dagger a_2+1/2)
+g \hbar(\omega_1+\omega_2)e^{\a (a^\dagger_1a_2+a^\dagger_2 a_1)}\;, \label{coupHamexpMod}
\ee
where (dropping $m$ and $\omega$) $x\sim (a+a^\dagger)/\sqrt{2}$ relates to (\ref{coupHamexp}).
%In this form we can check explicitly that, starting with the vacuum state, $|\psi(0)\rangle=|0\rangle$, 
%the exponential coupling starts small, then grows with time.\\
In appendix \ref{ModifedToyModelAppendix} we present some analysis on this modified model and show 
that it provides at least circumstantial evidence for our proposal.\\
%We will consider two versions of this model. One version, for the use at early times, with commutators, 
%is written in terms of oscillators $a_i,a_i^\dagger$, $i=1,2$, with Hamiltonian
%\be
%H=\hbar\omega_1(a_1^\dagger a_1+1/2)+\hbar \omega_2(a_2^\dagger a_2+1/2)
%+g \hbar(\omega_1+\omega_2)e^{\a (a^\dagger_1a_2+a^\dagger_2 a_1)}\;,
%\ee
%with $g\ll 1$.

%Another version, for use in the path integral, is written in phase space %$x_i,p_i$, with Hamiltonian
%\be
%H=\frac{p_1^2+p_2^2}{2m}+m\frac{\omega_1^2x_1^2}{2}+m\frac{\omega_2^2x_2^2}{2}
%+g\hbar(\omega_1+\omega_2)e^{\a m^2x_1x_2}
%\ee
%with $g\ll 1$.

%Note that, for going from one form to the other, we can write (dropping the factors of $m$ and $\omega$ 
%for simplicity) $x=(a+a^\dagger)/\sqrt{2}$.

%{\em relation between the two, regularization issues}

%In principle, dealing numerically with the above via the Lanczos algorithm would lead to regularization issues, but we describe in Appendix A a modification of the algorithm that avoids it, without the need to compute the Lanczos coefficients, and in Appendix B we show a numerical implementation in position space for the second version of the model.
\noindent
To align with our earlier assumptions and allow for instanton solutions, we require the potential to have at least 
two distinct minima. Since the potential is positive definite, it necessarily has minima; the question is whether 
more than one such minimum exists. These can be found by solving the stationarity conditions,
\bea
   &&\frac{1}{m}\frac{\d V}{\d x_1}=\omega_1^2 x_1+\a mx_2 g\hbar (\omega_1+\omega_2)
   e^{\a m^2 x_1 x_2}=0\,,\nonumber\\
   \\
   && \frac{1}{m}\frac{\d V}{\d x_2}=\omega_2^2 x_2+\a mx_1 g\hbar(\omega_1+\omega_2)e^{\a m^2 x_1
   x_2}=0\;.\nonumber 
\eea
Clearly, one solution is the trivial minimum at $x_1 = x_2 = 0$. To find additional extrema, consider the non-
trivial solutions satisfying
%\bea
%   \omega_1^2x_1^2&=&\omega_2^2x_2^2\Rightarrow x_2=\pm x_1\frac{\omega_1}{\omega_2}\;\;
%   {\rm and}\;\; x_1\left[\pm \omega_2\omega_1+\a m g\hbar(\omega_1+\omega_2)e^{\pm \a 
%   m^2 x_1^2\frac{\omega_1}
%   {\omega_2}}\right]=0\,,\nonumber\\
%   \Rightarrow x_1^2&=& \pm  \frac{\omega_2}{\a m^2\omega_1}\ln \mp\frac{\omega_2\omega_1}{\a
%   mg\hbar(\omega_1+\omega_2)}\, .
%\eea
\begin{eqnarray}
   x_2 & = & \pm x_1\frac{\omega_1}{\omega_2}    \nonumber \\
   0 & = & \left[\pm \omega_2\omega_1+\a m g\hbar(\omega_1+\omega_2)e^{\pm \a 
   m^2 x_1^2\frac{\omega_1}
   {\omega_2}}\right],\nonumber \\
   \Rightarrow x_1^2&=& \pm  \frac{\omega_2}{\a m^2\omega_1}\ln \mp\frac{\omega_2\omega_1}{\a
   mg\hbar(\omega_1+\omega_2)}\, .
\end{eqnarray}
%\begin{eqnarray}
%x_1 x_2 & = & \frac{1}{\alpha m^2} \log| \frac{\omega_1 \omega_2 }{\alpha m g \hbar (\omega_1 + \omega_2) }    |   \nonumber \\
%\frac{x_1}{x_2} & = & \pm \frac{\omega_2}{\omega_1}
%\end{eqnarray}
For $\a>0$ the lower sign is relevant which means that, in order to have real solutions, we need
\be
\label{eq:potential condition}
\frac{\omega_2\omega_1}{\a
mg\hbar(\omega_1+\omega_2)}<1\;.
\ee
If the condition is met (and if the matrix of second derivatives has positive eigenvalues, see subsection below), 
then we have a local maximum  at $x_1 = x_2 = 0$ and two minima at
%Note that if the conditions are not met, the only minimum is $x_1=x_2=0$, so the solution is for a $x_1(t)=x_2(t)=0$ for all times, but that is a {\em classical} solution, so that was already excluded, as we already commented. %\textcolor{red}{But then, really, 
%this system is then not 
%of the chaotic type, with a plateau (so the classical solutions do not cancel out in the end), 
%but rather of the integrable type, with an oscillating complexity (the "plateau"
%is negligible).}
\be
x_1=\pm \sqrt{\frac{\omega_2}{|\a| m^2 \omega_1}}\sqrt{\ln \frac{|\a|mg\hbar (\omega_1+\omega_2)}
{\omega_1
\omega_2}}\;,\;\; x_2=-\sgn(\a)x_1\frac{\omega_2}{\omega_1}\;.
\ee
The existence of these additional minima implies that the model admits nontrivial instanton solutions 
interpolating between them. Since these instantons have Euclidean action satisfying $e^{-S_E} < 1$, they 
contribute to the long-time transition amplitudes and therefore also to the Krylov complexity plateau.\\

\noindent
Our analysis of this model will involve three steps.  First, we will compute the Lanczos coefficients and confirm that they exhibit linear growth which is consistent with chaotic dynamics.  Second, we will sketch how (for this model) the cancellation of the classical solutions for the path integral take place.  Finally, we will demonstrate that our proposed formula for the plateau (\ref{instantonApprox}) is not contradicted by the early time behavior of spread complexity.   
%{\em second derivative conditions}

\subsection{Early times or small $n$}

To compute the Krylov wavefunctions explicitly, we found it most natural to implement the Lanczos algorithm 
directly in position space, as described in Appendix~\ref{sec:lanczos position space}. The early-time data 
presented below is obtained using that method. In this section, we report the first few Lanczos coefficients and 
the early-time spread complexity for our toy model. We tested several values of the coupling constant g, while 
keeping all other parameters fixed: $\alpha = 10^{-3},
\quad \omega_1 = \omega_2 = m = \hbar = 1$.
The reference state $\psi(x_1, x_2)$ is taken to be a two-dimensional Gaussian wavepacket,
\begin{equation}
    \psi(x_1,x_2) \equiv
    \left( \frac{m\omega_1}{\pi\hbar} \right)^{\frac{1}{4}}
    \left( \frac{m\omega_2}{\pi\hbar} \right)^{\frac{1}{4}}
    e^{ -\frac{1}{2\hbar} \left( m\omega_1(x_1-c_1)^2 + m\omega_2(x_2-c_2)^2 \right) }\,.
\end{equation}
If $c_1 = c_2 = 0$, this form corresponds to the ground state of a standard two-dimensional harmonic oscillator with no coupling.\\

\noindent
The inequality (\ref{eq:potential condition}) delineates two regimes; one in which there is only a single critical 
point of the potential — the global minimum $x_1=x_2=0$, and another with three critical points — a saddle at the 
origin and two off-axis minima, allowing for an instanton solution. With the parameters chosen above, we find 
that the transition between these regimes occurs at $g = 500$: for $g > 500$, the potential admits three critical 
points. This confirms that our toy model, in the appropriate parameter regime, realizes the necessary structure for 
instanton-mediated transitions, and therefore allows a meaningful test of our theoretical framework.

\begin{figure}[hbt!]
    \centering
    \includegraphics[scale=0.4]{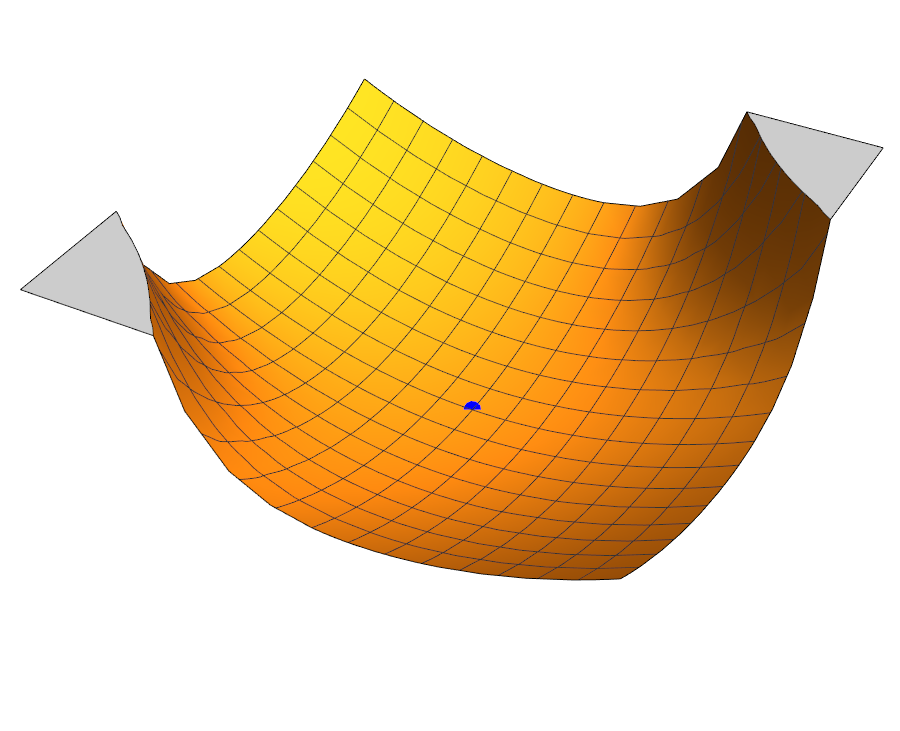}
    \includegraphics[scale=0.4]{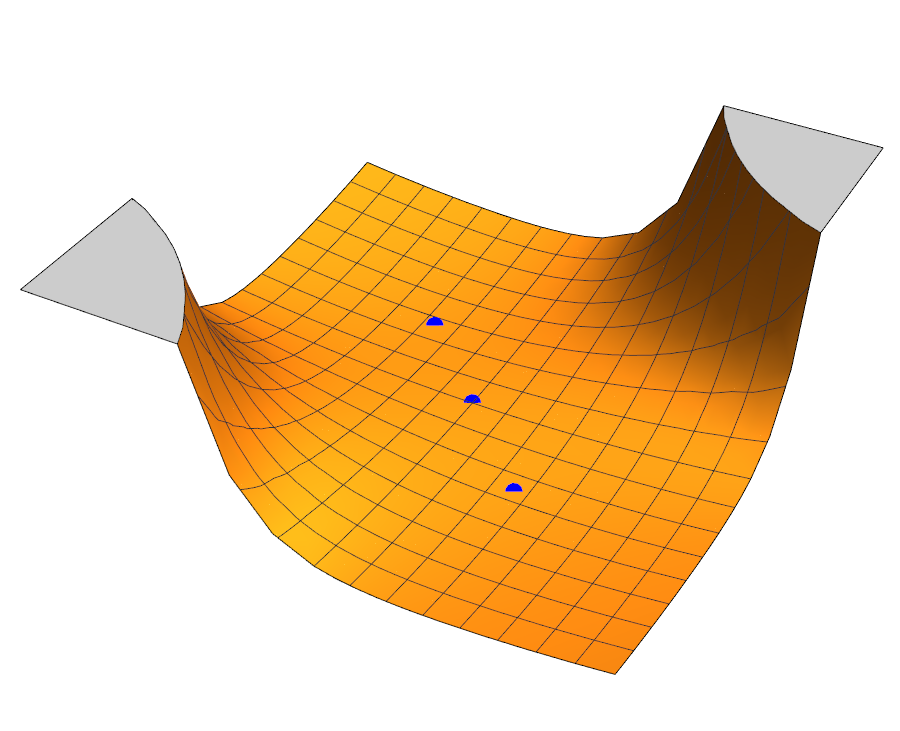}
    \caption{The toy model potential as a function of $x_1$ and $x_2$ with $\alpha=10^{-3}$, $\omega_1=\omega_2=m=\hbar=1$, $g=100$ on the left and $g=1000$ on the right. $x_1$ and $x_2$ ranging from $-80$ to 80. Critical points marked with blue dots.}
\end{figure}

\subsubsection{Lanczos coefficients and Krylov wave functions} \label{sec:lanczos coefficients small n}

\begin{figure}[]
    \centering
    \includegraphics[width=0.9\linewidth]{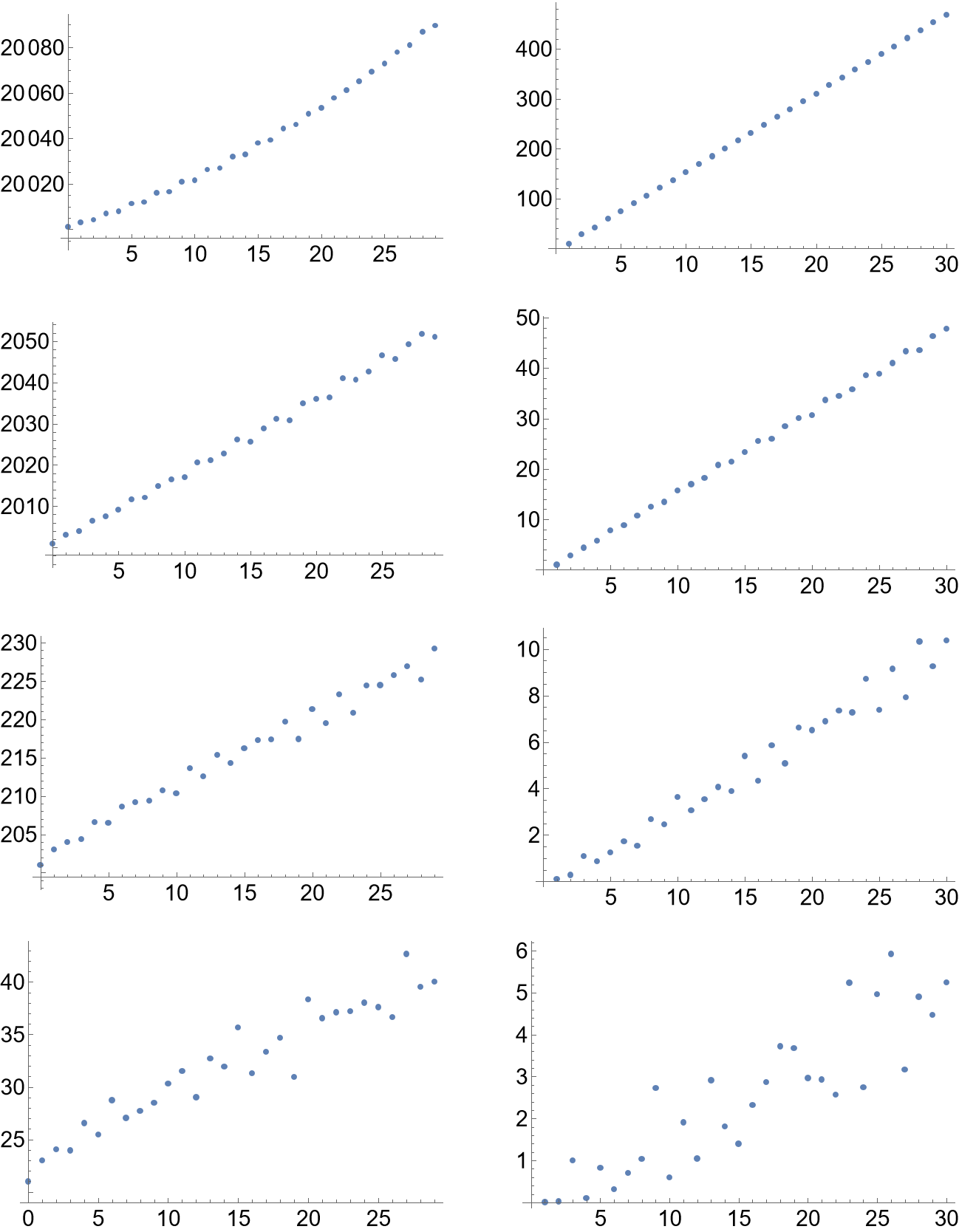}
    \caption{Lanczos coefficients obtained with a gaussian initial state centered at the origin. $a_n$ from 0 to 29 
    on the left column and $b_n$ from 1 to 30 on the right.}
    \label{fig:lanczos same omegas}
\end{figure}

We start our analysis with our reference state centered at the origin $(c_1=c_2=0)$. In Figure \ref{fig:lanczos 
same omegas}, we display the Lanczos coefficients obtained in the different potential regimes. As $g$ 
increases, both $b_n$'s and $a_n$'s tend to show more ordered growth patterns (reflected in the values of 
$R^2$). Additionally, the $a_n$'s seem to shift from linear growth to higher order growth, and the $b_n$'s tend 
to maintain a linear behavior but with increasing slope. Fits obtained for the coefficients from Figure 
\ref{fig:lanczos same omegas} are given in Tables \ref{tab:an fits} and \ref{tab:bn fits}.

\begin{table}[hbt!]
    \centering
    \begin{tabular}{c|c|c}
        $Log_{10}(g)$  &   Model for $a(n)$  &   $R^2$  \\
        \hline
        4   &   $\left(4.89908\cdot 10^{-2}\right) n^2+(1.65980)n+2.00011\cdot 10^4$  &  $0.99926288$\\
        4   &   $(3.08054)n+1.99944\cdot 10^4$  &  $0.98439119$\\
        3   &   $(1.77457)n+2.00048\cdot 10^3$  &  $0.99660155$\\
        2   &   $\left(8.90513\cdot 10^{-1}\right) n+2.02371\cdot 10^2$  &   $0.98270530$\\
        1   &   $\left(6.21240\cdot 10^{-1}\right) n+2.30315\cdot 10^1$  &   $0.92052565$
    \end{tabular}
    \caption{Linear regression fits obtained for $a(n)$ with linear or quadratic models. Reference state centered at the origin.\label{tab:an fits}}
\end{table}

\begin{table}[hbt!]
    \centering
    \begin{tabular}{c|c|c}
        \label{tab:linear bn}
        $Log_{10}(g)$  &   Model for $b(n)$  &   $R^2$  \\
        \hline
        4   &   $\left(1.57632\cdot 10^1\right) n-4.39648$  &  $0.99998620$\\
        3   &   $(1.60204)n-5.59445\cdot 10^{-1}$  &  $0.99905461$\\
        2   &   $\left(3.46899\cdot 10^{-1}\right) n-4.24383\cdot 10^{-1}$  &   $0.97261828$\\
        1   &   $\left(1.76307\cdot 10^{-1}\right) n-2.61818\cdot 10^{-1}$  &   $0.79734244$
    \end{tabular}
    \caption{Linear regression fits obtained for $b(n)$ with linear models. Reference state centered at the origin.\label{tab:bn fits}}
\end{table}

\begin{figure}[hbt!]
    \centering
    \includegraphics[width=0.9\linewidth]{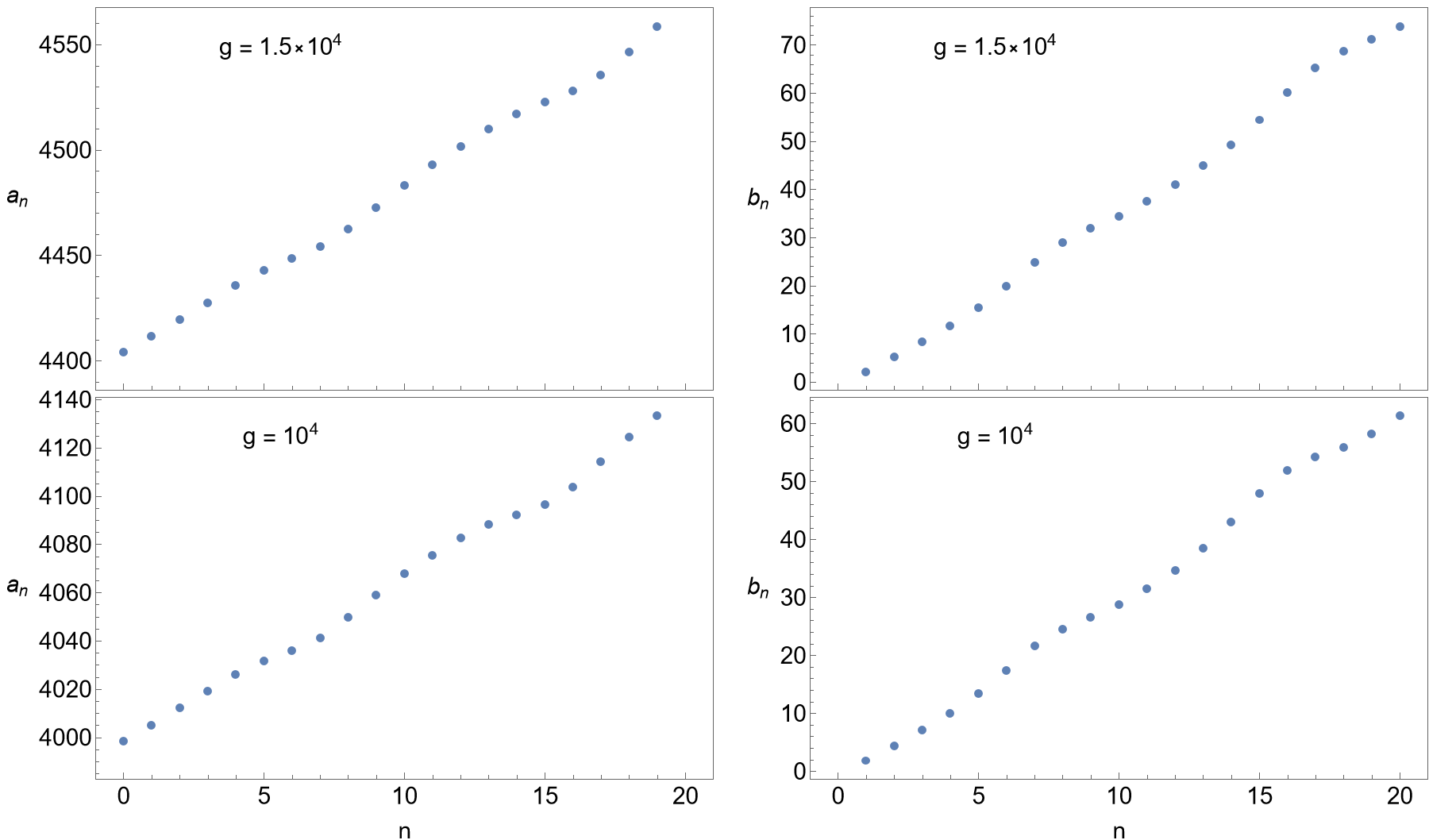}
    \caption{Lanczos coefficients obtained for the case of a gaussian reference state shifted from the origin.}
    \label{fig:lanczos shifted}
\end{figure}

\noindent
It is important to emphasize that the interpretation of the initial state’s location — specifically, being centered at 
the origin — depends crucially on the parameter regime. When $g < 500$, the origin corresponds to a global 
minimum of the potential. In contrast, for $g > 500$, the origin becomes a saddle point, and thus placing the 
initial state there means it is centered at an unstable configuration. To explore this further, we consider an 
alternative setup in which the initial Gaussian wavepacket is centered not at the origin, but instead at one of 
the nontrivial minima that emerge when $g > 500$. This corresponds to choosing $c_1 \neq 0$ and $c_2 \neq 
0$ in the initial state. The results for this shifted initial condition are shown in Figure~\ref{fig:lanczos shifted} and summarized in 
Table~\ref{tab:shifted fits}.\\

\noindent
As in the origin-centered case, we observe that both Lanczos coefficients $a_n$ 
and $b_n$ exhibit linear growth at early times — a key feature associated with complexity generation. 
Interestingly, this linear growth appears to be modulated by oscillations, suggesting richer underlying dynamics 
possibly related to the nontrivial geometry of the potential landscape near the minima.

\begin{table}[hbt!]
    \centering
    \begin{tabular}{c|c|c}
        \label{tab:linear bn}
        $g$  &   Model for $a(n)$ or $b(n)$   &   $R^2$  \\
        \hline
        $1.5\cdot10^{4}$   &   $a(n)=8.02136 n+4.40248\cdot 10^3$  &  $0.997318$\\
        $10^{4}$   &   $a(n)=6.91662 n+3.99705\cdot 10^3$   &   $0.996152$\\
        $1.5\cdot10^{4}$   &   $b(n)=3.88673 n-3.39219$  &  $0.996271$\\
        $10^{4}$   &   $b(n)=3.22985 n-2.31652$   &   $0.996049$
    \end{tabular}
    \caption{Linear regression fits obtained for $a(n)$ and $b(n)$ for the case of a gaussian reference state 
    shifted from the origin.\label{tab:shifted fits}}
\end{table}

\noindent
With our method, we have also been able to obtain the wavefunctions of the Krylov basis vectors. Contour 
plots of the first Krylov wave functions for the different cases are given in appendix \ref{sec:Krylov basis wave 
functions}.

\subsubsection{Early times complexity growth}

\begin{figure}[hbt!]
    \centering
    \includegraphics[width=0.8\linewidth]{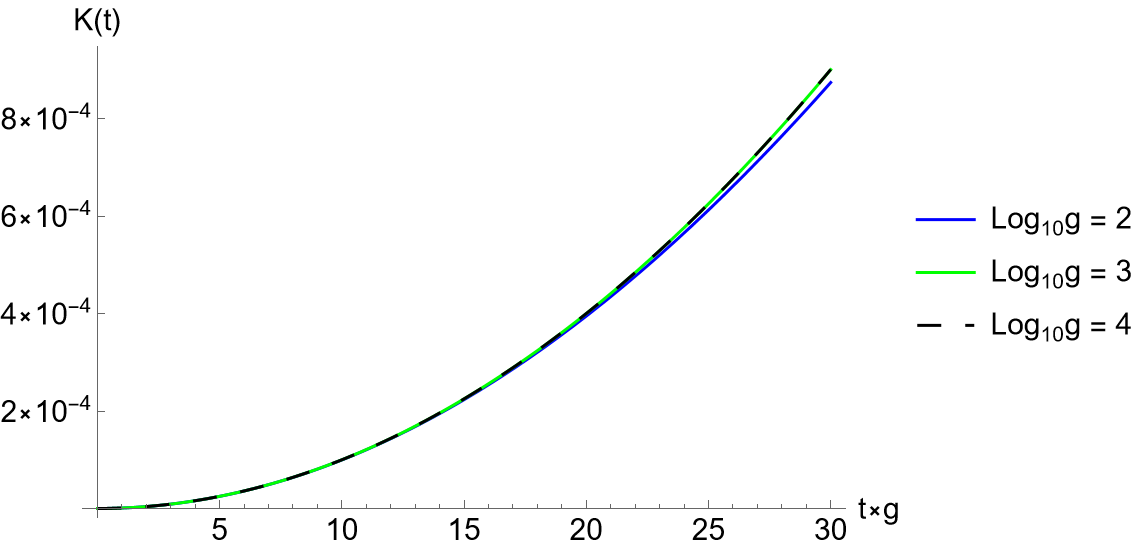}
    \caption{$K(t)$ vs $(t\cdot g)$ for the cases $Log_{10}(g)=2,3,4$. The curves corresponding to $Log_{10}
    (g)=3,4$ appear superimposed. The reference state is a gaussian centered at the origin.}
    \label{fig:Kt}
\end{figure}

\begin{figure}[hbt!]
    \centering
    \includegraphics[width=0.8\linewidth]{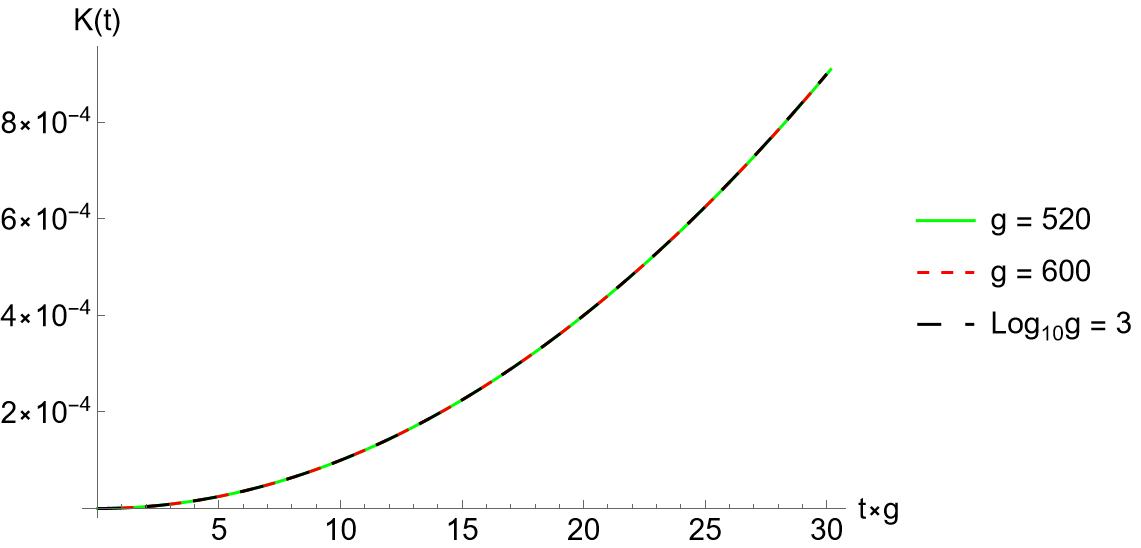}
    \caption{$K(t)$ vs $(t\cdot g)$ for the cases $g = 520, 600, 1000$. All curves appear superimposed. The 
    reference state is a gaussian centered at the origin.}
    \label{fig:Kt2}
\end{figure}

In order to obtain the complexity for our model, we calculate the survival amplitude using perturbation theory in 
time. That is, we calculate
\begin{equation}
    \psi_0(t) \equiv \langle \psi| \psi(t)\rangle = \bra{\psi} \sum_{n=0}^{j} \frac{1}{n!}\left(\frac{-iHt}{\hbar} 
    \right)^n \ket{\psi},
\end{equation}
where, unless stated otherwise, we set $j=200$, so 200th order. Then, by employing the known relation 
between the Krylov basis and the survival amplitude, we obtain the Krylov basis coefficients $\psi_n$, from \cite{Beetar:2023mfn},
\begin{align}
    \psi_n &\equiv \braket{K_n}{\psi(t)}, \\
    \psi_1 &= \frac{ i\partial_t \psi_0 - a_0 \psi_0 }{b_1}, \\
    \psi_{n+1} &= \frac{ i\partial_t \psi_{n} - a_{n} \psi_{n} - b_{n}\psi_{n-1} }{b_{n+1}}.
\end{align}
The results reveal a similar growth pattern at early times for the curves corresponding to $g\ge 100$. Figures 
\ref{fig:Kt} and \ref{fig:Kt2} show complexity curves with a scaled time axis. Their behavior is very well fitted by 
quadratic models given in Table \ref{tab:Kt fits}.\\

\begin{table}[hbt!]
    \centering
    \begin{tabular}{c|c|c}
        $g$  &   Model for $K(t)$   &   $R^2$  \\
        \hline
        $10^{4}$   &   $\left(1.00154\cdot 10^2\right) t^2-\left(2.45535\cdot 10^{-4}\right) t+6.74629\cdot 10^{-8}$  &  $0.999999986384570$\\
        $10^{3}$   &   $(1.00103) t^2-\left(1.63994\cdot 10^{-5}\right) t+4.50608\cdot 10^{-8}$   &   $0.999999993921361$\\
        $600$   &   $\left(3.60041\times 10^{-1}\right) t^2-\left(1.07644\times 10^{-6}\right) t+4.93672\times 10^{-9}$   &   $0.99999999992650$\\
        $520$   &   $\left(2.70301\times 10^{-1}\right) t^2+\left(3.04404\times 10^{-6}\right) t-1.61582\times 10^{-8}$   &   $0.999999999236404$\\
        $10^{2}$   &   $\left(9.50643\cdot 10^{-3}\right) t^2+\left(7.84158\cdot 10^{-5}\right) t-2.15056\cdot 10^{-6}$  &  $0.999985543858164$ \\
        $(10^{4})^*$   &   $(3.24672) t^2+\left(1.02673\cdot 10^{-7}\right) t-2.53104\cdot 10^{-11}$  &  $0.99999999999729$
    \end{tabular}
    \caption{Linear regression fits obtained for $K(t)$. The reference state is centered at the origin except for the $(10^{4})^*$ row, corresponding to a reference state shifted from the origin and centered at a minimum of the potential.\label{tab:Kt fits}}
\end{table}

\begin{figure}[hbt!]
    \centering
    \includegraphics[width=0.8\linewidth]{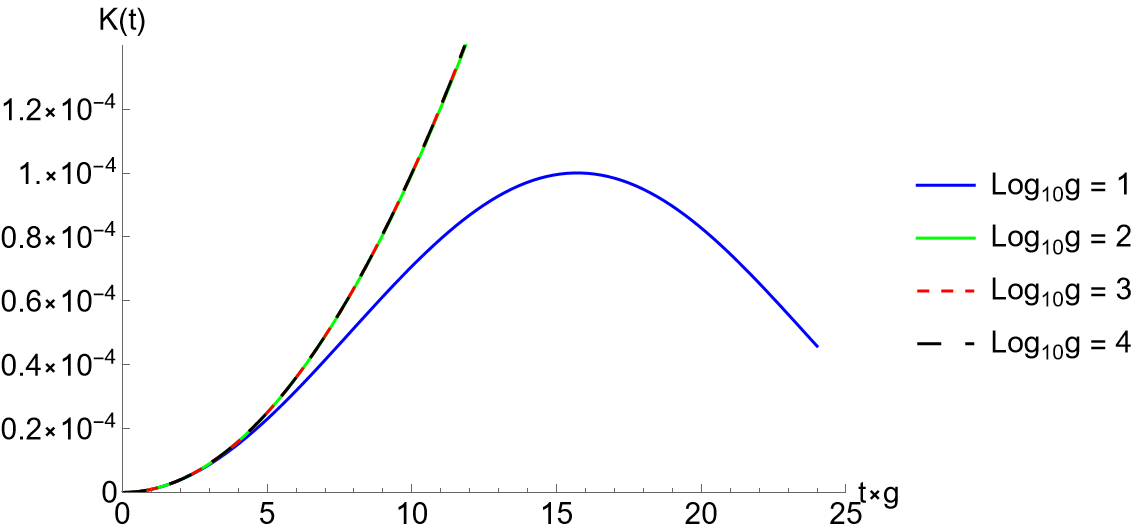}
    \caption{$K(t)$ vs $(t\cdot g)$ for the cases $Log_{10}(g)=1,2,3,4$. The curves corresponding to $Log_{10}
    (g)=2,3,4$ appear superimposed. The reference state is a gaussian centered at the origin.}
    \label{fig:Kt with g1}
\end{figure}

\begin{figure}[hbt!]
    \centering
    \includegraphics[width=0.9\linewidth]{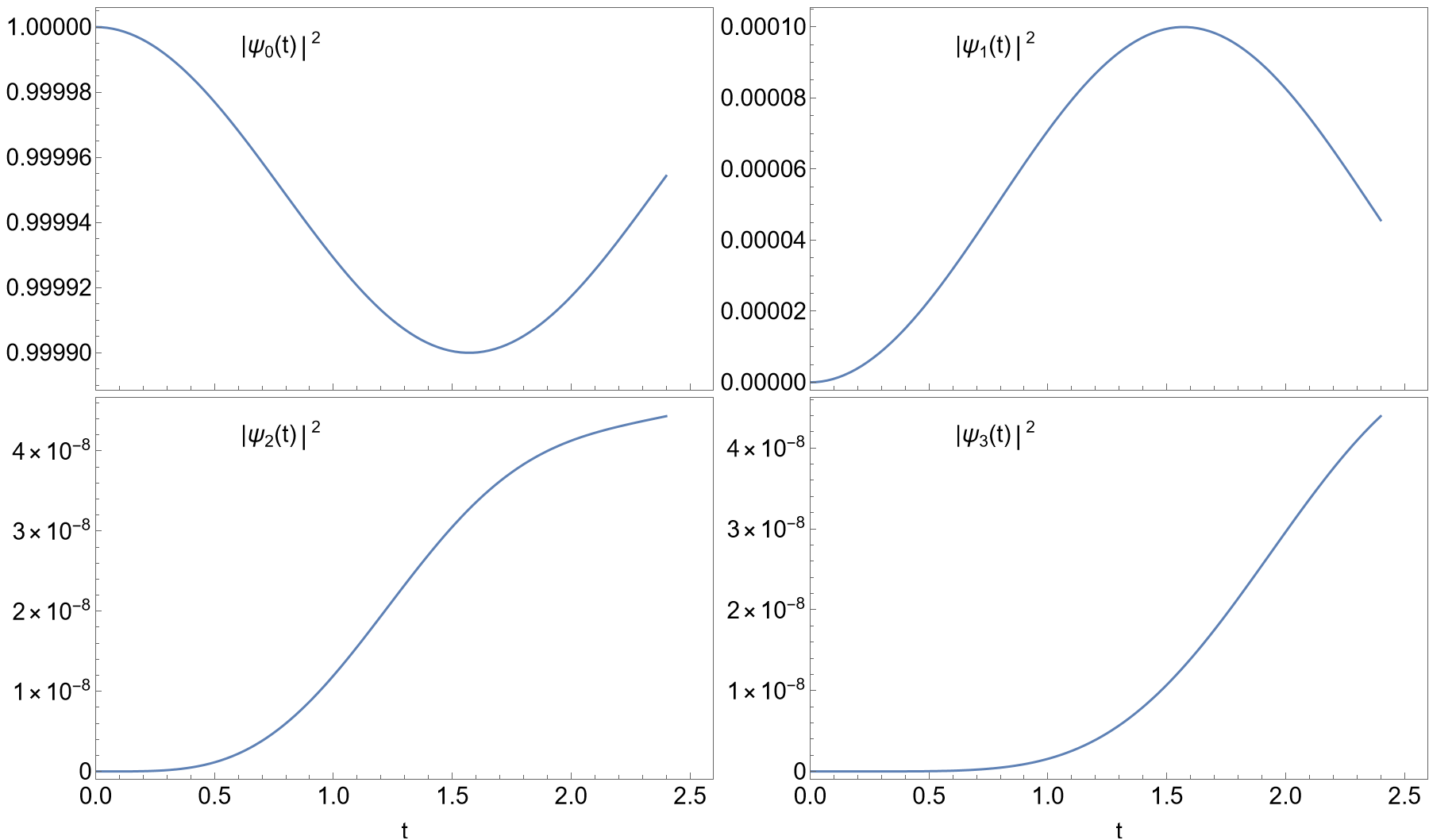}
    \caption{Probabilities of finding the time-evolved state at the first four Krylov basis vectors ($|\psi_n(t)|^2$) as a function of time. The reference state is a gaussian centered at the origin and $g=10$.}
    \label{fig:g1 probs}
\end{figure}

\noindent
Interestingly, the behavior of $K(t)$ for $g = 10$, shown in Figure~\ref{fig:Kt with g1}, differs markedly from the 
larger coupling cases. In this regime, the complexity initially grows, reaches a peak, and then decreases — a 
pattern reminiscent of systems with finite-dimensional Hilbert spaces \cite{Caputa:2021sib}. This is a consequence of the behavior of the survival probability, as made evident in Figure \ref{fig:g1 probs}. Unfortunately, due to the limited time 
range accessible in our computations, we are unable to verify whether the complexity remains oscillatory 
at late times or eventually saturates to a plateau.\\

\noindent
For small $g$, the potential closely resembles that of a two-dimensional harmonic oscillator, and so it is reasonable to expect non-chaotic behavior near the origin — at 
least for early times. While the Hilbert space of the harmonic oscillator is infinite, in the limit $g \to 0$, the 
reference state becomes the ground state of the uncoupled system, in this case, no complexity growth is 
expected, as the dynamics is trivial.\\

\begin{figure}[hbt!]
    \centering
    \includegraphics[width=0.6\linewidth]{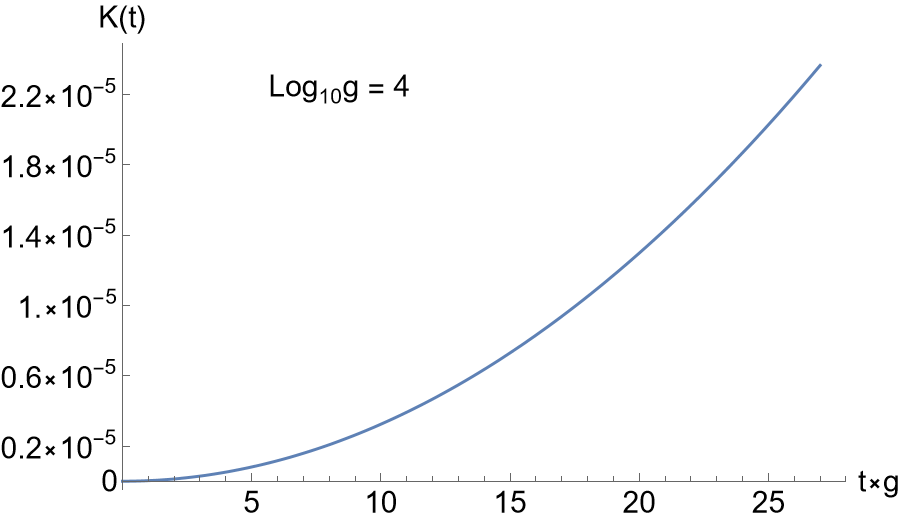}
    \caption{$K(t)$ vs $(t\cdot g)$ for the case with $Log_{10}(g)=4$ and a gaussian reference state shifted 
    from the origin and centered at a minimum of the potential.}
    \label{fig:Kt g4 shifted}
\end{figure}

\noindent
Before concluding this section, we also consider the case with a shifted reference state and strong coupling ($
\log_{10}(g) = 4$). Due to computational constraints, we truncate the time evolution at 55 orders in the 
perturbative expansion. The resulting complexity curve, shown in Figure~\ref{fig:Kt g4 shifted}, fits a quadratic 
form (see Table~\ref{tab:Kt fits}), much like the case with a reference state centered at the origin — albeit with a slower 
growth rate, likely reflecting the different local structure of the potential around the new center.

\FloatBarrier
\subsection{Path integral cancellations, instanton and comparison}

Since we want to understand how the cancellation of the classical solutions work, it will prove instructive to 
start with a sketch of that. Specifically, we would like to {\it perturbatively} compute  
\be
   \langle x_i(t)|x'_j(0)\rangle =\langle x_i |e^{-\frac{i}{\hbar} H t}|x'_j\rangle =\int_{x_i(t), x'_j(0)}{\cal D}x_k
   (t')e^{\frac{i}{\hbar}S[x_k(t')]}\;,\;\; i,j,k=1,2\;,
\ee
in the case that there exists a classical solution that interpolates between $x'(0)$ and $x(t)$.\\

\noindent
Expanding $e^{iS}$ in $g$, gives, at order $g^0$ just the free harmonic oscillator propagators, 
\be
\langle x_i(t)|x'_j(0)\rangle^{(0)} =\sum_{n_i} \langle x_i|n_i\rangle e^{-\frac{i}{\hbar}\omega_i (n_i+1/2)t}\langle n_i|x'_j\rangle 
=\delta_{ij}\sum_{n_i=0}^\infty \psi_{n_i}(x_i)\psi_{n_i}^*(x'_i)e^{-\frac{i}{\hbar}\omega_i (n+1/2)t}\,,
\ee
coming from $e^{\frac{i}{\hbar}S_0}$. This is, of course, the  exact result at order $g^0$, but since we are 
actually interested in the {\em classical 
approximation} we can write
\be
   \langle x_i(t)|x'_j(0)\rangle^{(0)}\simeq \delta_{ij}e^{\frac{i}{\hbar}S_{\rm cl}[x_i^{\rm cl}(t)]}\;,
\ee
%\be
%   \langle x_i(t)|x'_j(0)\rangle^{(0)}\simeq \delta_{ij}e^{\frac{i}{\hbar}S_{\rm cl}[x_i^{\rm cl}(t)]}=\delta_{ij}
%   e^{\frac{i}{\hbar}\sum_{i=1,2}\left[\int dt \frac{dx_i(t)}{dt}\sqrt{2m\left(E_i-m\frac{ \omega_i^2}{2}x_i^2(t)\right)}-
%   E_it\right]}\;,
%\ee
where $x_i^{\rm cl}(t)$ is the classical path satisfying
\be
   x^{\rm cl}_i(t)=A_i \cos \left(\omega_i t+\delta_i\right)\;,\;\;
   E_i=\frac{m}{2}(\dot x_i^2+\omega_i^2x_i^2)=\frac{m}{2}A_i^2\omega_i^2\;,
\ee
with 
\be
   A_i\cos \delta_i=x'_i\;,\;\;
   A_i\cos(\omega_i t+\delta_i)=x_i\,.
\ee
Solving for $A_i$ and $E_i$, we find that
\begin{eqnarray}
    A_i = \sqrt{ (x’_i)^2 + \frac{(x’_i \cos(\omega_i t) - x_i)^2}{\sin^2(\omega_i t)} }, \quad E_i = \frac{m}{2} 
    \omega_i^2 \left[ (x’_i)^2 + \frac{(x’_i \cos(\omega_i t) - x_i)^2}{\sin^2(\omega_i t)} \right]\,.
\end{eqnarray}

\noindent
The classical action along such a path is given by
\begin{eqnarray}
   S_{\rm cl}[x_i^{\rm cl}(t)] = \sum_{i=1,2} \left[ \int dt \, \dot{x}_i(t) \sqrt{ 2m \left( E_i - \frac{1}{2} m \omega_i^2 
   x_i^2(t) \right) } - E_i t \right]\,,
\end{eqnarray}
and using the identity
\begin{eqnarray}
\int dx \sqrt{a - c x^2} = \frac{a}{2\sqrt{c}} \arcsin \left( x \sqrt{ \frac{c}{a} } \right) + \frac{x}{2} \sqrt{ a - c x^2 }\,,
\end{eqnarray}
we obtain an explicit expression,
\begin{eqnarray}
   \langle x_i(t) | x’_j(0) \rangle^{(0)}
   \simeq \delta_{ij} e^{ \frac{i}{\hbar} \sum_{i=1,2} \left\{
   \left. \sqrt{2m} \left[ \frac{E_i}{\sqrt{2m \omega_i^2}} \arcsin \left( x \sqrt{ \frac{m \omega_i^2}{2 E_i} } \right) + 
   \frac{x}{2} \sqrt{ E_i - \frac{1}{2} m \omega_i^2 x^2 } \right] \right|_{x’_i}^{x_i} - E_i t \right\} }\,.
\end{eqnarray}\\
At first order in $g$, the path 
integral is a discretized sum in $N$ points ($\Delta 
t=N\delta t$).  Since we are interested in studying the contribution from the classical solutions  we can always
find $n\gg 1$ with $n/N\rightarrow 0$, 
so $n \delta t\rightarrow 0$, such that
\be
g\delta t \sum_{t=t_1}^{t_N}e^{\a m^2x_1(t)x_2(t)}
\geq g\delta t \prod_{t=t_1}^{t_N} e^{\a m^2 x_1(t) 
x_2(t)/n}=g \delta t e^{\int dt \a m^2
\frac{x_1(t)x_2(t)}{n\delta t}}\;.
\ee
The order $g$ path integral thus yields the inequality
(where the inequality sign refers only to the 
modulus of the expression)
\bea
\langle x_i(t)|x'_j(0)\rangle^{(1)}&\geq&ig(\omega_1+\omega_2)
\int_{x_i(t),x'_j(0)}{\cal D}x_k(t)\Pi_t\delta te^{\frac{i}{\hbar}\delta t\left[\sum_{i=1,2}\left(\frac{m\dot x_k^2(t)}{2}
-m\frac{\omega_k^2x_k^2(t)}{2}\right)+\frac{\hbar\a m^2}{in\delta t}x_1(t)x_2(t)\right]}\cr
&=&ig [(\omega_1+\omega_2)\delta t ]\int_{x_i(t),x'_j(0)}{\cal D}x_k(t)e^{\frac{i}{\hbar}\int dt\left[\sum_{i=1,2}
\left(\frac{m\dot x_k^2(t)}{2}-m\frac{\omega_k^2x_k^2(t)}{2}\right)+\frac{\hbar\a m^2}{in\delta t}x_1(t)x_2(t)\right]} .\cr
&&
\eea
One can perform a change of variables $x_i \to y_i$ that diagonalizes the coupled potential, where $\tilde{\omega}_i^2 
\approx \pm i \hbar\alpha m / n\delta t$. In this rotated basis, the propagator at order g becomes
\begin{eqnarray}
   \langle x_i(t) | x’_j(0) \rangle^{(1)} \simeq i g [(\omega_1 + \omega_2) \delta t ]\, 
   e^{ \frac{i}{\hbar} S_{\text{cl}}^{(1)}[y_i^{\rm cl}(t)] },
\end{eqnarray}
with $y_i^{\rm cl}(t) = B_i \cos( \tilde{\omega}_i t + \tilde{\delta}_i ),
\quad \tilde{E}_i = \frac{m}{2} B_i^2 \tilde{\omega}_i^2$. 
Combining the two contributions, 
the full propagator to first order gives
(where again the inequality only refers to the 
modulus of the second term)
\bea
   \langle x_i(t)|x'_j(0)\rangle^{(0)+(1)}&\geq&\sum_{n_i=0}^\infty\Biggl[\delta_{ij}
   \psi_{n_i}(x_i)\psi_{n_i}^*(x'_i)e^{-i\omega_i (n+1/2)t}\nonumber\\
   &+&ig [(\omega_1+\omega_2)\delta t ]
   \psi_{n_i}(y_i)\psi_{n_i}^*(y'_i)e^{-i\tilde \omega_i (n+1/2)t}\Biggr]\nonumber\\
   &\simeq&\left[\delta_{ij}e^{\frac{i}{\hbar}\sum_{i=1,2}\left\{
   \sqrt{2m}\left.\left[\frac{E_i}{\sqrt{2m
   \omega_i^2}}\arcsin \left(x\sqrt{\frac{m\omega_i^2}{2E_i}}\right)+\frac{x}{2}\sqrt{E_i-m\frac{\omega_i^2}{2}x^2}
   \right]\right|_{x'_i}^{x_i}-E_it\right\}}\right.\cr
   &+&\left.ig [(\omega_1+\omega_2)\delta t ]
   e^{\frac{i}{\hbar}\sum_{i=1,2}\left\{
   \sqrt{2m}\left.\left[\frac{\tilde E_i}{\sqrt{2m
   \tilde \omega_i^2}}\arcsin \left(y\sqrt{\frac{m\tilde \omega_i^2}{2\tilde E_i}}\right)+\frac{y}{2}\sqrt{\tilde E_i-
   m\frac{\tilde \omega_i^2}{2}y^2}
   \right]\right|_{y'_i}^{y_i}-\tilde E_it\right\}}\right]\,.\nonumber\\
\eea
Note that, since $\tilde \omega_i$ are complex, the second term is not a pure phase, but is a phase 
multiplied by an exponentially increasing factor 
{\em as a function of time} 
(plus an exponentially decreasing one, that can be 
neglected), so as time goes by, it becomes important, despite its infinitesimal coefficient.
This in turn implies that the Krylov amplitudes satisfy
\begin{eqnarray} \label{psixixj}
  \psi_n(t)&=&\left(\prod_{i,j}\int dx_i \int dx'_j\right) K_n^*(\{x_i\})\psi(\{x'_j\})\langle \{x_i(t)\}|\{x'_j(0)\}\rangle\cr
  &=&\left(\prod_i \int dx_i\int dx'_i\right) K_n^*(\{x_i\})\psi(\{x'_i\})\langle \{x_i(t)\}|\{x'_i(0)\}\rangle\cr
  &\gsim &\left(\prod_i \int dx_i\int dx'_i\right)K_n^*(x_1,x_2)\psi(x'_1,x'_2)\times\cr
  &&\times e^{\frac{i}{\hbar}\sum_{i=1,2}\left\{
 \sqrt{2m}\left.\left[\frac{E_i}{\sqrt{2m
 \omega_i^2}}\arcsin \left(x\sqrt{\frac{m\omega_i^2}{2E_i}}\right)+\frac{x}{2}\sqrt{E_i-
m\frac{\omega_i^2}{2}x^2}
 \right]\right|_{x'_i}^{x_i}-E_it\right\}}\cr
 &&+ig [(\omega_1+\omega_2)\delta t ]\left(\prod_i \int dy_i\int dy'_i\right)K_n^*(x_1,x_2)\psi(x'_1,x'_2)\times\cr
 &&\times e^{\frac{i}{\hbar}\sum_{i=1,2}\left\{
 \sqrt{2m}\left.\left[\frac{\tilde E_i}{\sqrt{2m
 \tilde \omega_i^2}}\arcsin \left(y\sqrt{\frac{m\tilde \omega_i^2}{2\tilde E_i}}\right)+\frac{y}{2}\sqrt{\tilde E_i-
 m\frac{\tilde \omega_i^2}{2}y^2}
 \right]\right|_{y'_i}^{y_i}-\tilde E_it\right\}}\,.
\end{eqnarray}\\
Since the transformation $(x_1, x_2) \to (y_1, y_2)$ is a rotation (with unit Jacobian), the measure is 
unchanged, but the phases in the integrands differ. The two terms thus represent out-of-phase oscillations, and 
higher-order terms in $g$ continue this trend, and we will get an infinite set of {\em essentially 
arbitrary} phases, multiplied by $(g\delta t)^n$ times increasing exponentials. Then, 
as $t$ increases, these phases increasingly destructively 
interfere, leading to suppression of the classical contribution. If the contributions from classical (real-time) 
paths cancel out at late times due to phase interference, as argued above, then the quantum (imaginary-time) 
contributions — i.e., instantons — dominate. In that case, the long-time complexity is approximated by
\begin{eqnarray}
   K(t \to \infty)
   \simeq e^{-\frac{2}{\hbar} S_E(\text{instanton})}
   \sum_n n \left| K_n^* (\{x_2\}) \psi(\{x_1\}) + K_n^* (\{x_1\}) \psi(\{x_2\}) \right|^2\,,
\end{eqnarray}   
as derived in equation~\eqref{compplateau}. Given that we have explicit expressions for the first few Krylov 
basis elements $K_n(x_1, x_2)$, and a numerical estimate of the instanton action $S_E$, we can now 
compute the plateau value of the complexity and compare it with the late-time behavior observed numerically.

\subsubsection{Toy model instanton test}

The calculation of the instanton action for our model can be cast in terms of the one-dimensional motion of a 
particle in the line connecting the three critical points of the potential. Here we consider $\alpha>0$ and set the 
minimum of the potential to 0.\\

\noindent
The line joining the critical points is given by $x_2 = -x_1 \frac{\omega_1}
{\omega_2}$ and the normalized potential along this line is
\begin{equation}
    U(x) \equiv m x^2 \omega_1^2 + g \hbar(\omega_1+\omega_2) e^{-\alpha m^2 x^2 \frac{\omega_1}
    {\omega_2} }-\min(V(x_1,x_2))\,,
\end{equation}
given as a function of $x_1$ only. For ease of notation, we set $x\equiv x_1$ in what follows. In terms of $x$, 
the critical points can now be written as
\begin{equation}
    0, \quad x_0 \equiv -\sqrt{ \frac{-\omega_2}{\alpha m^2 \omega_1} \ln{ \frac{\omega_1 \omega_2}{g 
    \hbar(\omega_1+\omega_2) \alpha m} } },
    \quad x_f \equiv \sqrt{ \frac{-\omega_2}{\alpha m^2 \omega_1} \ln{ \frac{\omega_1 \omega_2}{g 
    \hbar(\omega_1+\omega_2) \alpha m} } }.
\end{equation}
With this set up, the instanton action takes the form
\begin{equation}
    S_E(instanton) = \int_0^\infty dt \left( \frac{m}{2}\dot{x}^2 + U(x)\right),
\end{equation}
\noindent
with boundary conditions: $\dot{x}(0) = \dot{x}(\infty) = 0$. Following the the usual manipulations for one-dimensional instantons (see \cite{Nastase:2022,coleman1988aspects,rajaraman1982solitons}), we obtain
\begin{equation}
    S_E(instanton) =\int_{x_0}^{x_f} dx \; \sqrt{2m U(x)},
\end{equation}
\noindent
which can than be integrated numerically. In Table \ref{tab:actions} we give results for $S_E(instanton)$ computed for different values of $g$.\\

\begin{table}[hbt!]
    \centering
    \begin{tabular}{c|c|c}
        $g$  &  $S_E(instanton)$  &  $e^{-2S_E(instanton)}$\\
        \hline
        $10^4$  &   $1.14278\cdot 10^4$  &  $9.16644\cdot 10^{-9927}$\\
        $10^{3}$  &   $8.48216\cdot 10^2$  &  $1.77538\cdot 10^{-737}$\\
        $600$  &   $1.06390\times 10^2$  &  $3.89695\times 10^{-93}$\\
        $520$  &   $1.04109\times 10^1$  &  $9.06095\times 10^{-10}$\\
        $500(1+10^{-10})$  &   $1.33333\cdot 10^{-12}$  &   $1.00000$
    \end{tabular}
    \caption{Numerical results obtained for the instanton action with different values of $g$.\label{tab:actions}}
\end{table}

\noindent
With these results and the wave functions of the initial Krylov vectors, we can then use 
eqs. (\ref{eq:conjecture psi}) and (\ref{compplateau}), to obtain values for the following partial sums,
\begin{align}
    K_\infty(n) &\equiv
    e^{-\frac{2}{\hbar}S_E({\rm instanton})}\sum_{j=0}^n j \left|K_j^*(\text{min}_2)\psi(\text{min}_1)+K_j^*(\text{min}_1)
    \psi(\text{min}_2)\right|^2, \\
    \quad\psi_\infty(n) &\equiv
    e^{-\frac{2}{\hbar}S_E({\rm instanton})}\sum_{j=0}^n \left|K_j^*(\text{min}_2)\psi(\text{min}_1)+K_j^*(\text{min}_1)
    \psi(\text{min}_2)\right|^2,
\end{align}
which give, respectively, the contribution of the first $n$ Krylov basis vectors to the complexity and the total probability of the time evolved state at large times.  
When summing over the full Krylov basis, i.e., when $n \rightarrow \infty$ our prediction is that the above sums will constitute the main contribution to these quantities.\\

\begin{figure}[hbt!]
    \centering
    \includegraphics[width=0.9\linewidth]{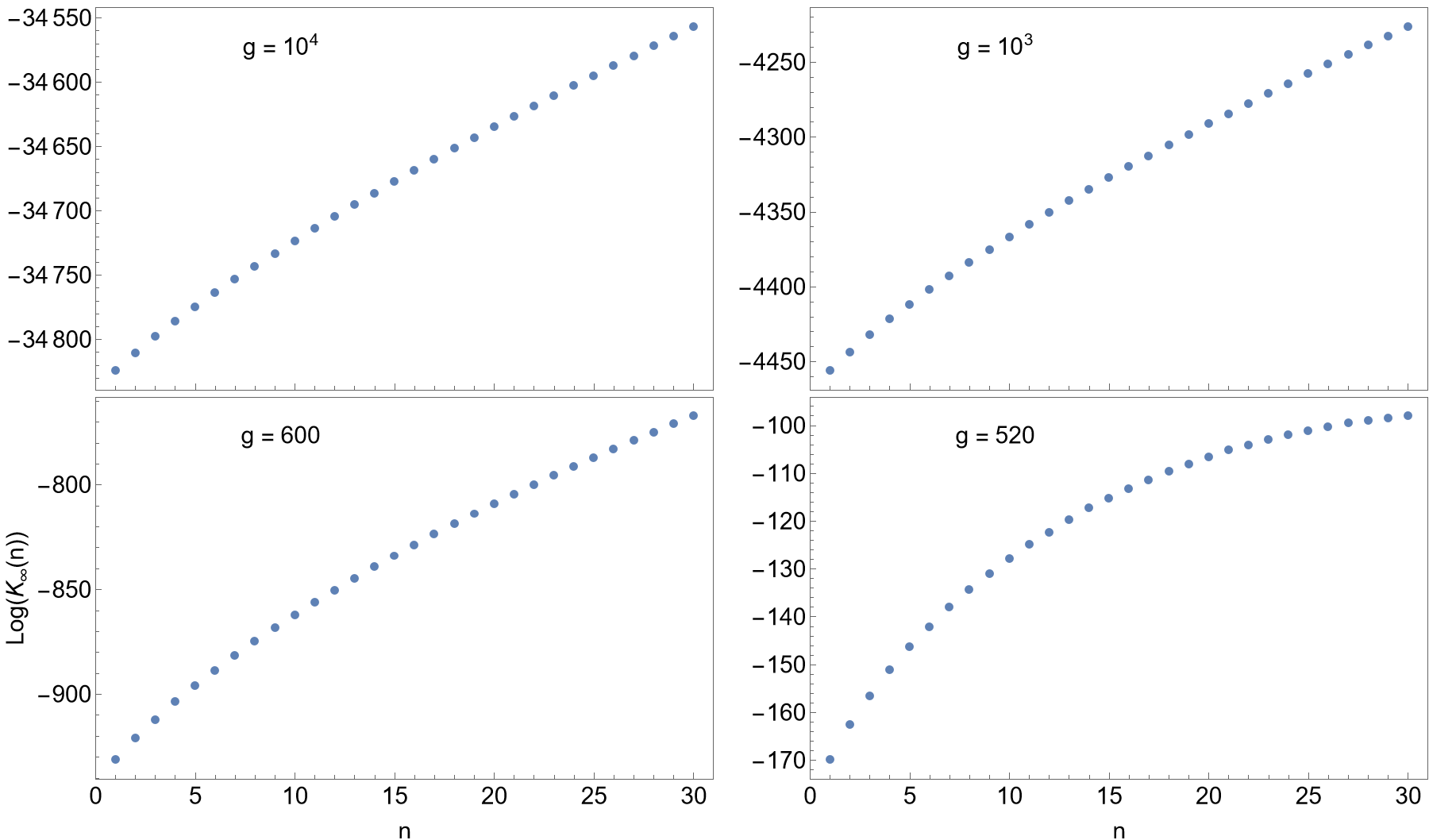}
    \caption{Log plots of the partial complexity sums $K_\infty(n)$, for different values of $g$.}
    \label{fig:K infty}
\end{figure}

\noindent
We observe that, for small $n$, the partial sums contributing to the complexity plateau are exceedingly 
small but exhibit steady growth with increasing $n$ (Figure \ref{fig:K infty}). This behavior arises from two key factors: the exponential 
suppression by the instanton factor $e^{-\frac{2}{\hbar}S_E(\text{instanton})}$, and the exponentially decaying profile of the 
reference state wave function at the potential minima. Since the reference state is a Gaussian centered at the 
origin, its overlap with the instanton-supporting regions is initially negligible. This observation is not in 
contradiction with our theoretical expectations. Rather, it indicates that a significant plateau value for the 
complexity requires access to a sufficiently large number of Krylov basis states. Importantly, the semiclassical 
approximation employed in deriving the instanton formula is valid precisely in the regime where the action 
$S_E$ is large, which naturally correlates with the suppression of contributions from low-$n$ Krylov states.\\

\noindent
From our early-time numerical analysis (see Figures~\ref{fig:Kt} and \ref{fig:Kt2}), we estimate that the plateau 
value of the complexity should be at least\footnote{Though this bound for complexity may seem tiny, recall that our initial wavefunction is either an eigenstate of the harmonic oscillator Hamiltonians or localised around one of the minima of the potential.  If this is a good approximation to an eigenstate of the full Hamiltonian, the complexity will remain small.  } on the order of $10^{-3}$. In Figure~\ref{fig:K infty}, we display $\log 
K_\infty(n)$ for various coupling values $g=10^4, 10^3, 600, 520$. In these cases, the data suggest a nearly linear 
growth in $\log K_\infty(n)$ at large $n$, consistent with the possibility of eventually reaching $\log K_\infty(n) 
\geq -3$. For $g=520$, the data instead follow a more parabolic trend in $\log K_\infty(n)$, but even here 
extrapolation hints that a value above -3 may be achieved at sufficiently large $n$. However, due to 
computational limitations, we are unable to reliably access Krylov states at such large $n$, preventing a 
definitive test of the instanton formula in this regime. We therefore conclude that, while the current data do not 
confirm the formula, they are not in contradiction with it either.\\

\begin{figure}[hbt!]
    \centering
    \includegraphics[width=0.45\linewidth]{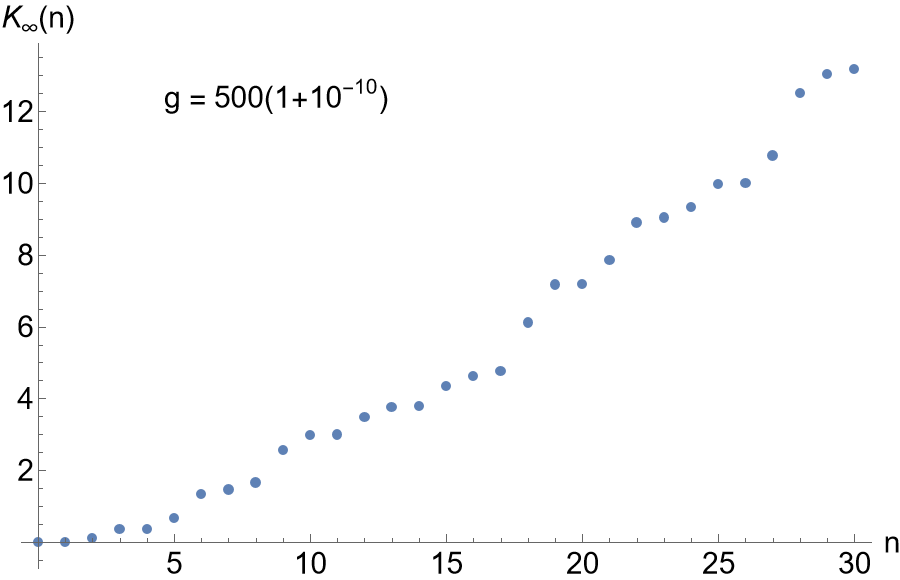}
    \includegraphics[width=0.45\linewidth]{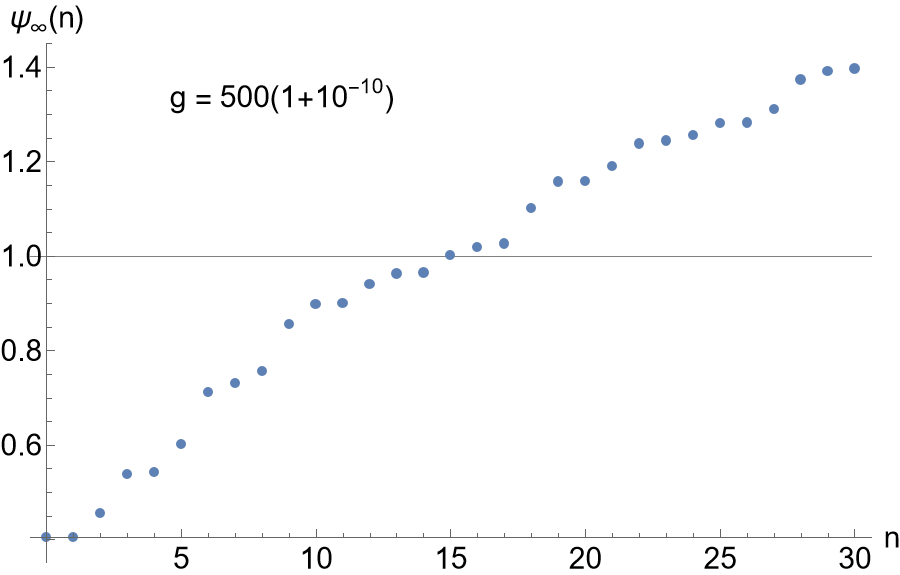}
    \caption{Plots of the partial sums $K_\infty(n)$ and $\psi_\infty(n)$ in the limit of small instanton action.}
    \label{fig:psi infty}
\end{figure}

\noindent
We have also investigated the limiting behavior as $g \to \left(\frac{\omega_1 \omega_2}{m \alpha \hbar 
(\omega_1 + \omega_2)}\right)^+$, in which the two minima of the potential coalesce near the origin. In this 
limit, the instanton solution ceases to exist and the instanton action $S_E$ approaches zero. As expected, the 
instanton complexity formula fails in this regime. This is confirmed by numerical results shown in 
Figure~\ref{fig:psi infty}, where the computed sum of squared overlaps of the time-evolved state with just the 
first few Krylov vectors exceeds unity — an unphysical result. This provides an important consistency check, 
affirming the limited domain of validity of the semiclassical instanton-based expression for the late-time 
complexity.

\section{Conclusions and Outlook}\label{sec:conclusions}

\noindent
Krylov complexity was conceived as a purely algorithmic observable, yet the results of this paper show that it 
admits a fully semiclassical description.  By mapping the Lanczos chain onto a one-dimensional path integral,
 expressing the projections of the time evolved state onto the Krylov basis $\psi_n(t)$ in
terms of path integrals and exploiting the cancellation of the contribution from classical paths,
we identified the late-time plateau of quantum-chaotic systems with a formula involving 
the action of a single Euclidean instanton.  This 
identification does more than provide intuition: formula~\eqref{eq:intro-plateau} turns the plateau height into a 
calculable, essentially topological, quantity determined by the values of the Krylov wave functions at the 
minima of a classical potential $V(x)$. The argument we propose is actually quite general, extending 
beyond our interest in quantum complexity. It allows for the expression of any late-time transition 
amplitude in terms of instanton solutions.
\\

\noindent
Our model of coupled harmonic oscillators with exponential coupling offers a  test of the proposal. 
We have explored various parameters and initial states. We turned analytic early-times series into a numerical 
evaluation bounded by the Krylov basis index $n$, and compared with the late-times
semiclassical instanton picture. 
%and our results demonstrate varied dynamical behaviors generated by the model Hamiltonian. 
While we could not readily {\em test} our formula at all time regimes, we showed that at least it is 
{\em not inconsistent} with early times, from our numerical results.
The path integral framework further 
clarifies why Krylov complexity grows without bound in unstable potentials and 
why integrable dynamics remain confined to an 
$O(1)$ window: in both cases the instanton saddle is absent.
\\

\noindent
Two technical developments were found during this analysis. 
The return-amplitude Lanczos algorithm circumvents 
divergent recursions in bosonic theories, while the position-space 
Lanczos iteration provides a practical route to 
early-time Krylov complexity in genuinely infinite-dimensional Hilbert spaces. 
Both methods are not restricted to our particular model and ready 
for deployment in different settings.\\

\noindent
Several avenues for future research stand out. First of all, better understanding to what extent and under
which conditions the proposed cancellation of classical contributions to the path integral actually occur 
would be an important next step. A possible extension of our proposal to the complexity of operators 
may give us interesting new insights. 
Extending the instanton calculus to operators with multiple 
Lanczos directions would shed light on higher-dimensional “Krylov manifolds’’ and their possible 
kink- or vortex-like saddles.
Translating the construction to thermal or thermo-field-double states could reveal a quantitative link 
between Krylov complexity, the spectral form factor and semiclassical wormholes in holography. 
On the computational front, 
applying the path integral recipe to lattice spin chains and SYK-type models 
may illuminate the universal features 
of operator spreading at strong coupling. Finally, because the plateau height is now tied to the 
action of instanton solutions, near-term quantum simulators might extract semiclassical tunneling 
parameters by measuring Krylov complexity alone.\\

\noindent
Krylov complexity thus emerges not merely as a numerical diagnostic but as an analytically tractable probe of 
late-time quantum dynamics—one that bridges algorithmic constructions and semiclassical physics in an 
unexpectedly direct way.

%In this paper we have proposed a form of the Krylov complexity in terms of path integrals, and argued that in the quantum chaotic case we found a plateau, that was given in terms of Euclidean instanton solutions between mimima of the potential $V$, and the Krylov basis $K_n$ and the initial wave function $\psi$, evalued at these minima. We have then tested these results in terms of a toy model of two coupled harmonic oscillators with an exponential coupling and a small prefactor coupling constant. We found...

\section*{Acknowledgments}
We would like to thank Javier Magan and Nitin Gupta for useful discussions.
The work of HN is supported in part by  CNPq grant 304583/2023-5 and FAPESP grant 2019/21281-4.
HN would also like to thank the ICTP-SAIFR for their support through FAPESP grant 2021/14335-0. 
ELG is supported by FAPESP grant 2024/13362-2.
JM and HJRVZ are supported in part by the “Quantum Technologies for Sustainable Development” grant 
from the National Institute for Theoretical and Computational Sciences of South Africa (NITHECS).
CB is supported by the Oppenheimer Memorial Trust Research Fellowship and the Harry Crossley 
Research Fellowship. JM and CB would like to acknowledge support from the
ICTP, Trieste through the Associates Programme and from the
Simons Foundation through grant number 284558FY19.

\appendix 

\section{A reformulated Lanczos algorithm}

\label{Lanczos2}
Our starting point is, as usual, the Lanczos algorithm,
\begin{equation}
H|K_n\rangle = a_n |K_n\rangle + b_n |K_{n-1}\rangle + b_{n+1} |K_{n+1}\rangle.
\end{equation}
The probability amplitudes can be obtained from the above as 
\begin{equation}
\psi_{n}(t) = \langle K_n | e^{-\frac{i}{\hbar} t H} |K_0\rangle,
\end{equation}
which satisfies
\begin{equation}
i\partial_t \psi_n(t) = a_n \psi_n(t) + b_n \psi_{n-1}(t) + b_{n+1} \psi_{n+1}(t).     \label{SchrEq}
\end{equation}
In a typical system one would first obtain the Lanczos coefficients and then compute the probability amplitudes 
using the above equations.  In the system we are studying there are two issues:  Firstly, the number of Lanczos 
coefficients is unbounded since we are studying a bosonic system that is not restricted to a finite-dimensional 
symmetry sector.  Secondly, the Lanczos coefficients are ill-defined since they rely on the perturbative expansion 
in $t$ of the return amplitude which gives rise to non-normalisable states\footnote{Even after regularisation we 
find that the Lanczos coefficients diverge as the regularisation parameter is sent to $0$. However, this can be circumvented if one wants to obtain only the first few coefficients, as we've done in section \ref{sec:lanczos coefficients small n} by picking a small value for $\alpha$.}. In this appendix we 
demonstrate that the algorithm can be reformulated entirely in terms of the return amplitude, without the 
intermediate step of computing the Lanczos coefficients.\footnote{This might be useful in light of the 
potential problems with numerical evaluations of the Lanczos algorithm in finite precision recently found in 
\cite{Eckseler:2025cgy}.}
This algorithm is essentially a Fourier transformed 
version of that used in \cite{Muck:2022xfc} with the generalisation of non-zero $a_n$ coefficients.\\

\noindent
The starting point of this reformulation is to define the following two functions,
\begin{eqnarray}
\tilde{K}_n(t) & \equiv &  b_n\frac{\psi_n(t)}{\psi_{n-1}(t)},  \nonumber \\
 \tilde{L}_n(t) &  \equiv & i b_n\frac{\partial_t \psi_n(t)}{\psi_{n-1}(t)}. \nonumber
\end{eqnarray}
The functions $L_{n}(t)$ satisfy the following recursive equation involving a derivative with respect to time,
\begin{equation}
\tilde{L}_n(t) = i\partial_t \tilde{K}_n + \frac{\tilde{K}_n \tilde{L}_{n-1}}{\tilde{K}_{n-1}}.
\end{equation}
The Schr\"odinger-like equation (\ref{SchrEq}) can also be recast in terms of the above as 
\begin{eqnarray}
i b_n \frac{\partial_t \psi_n(t)}{\psi_{n-1}(t)} & = & a_n b_n  \frac{\psi_n(t)}{\psi_{n-1}(t)} + b_n^2 + b_{n+1} b_n \frac{\psi_{n+1}(t)}{\psi_{n-1}}    \nonumber \\
\Rightarrow \tilde{L}_{n}(t) & = & a_n \tilde{K}_{n}(t) + b_{n}^2 + b_{n+1}b_n \frac{\psi_{n+1}(t)}{\psi_{n}(t)} \frac{\psi_{n}(t)}{\psi_{n-1}(t)}    \nonumber \\
\Rightarrow \tilde{L}_{n}(t) & = & a_n \tilde{K}_{n}(t) + b_{n}^2 + \tilde{K}_{n+1}(t) \tilde{K}_n(t)\;,    \nonumber
\end{eqnarray}
so that
\begin{equation}
\tilde{K}_{n+1} = \frac{\tilde{L}_n}{ \tilde{K}_n} - \frac{b_n^2}{\tilde{K}_n } - a_n .   \label{Recursive2}
\end{equation}
The above still depends on the Lanczos coefficients.  However, they can be eliminated by studying the boundary behavior of the functions $\tilde{K}_n(t)$ and $\tilde{L}_n(t)$.   Explicitly, we have that 
\begin{eqnarray}
\tilde{K}_n(0) & = & 0     \nonumber \\
\left.  \partial_t K_{n}(t)   \right|_{t=0} & = & \textnormal{constant} .    \nonumber
\end{eqnarray}
The above conditions are responsible for encoding the ordering of the $\psi_n(t)$ probability amplitudes.  Using this along with (\ref{Recursive2}) we find that
\begin{eqnarray}
b_n^2 & = & \tilde{L}_n(0),     \nonumber  \\
a_n & = & \left.  \frac{\tilde{L}_n(t) - \tilde{L}_n(0)}{\tilde{K_n}(t)}   \right|_{t \rightarrow 0}   .  \nonumber
\end{eqnarray}
Since we know what the boundary condition is on both the $\tilde{K}_n$ and $\tilde{L}_n$ we can fix the Lanczos coefficients in terms the functions $\tilde{L}_n$ and $\tilde{K}_n$ as we send $t\rightarrow 0$.\\

\noindent
This now leads to a set of recursive equations in terms of the survival amplitude (\ref{SurvivalAmplitude}). These are given by
\begin{eqnarray}
\tilde{\kappa}_{1}(t) & = & i\frac{S'(t) - S'(0) S(t)}{S(t)},     \nonumber \\
\tilde{L}_1(t) & = & -\frac{S''(t) - S'(0) S'(t)}{S(t)},    \nonumber   \\
\tilde{\kappa}_{n+1}(t) & = & \frac{\tilde{L}_n(t) - \tilde{L}_{n}(0)}{\tilde{\kappa}_n(t) - \tilde{\kappa}_n(0)},   \nonumber  \\
\tilde{L}_{n+1}(t) & =  & i \partial_t \tilde{\kappa}_{n+1}(t) + \frac{(\tilde{\kappa}_{n+1}(t) - \tilde{\kappa}_{n+1}(0)) \tilde{L}_{n}(t) }{(\tilde{\kappa}_{n}(t) - \tilde{\kappa}_{n}(0)}\;,
\end{eqnarray}
which is \textbf{independent} of the Lanczos coefficients. From the above we can obtain 
\begin{equation}
\tilde{K}_n(t) = \tilde{\kappa}_n(t) - \tilde{\kappa}_n(0)\;,
\end{equation}
and the probability amplitudes as 
\begin{equation}
|\psi_n(t)|^2 = |S(t)|^2 \left|\prod_{m=1}^n \frac{(\tilde{K}_m(t) )^2}{\tilde{L}_m(0)} \right| .
\end{equation}

\section{Numerical Method for Lanczos Algorithm in Position Space}
\label{sec:lanczos position space}

We have developed a methodology that implements the Lanczos algorithm directly in two-dimensional 
position space, allowing for the calculation of the Lanczos coefficients and of the wave functions of the 
Krylov basis vectors iteratively. Although here we focus on the application for the toy model with the exponential interaction, given by
\begin{equation}
    H = \frac{m\omega_1^2}{2}x_1^2 + \frac{m\omega_2^2}{2}x_2^2 + \frac{1}{2m}p_1^2 
    + \frac{1}{2m}p_2^2
    + g \hbar (\omega_1+\omega_2) e^{\alpha m^2 x_1 x_2} \equiv H_0 + H_I,
\end{equation}
the methodology could be adapted for use with many other potentials. The initial state, however, is required to 
be a Gaussian.\\

\noindent
We start from the definition of the Krylov basis,
\begin{align}
    \ket{K_n} &\equiv \frac{\ket{A_n}}{\sqrt{\braket{A_n}{A_n}}} = \frac{\ket{A_n}}{b_n}, \\
    \ket{A_n} &\equiv H \ket{K_{n-1}} - \left( \sum_{j=0}^{n-1} \ket{K_j}\bra{K_j} \right) H \ket{K_{n-1}},
\end{align}
and then rewrite it in the position space representation,
\begin{align}
\begin{split}
    \label{eq_An}
    A_n(x_1,x_2) &= H(x_1,x_2) K_{n-1}(x_1,x_2) \\
    &- \sum_{j=0}^{n-1} K_j (x_1,x_2) \int_{-\infty}^{\infty} dx_1 dx_2 K_j(x_1,x_2)^* H(x_1,x_2) K_{n-1}(x_1,x_2).
\end{split}
\end{align}
\noindent
In order to turn this into an efficient numerical algorithm, we rely on regularities in the form of the 
wave functions $A_n (x_1,x_2)$. We choose a Gaussian as initial state,
\begin{equation}
    \psi(x_1,x_2) \equiv
    \left( \frac{m\omega_1}{\pi\hbar} \right)^{\frac{1}{4}}
    \left( \frac{m\omega_2}{\pi\hbar} \right)^{\frac{1}{4}}
    e^{ -\frac{1}{2\hbar} \left( m\omega_1(x_1-c_1)^2 + m\omega_2(x_2-c_2)^2 \right) },
\end{equation}
so that all of the $A_n(x_1,x_2)$ wave functions are given by
\begin{equation}
    \phi=\sum_{\mu,\nu,z} c(\mu,\nu,z) x_1^{\mu} x_2^{\nu} H_I^z \psi.
\end{equation}
Here $c(\mu,\nu,z)$ indicates a set of numerical coefficients associated with the unspecified wave 
function $\phi$ 
and with a specific value for each term of the sum, and $H_I$ refers to the position space representation of the 
interacting part of the Hamiltonian (the term exponential in $x_1x_2$). This regular form is merely a 
consequence of 
the fact that the action of the Hamiltonian on a function of this form always generates another one with the 
same form.\\

\noindent
Therefore, knowing the coefficients $c(\mu,\nu,z)$ is enough to reconstruct the full wavefunction $\phi$. 
Additionally, this allows us to implement the operations that appear in the Lanczos algorithm simply as numerical 
operations on these coefficients. Sums or products of wave functions can be realized in this way, as 
well as the application of the Hamiltonian. As for the integrals appearing in the projections, every one of them takes the form
\begin{equation}
    \int_{-\infty}^{\infty} dx_1 dx_2 \; \phi_1^*\phi_2
    =\sum_{\mu,\nu,z} c(\mu,\nu,z)\int_{-\infty}^{\infty} dx_1 dx_2 \;x_1^{\mu} x_2^{\nu} H_I^z \psi^2.
\end{equation}
The integral on the right is related to the Gaussian integral and can be solved analytically for general $\mu, \nu, z$. The solution can then be defined as a function of these indices and applied once for each term in the sum.\\

\noindent
This concludes the description of our method. Naturally, the wave functions involved become more complex as the algorithm advances, and the sets of coefficients $c(\mu,\nu,z)$ become very big, leading to higher computational costs. Hence, this is only useful for obtaining the first few Krylov vectors, or early time Krylov complexity. All the numerical data shown in the text was obtained from an implementation in Mathematica.

\section{Krylov basis wave functions}
\label{sec:Krylov basis wave functions}

\begin{figure}[hbt!]
    \centering
    \includegraphics[width=0.3\linewidth]{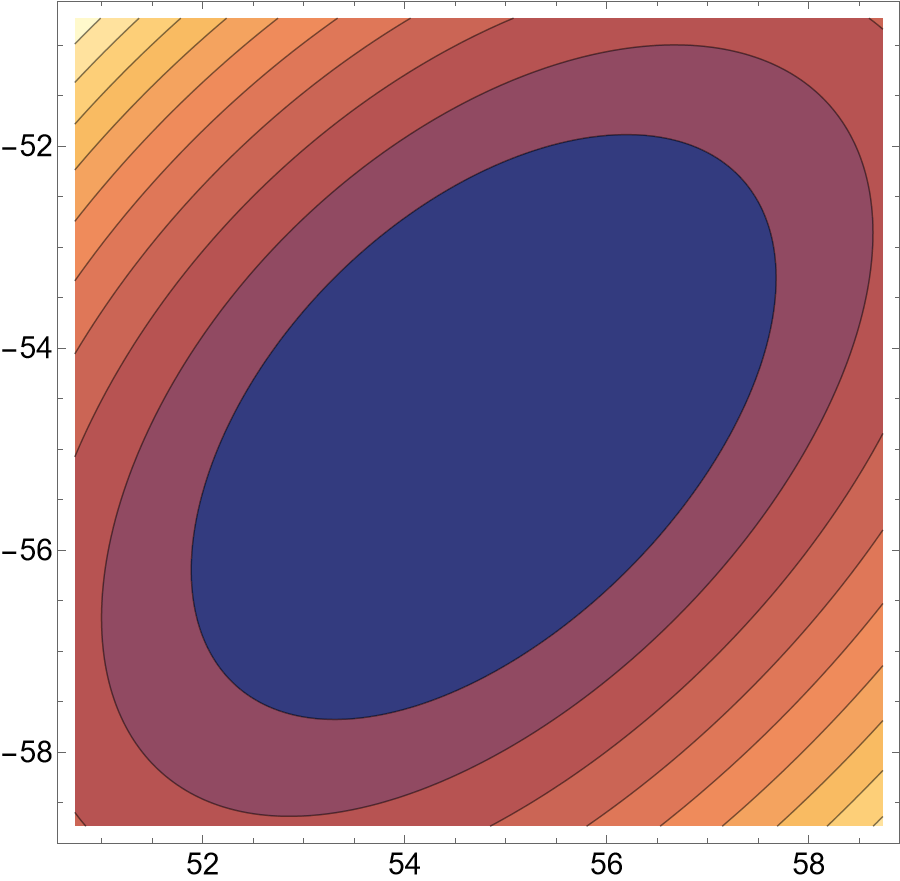}
    \includegraphics[width=0.3\linewidth]{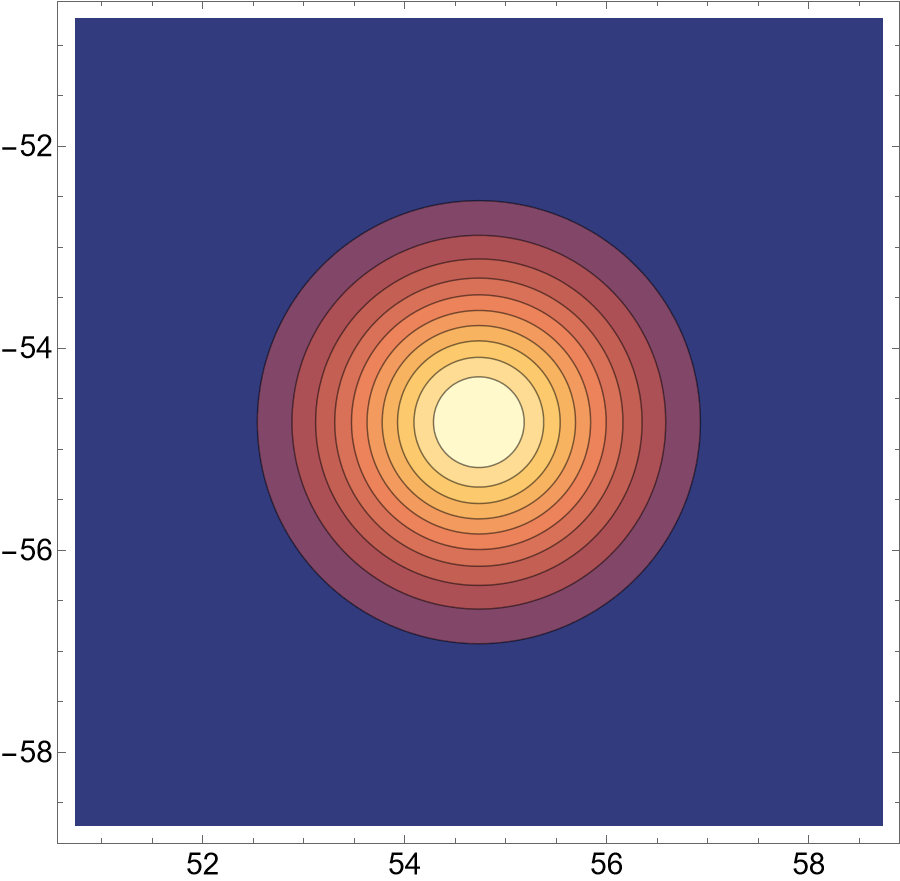}
    \includegraphics[width=0.3\linewidth]{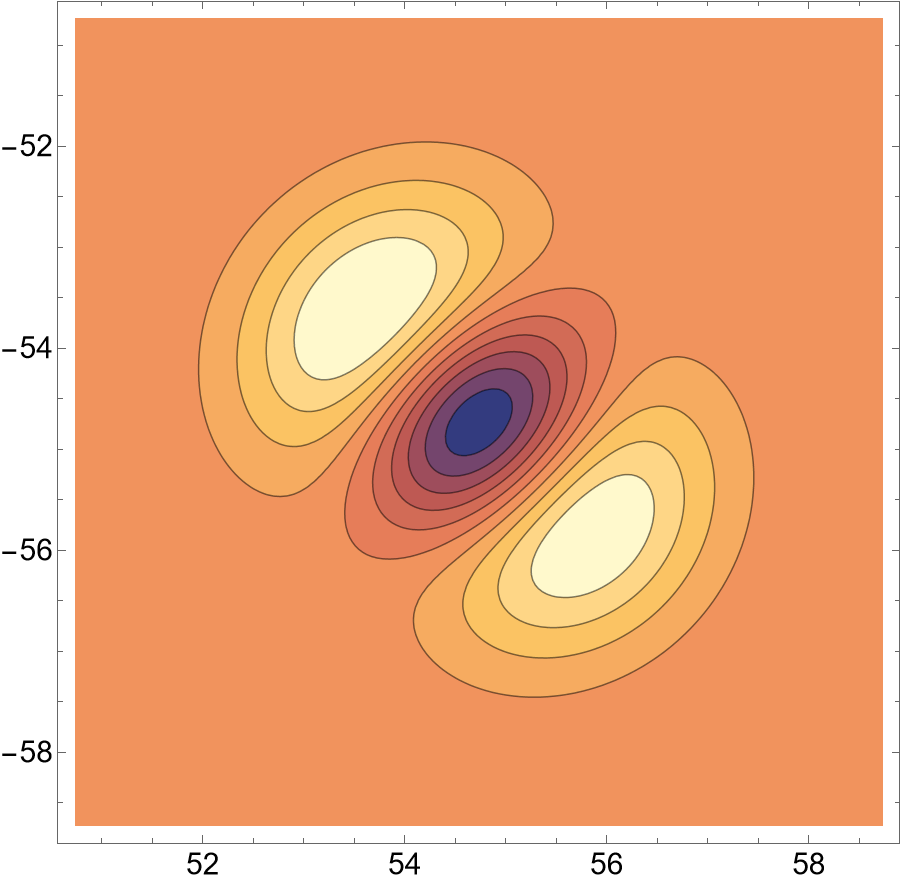}
    \includegraphics[width=0.3\linewidth]{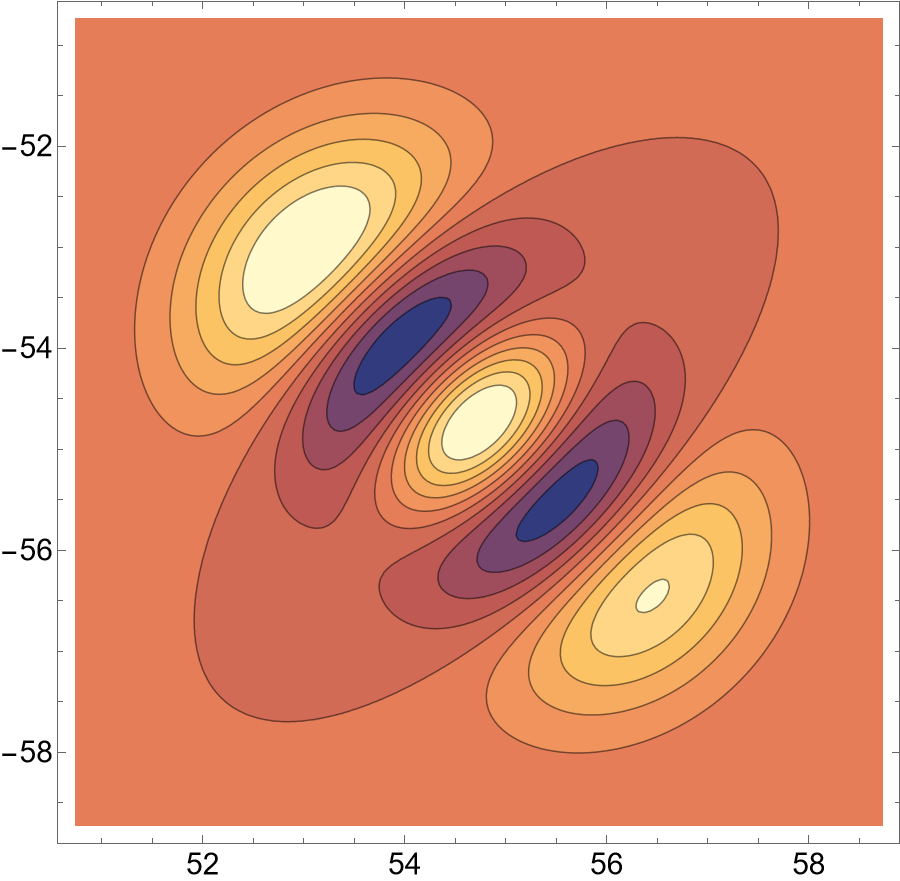}
    \includegraphics[width=0.3\linewidth]{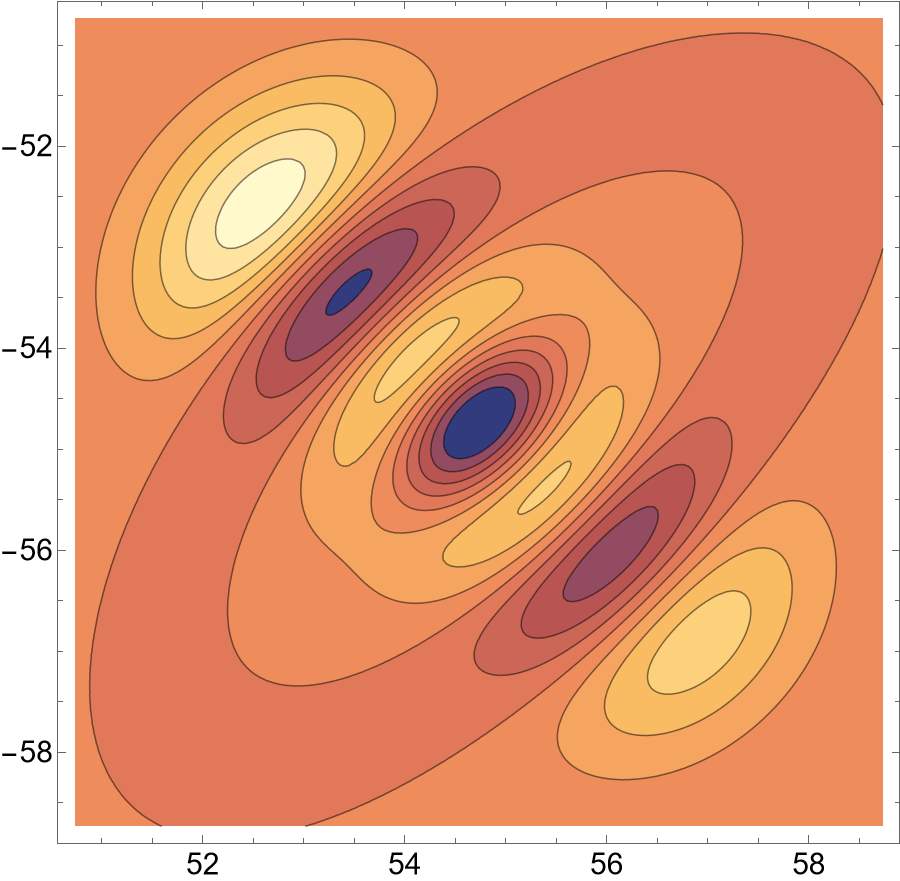}
    \includegraphics[width=0.3\linewidth]{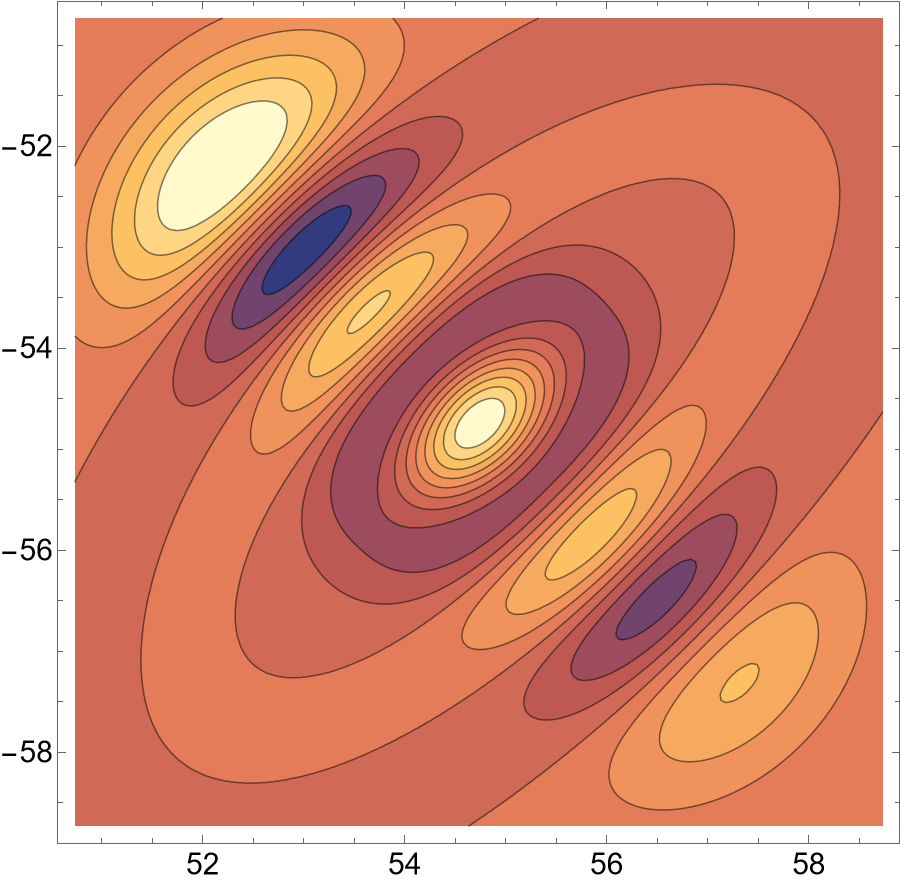}
    \caption{Contour plots of the potential and of the first Krylov wave functions for $g=10^4$. The reference state is a gaussian centered at one of the minima.}
    \label{fig:waves g1 shifted}
\end{figure}

\noindent
Here we include, as a complement to section \ref{sec:lanczos coefficients small n}, contour plots of the first Krylov wave functions obtained for our toy model with different values of $g$. Unless otherwise specified, the reference state is a gaussian centered at the origin (the only exception is Figure \ref{fig:waves g1 shifted}). In each figure are displayed, from top to bottom and left to right, the potential and the wave functions of the Krylov vectors: $K_0,K_1,K_2,K_3,K_4$. These are all real functions and lighter colors are used to indicate higher (more positive) values. The horizontal axis of each plot refers to $x_1$ and the vertical to $x_2$.

\begin{figure}[hbt!]
    \centering
    \includegraphics[width=0.3\linewidth]{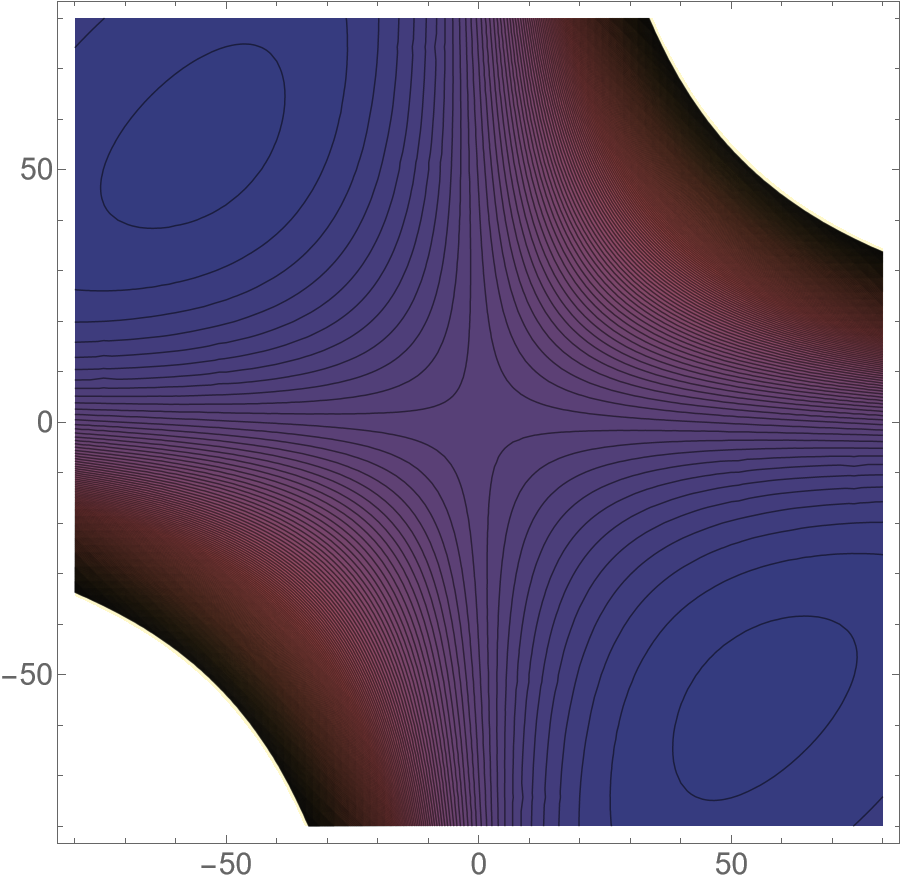}
    \includegraphics[width=0.3\linewidth]{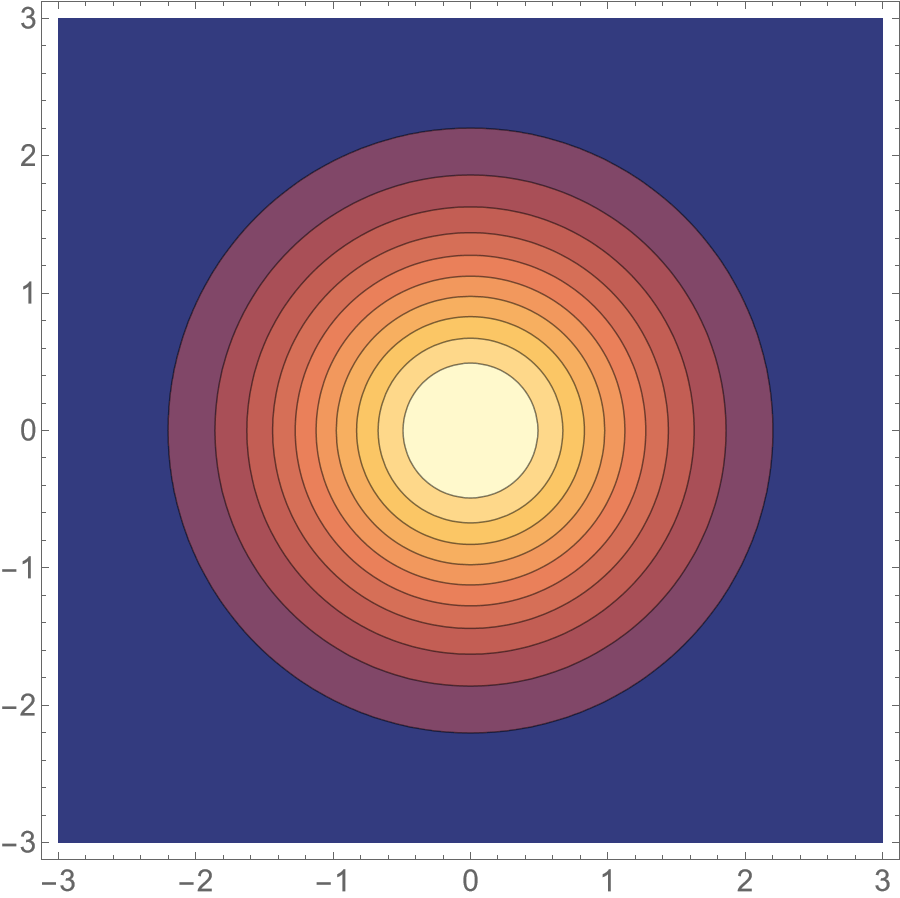}
    \includegraphics[width=0.3\linewidth]{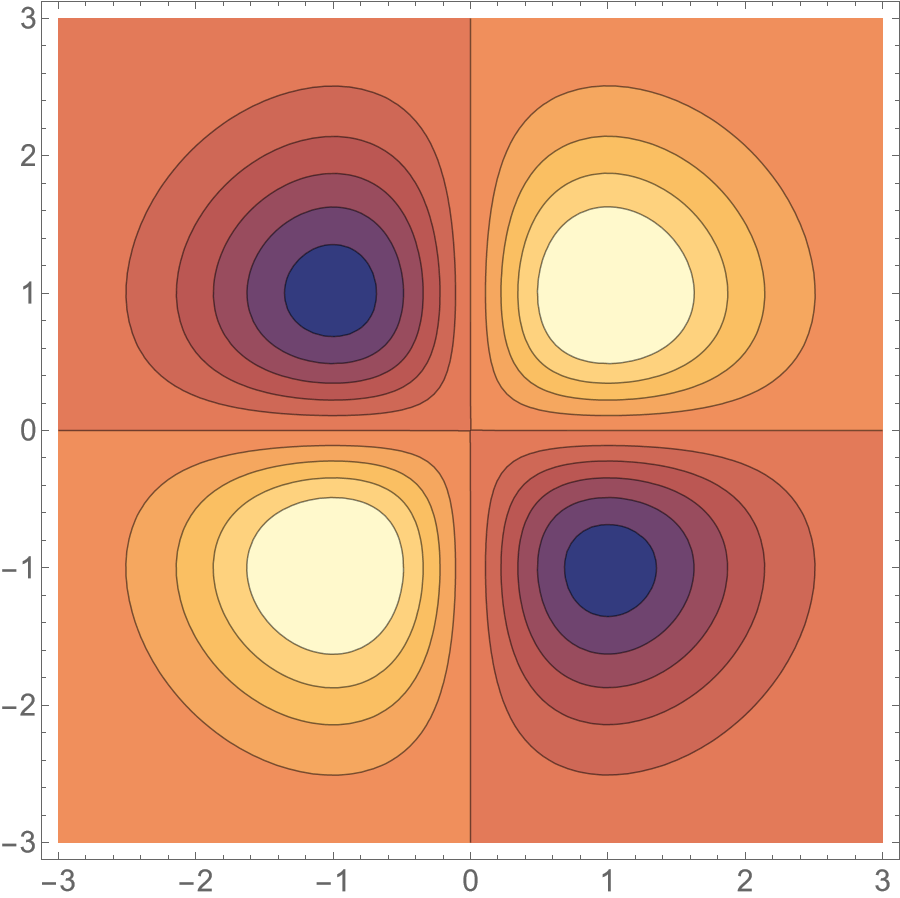}
    \includegraphics[width=0.3\linewidth]{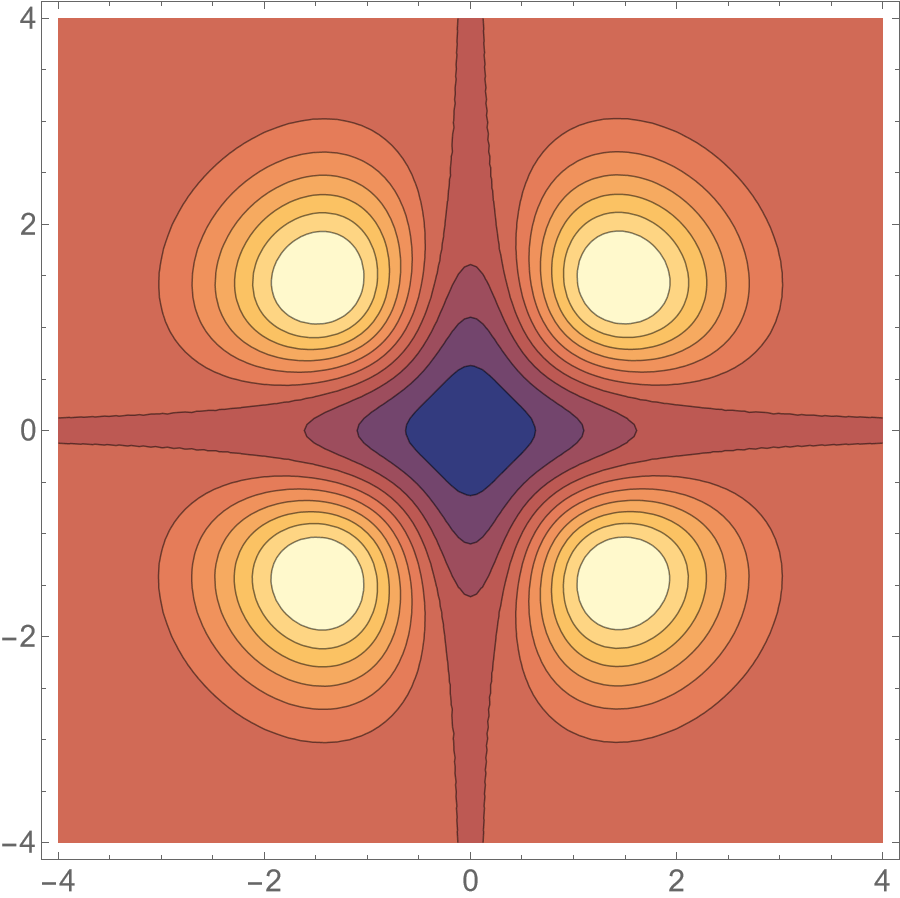}
    \includegraphics[width=0.3\linewidth]{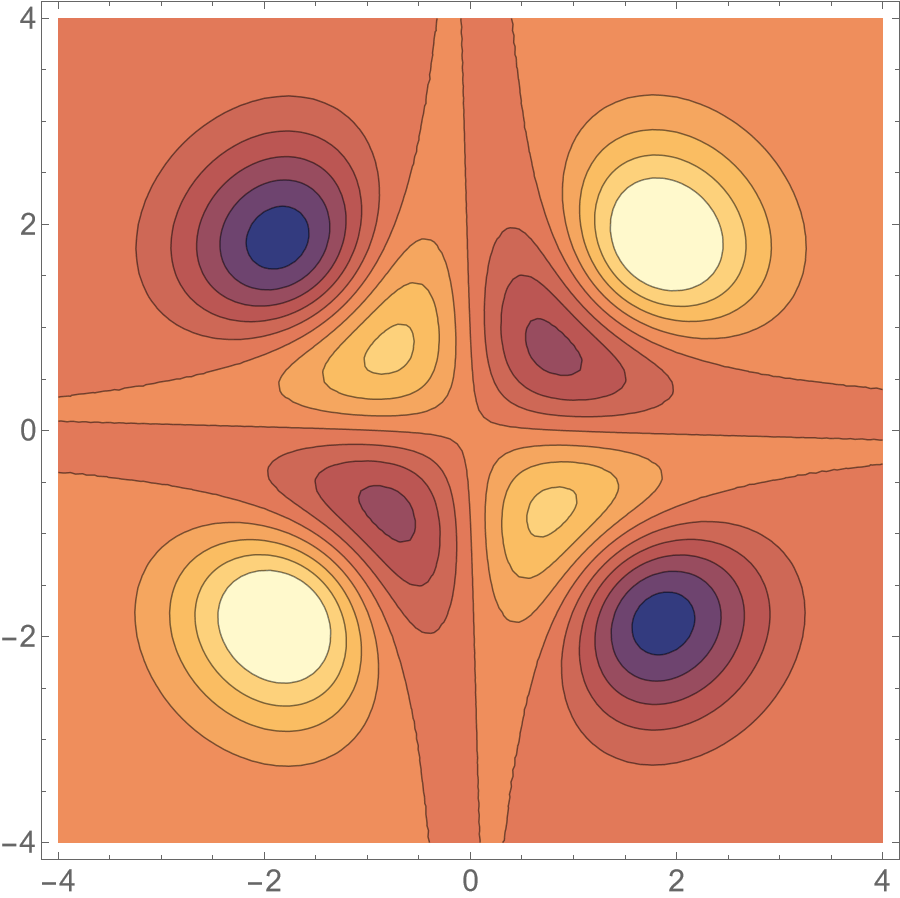}
    \includegraphics[width=0.3\linewidth]{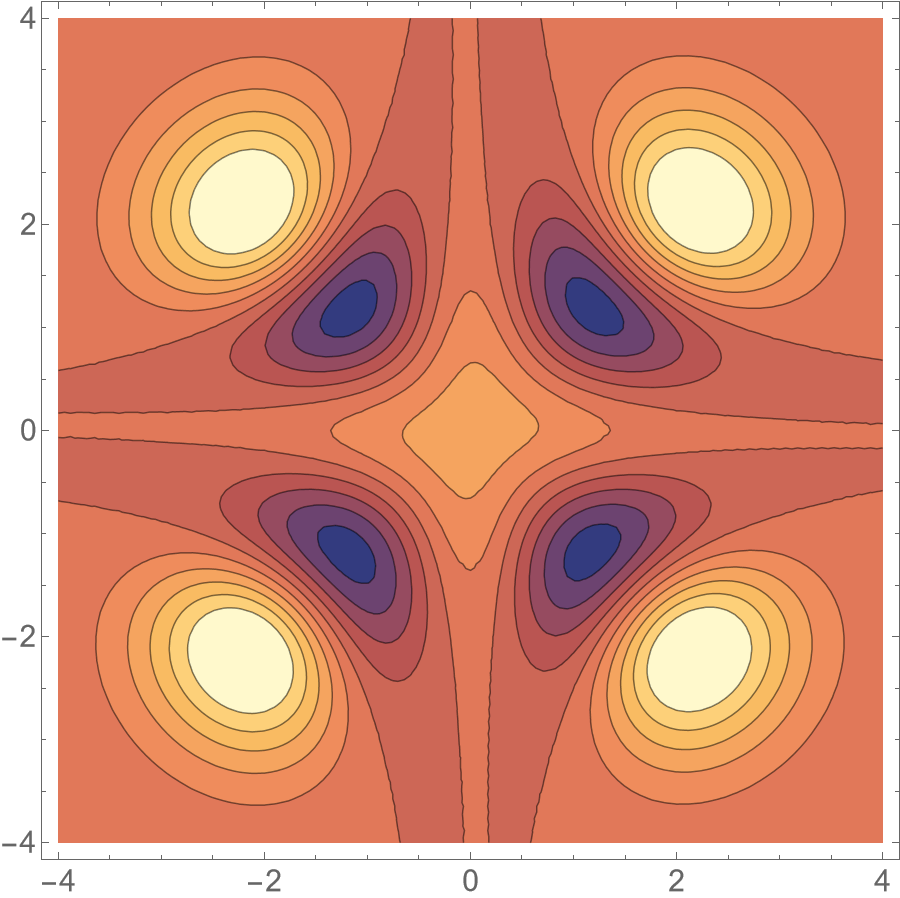}
    \caption{Contour plots of the potential and of the first Krylov wave functions for $g=10^4$.}
    \label{fig:waves g4}
\end{figure}

\begin{figure}[hbt!]
    \centering
    \includegraphics[width=0.3\linewidth]{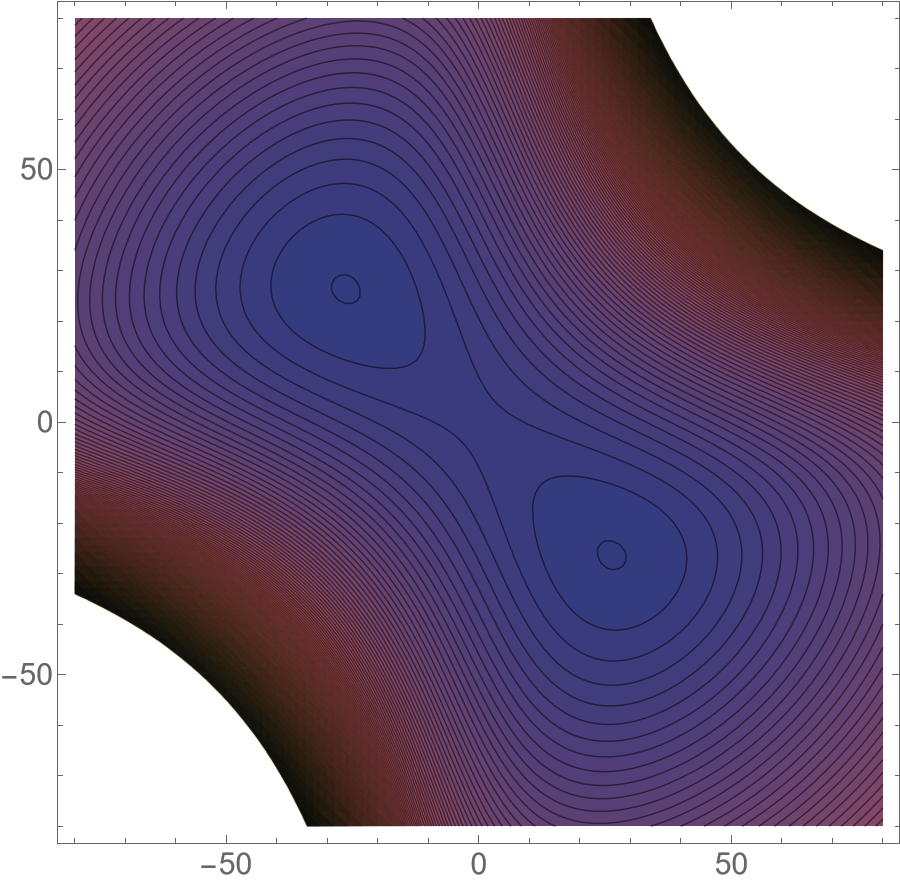}
    \includegraphics[width=0.3\linewidth]{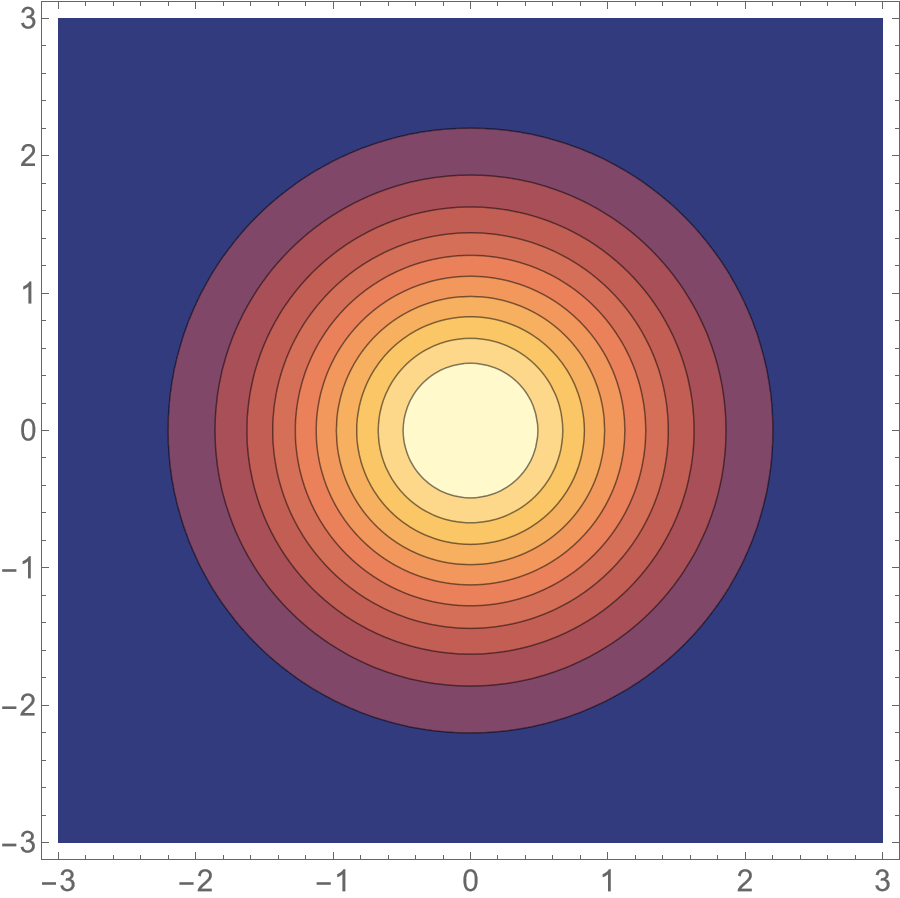}
    \includegraphics[width=0.3\linewidth]{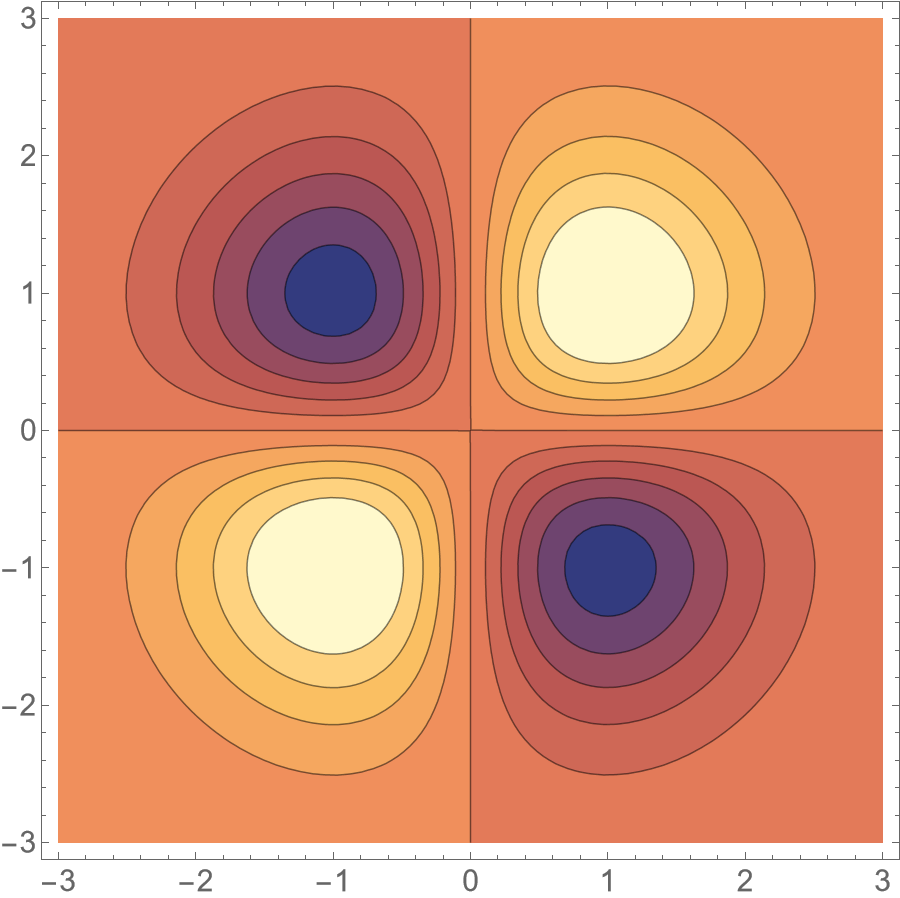}
    \includegraphics[width=0.3\linewidth]{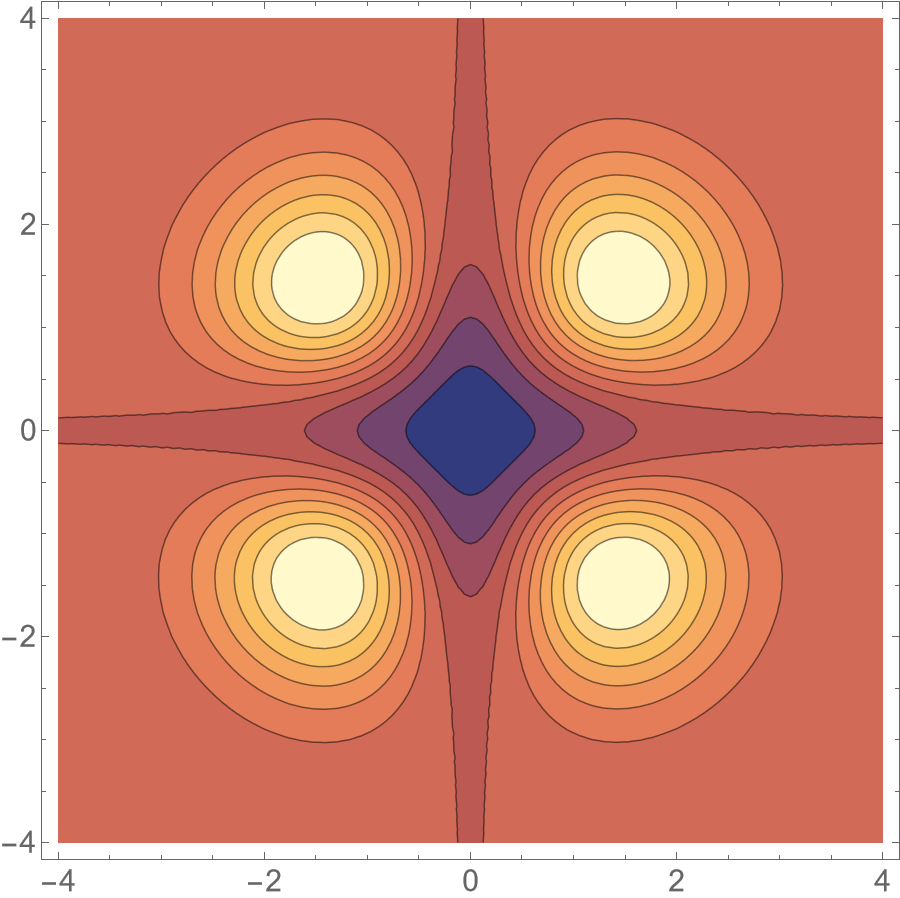}
    \includegraphics[width=0.3\linewidth]{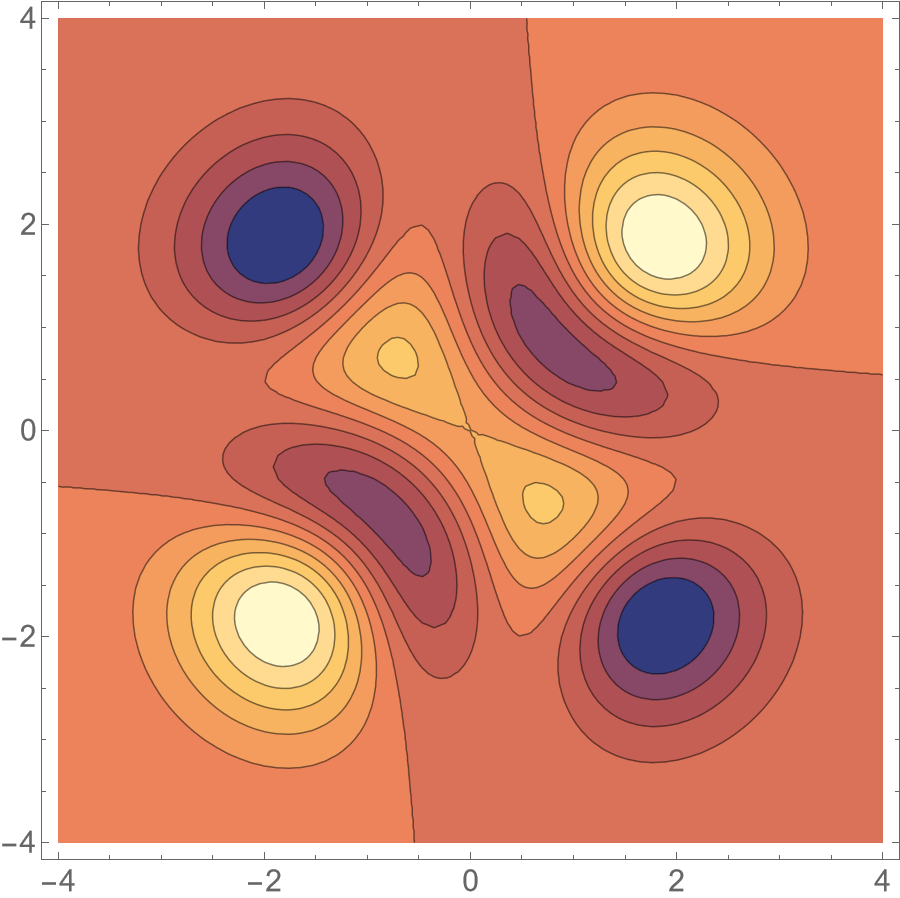}
    \includegraphics[width=0.3\linewidth]{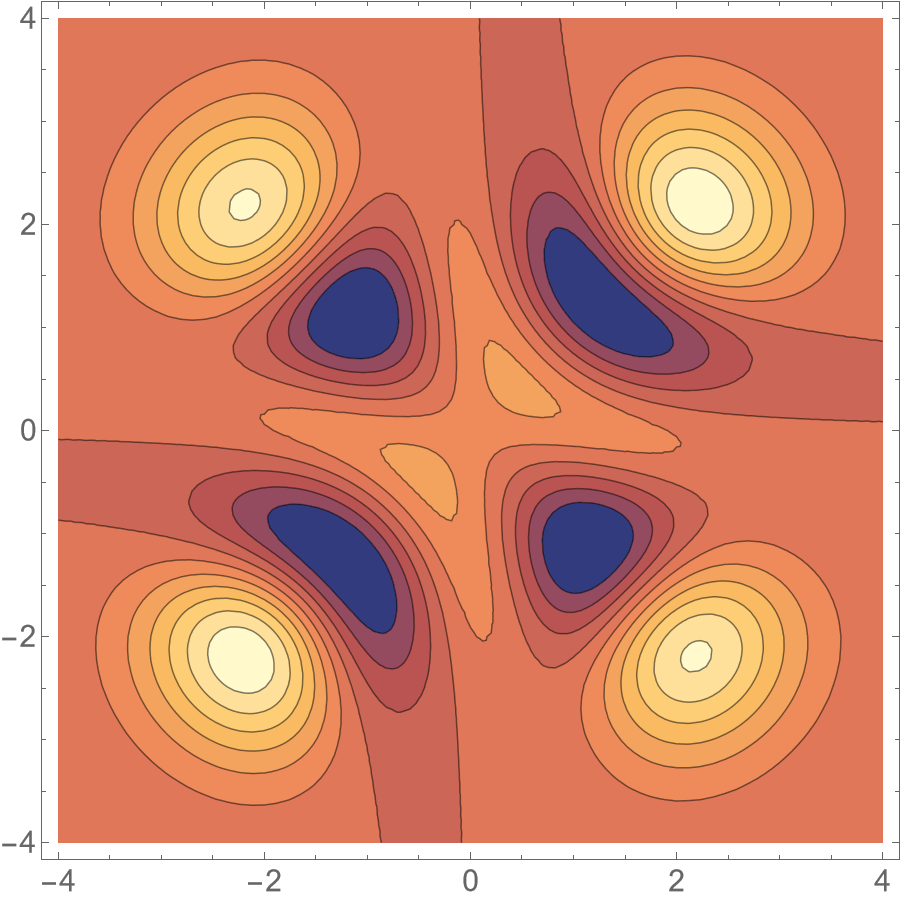}
    \caption{Contour plots of the potential and of the first Krylov wave functions for $g=10^3$.}
    \label{fig:waves g3}
\end{figure}

\begin{figure}[hbt!]
    \centering
    \includegraphics[width=0.3\linewidth]{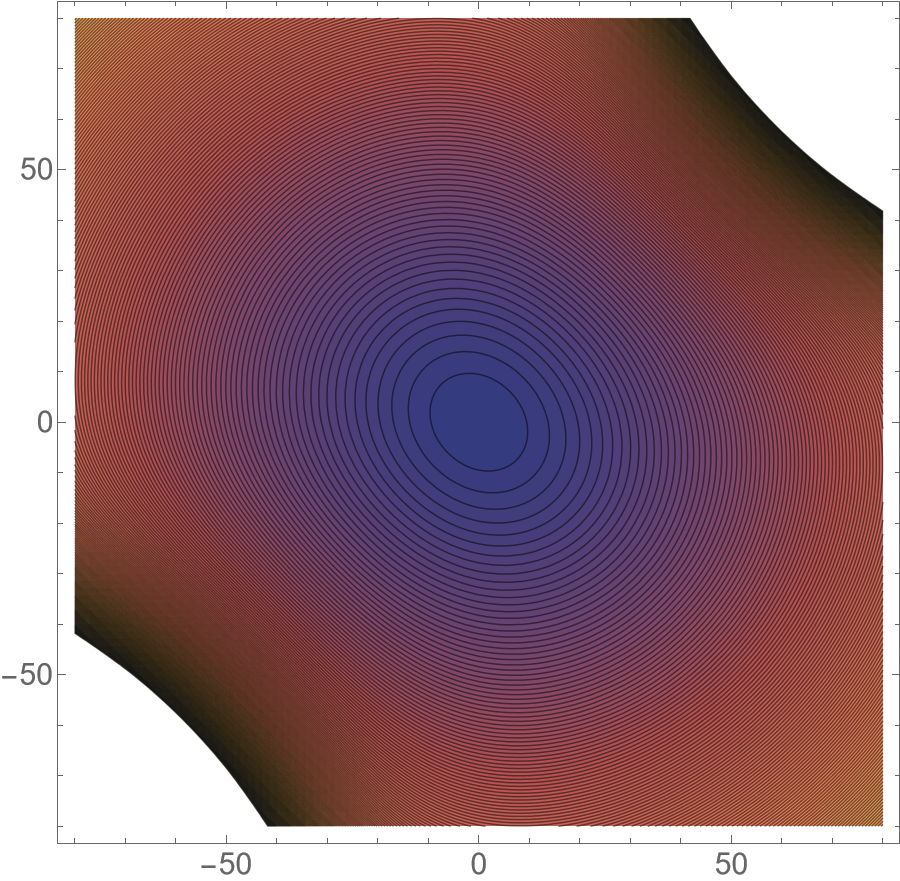}
    \includegraphics[width=0.3\linewidth]{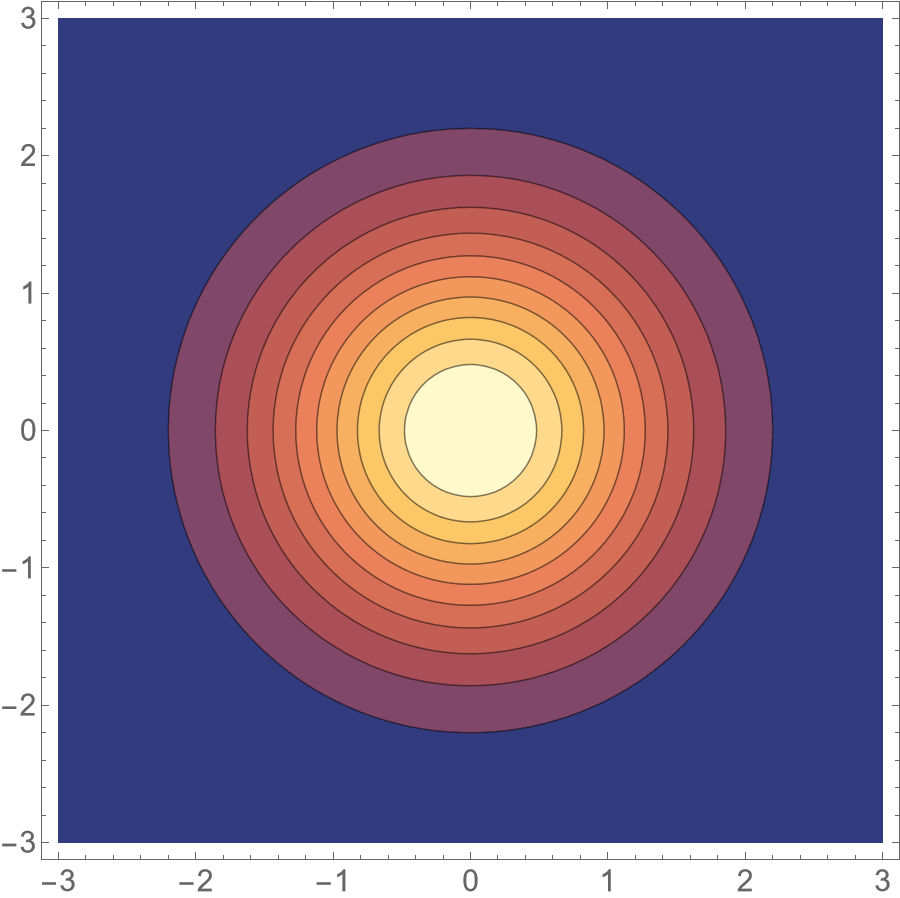}
    \includegraphics[width=0.3\linewidth]{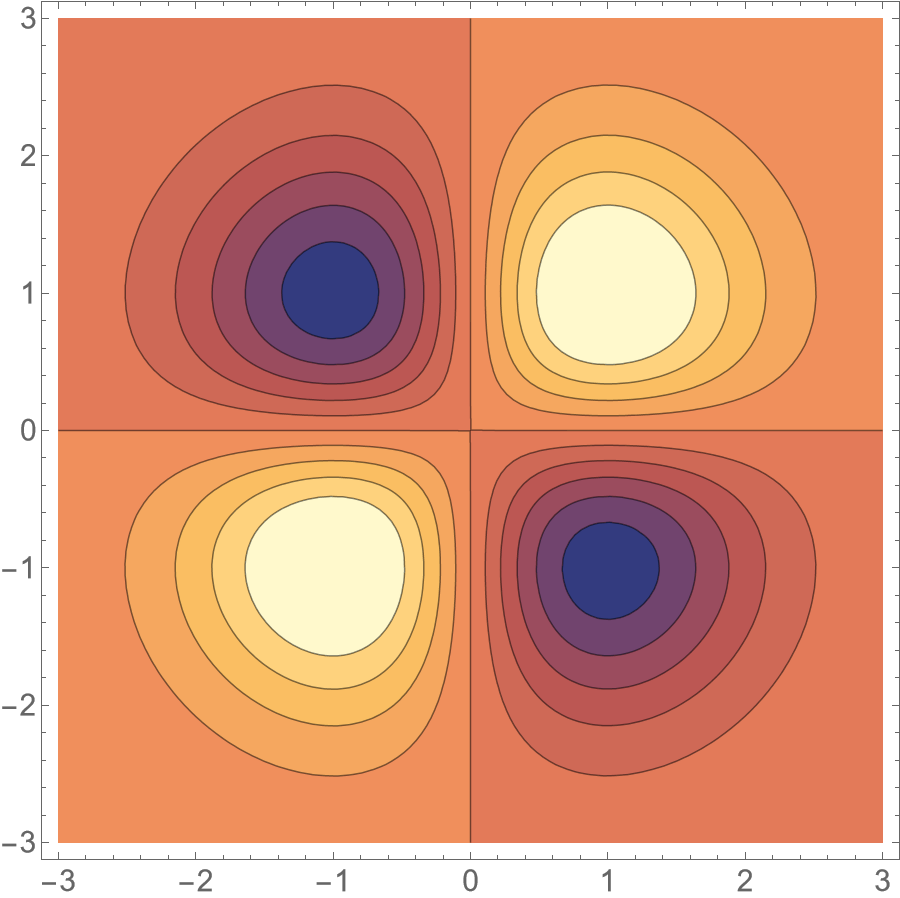}
    \includegraphics[width=0.3\linewidth]{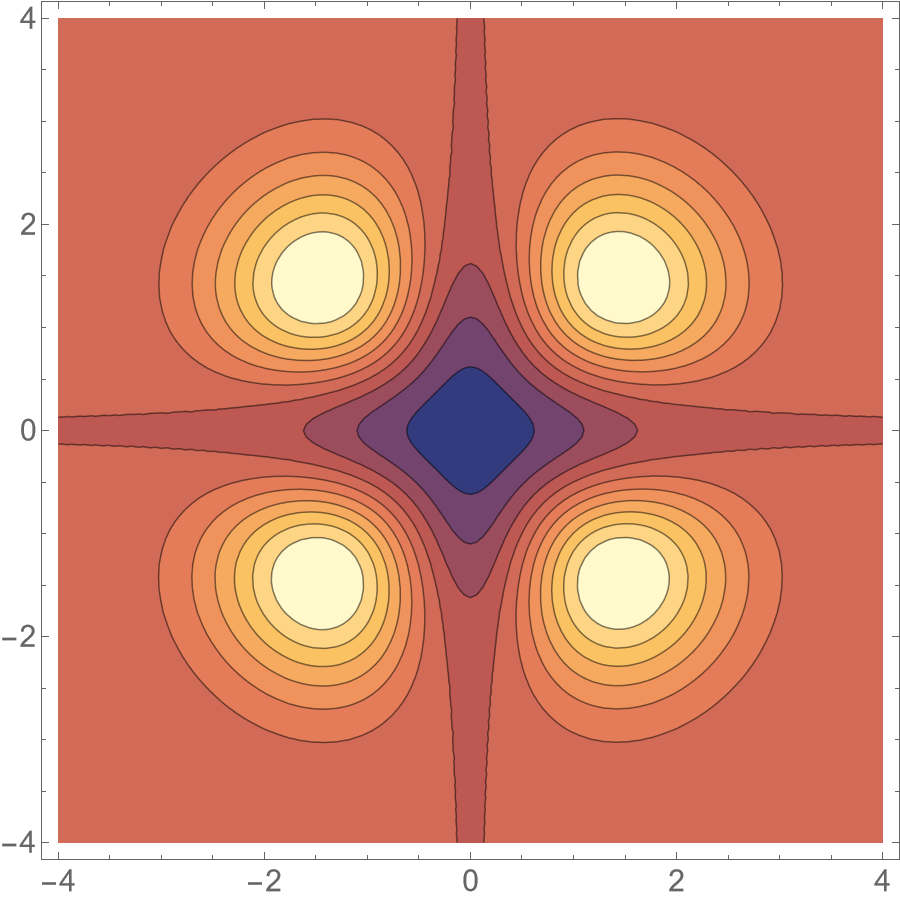}
    \includegraphics[width=0.3\linewidth]{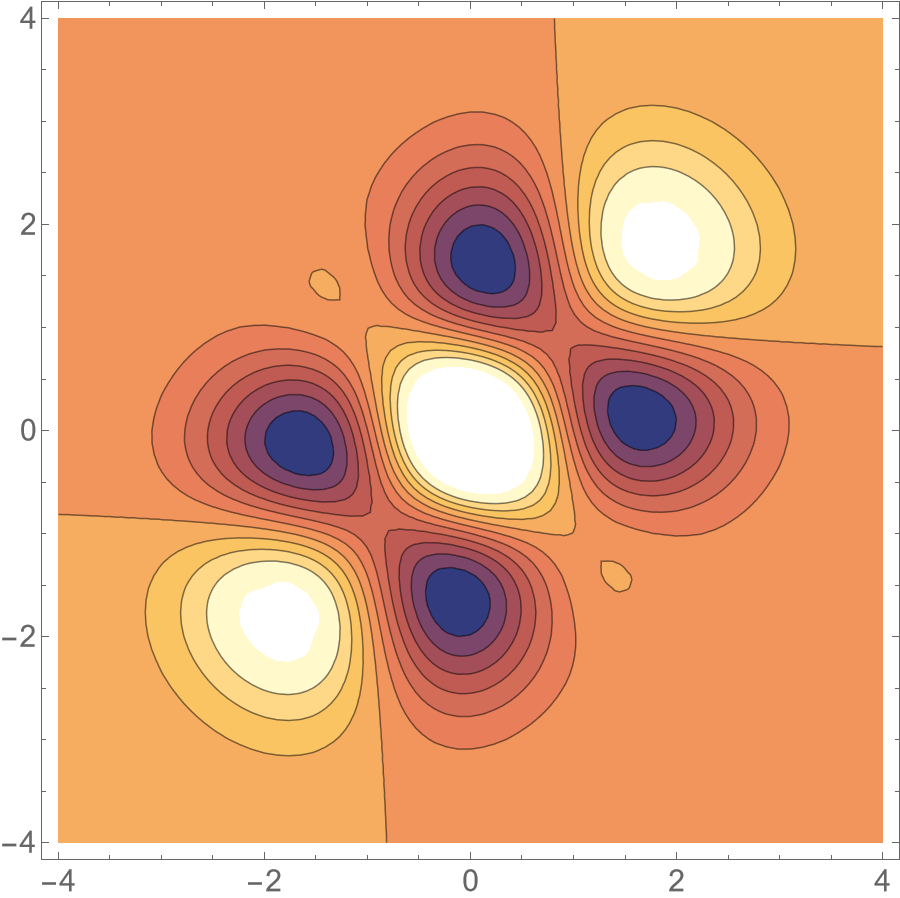}
    \includegraphics[width=0.3\linewidth]{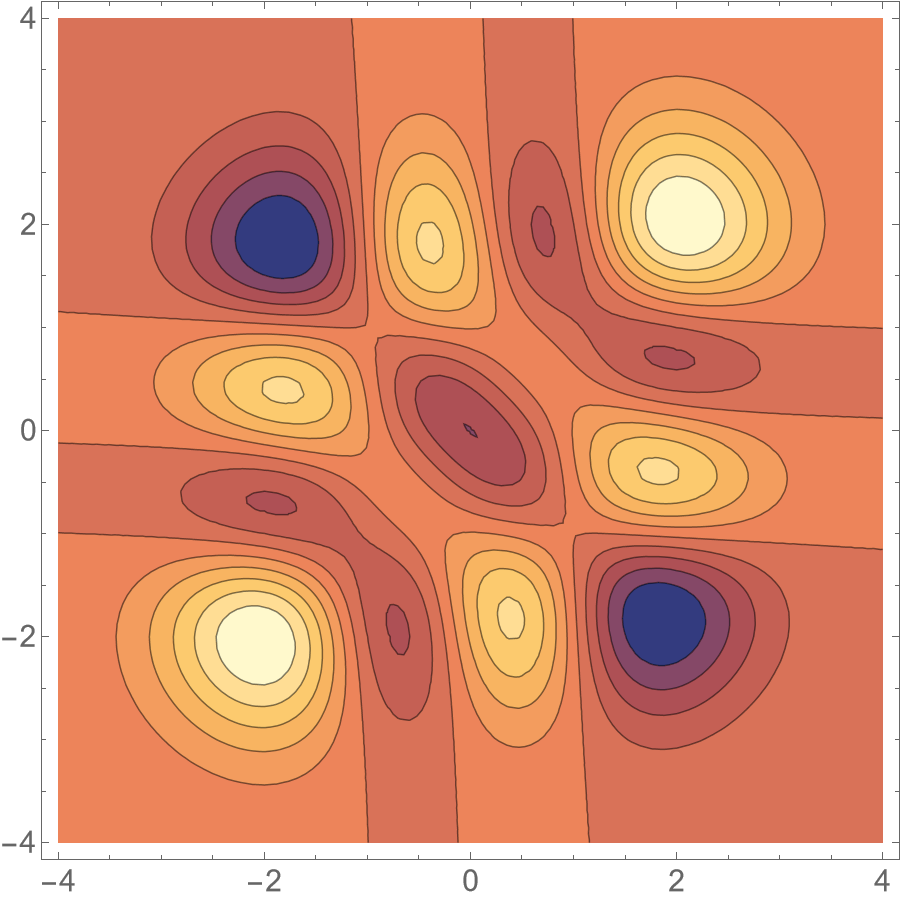}
    \caption{Contour plots of the potential and of the first Krylov wave functions for $g=10^2$.}
    \label{fig:waves g2}
\end{figure}

\begin{figure}[hbt!]
    \centering
    \includegraphics[width=0.3\linewidth]{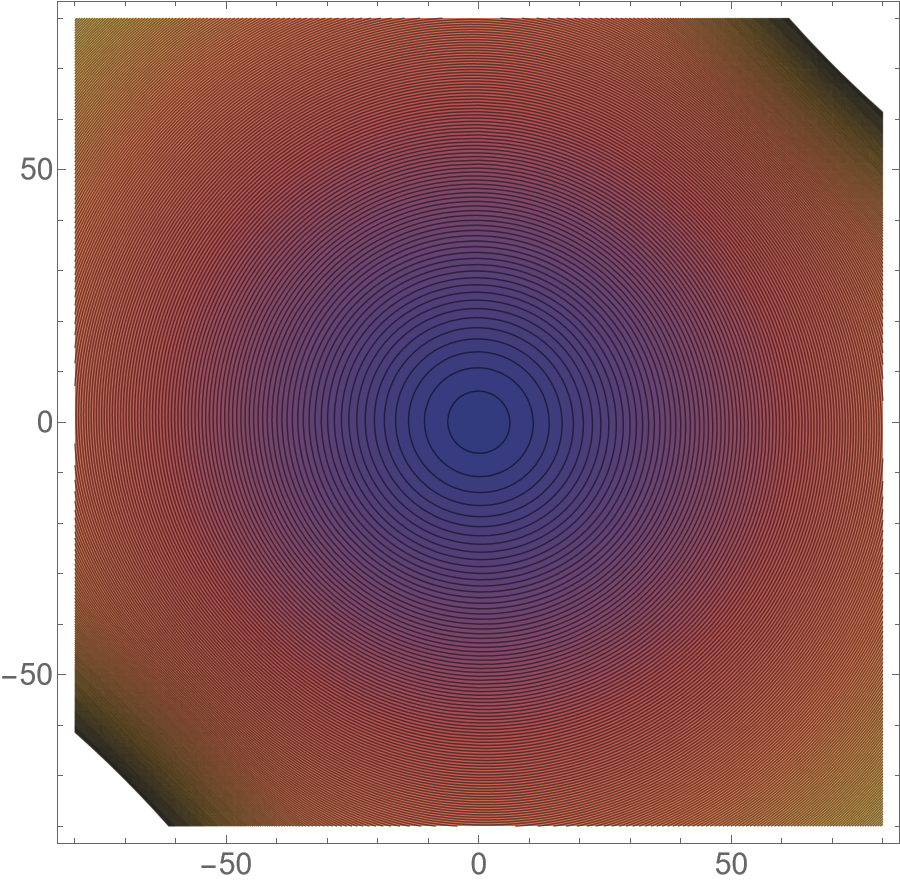}
    \includegraphics[width=0.3\linewidth]{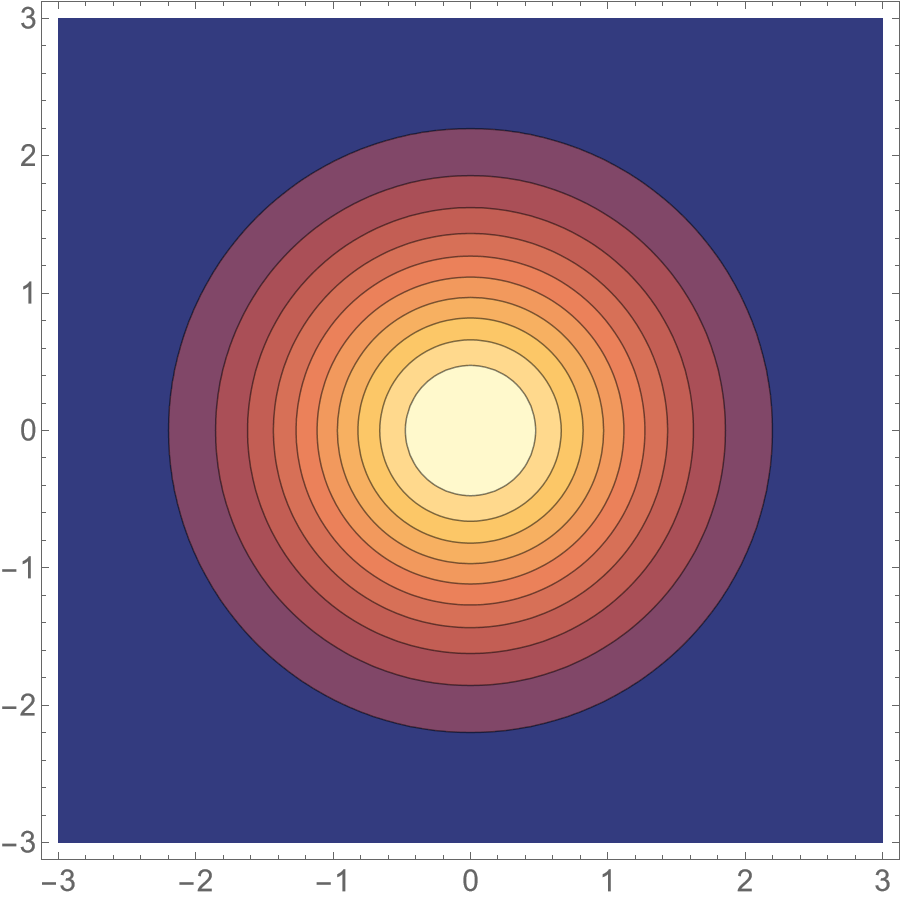}
    \includegraphics[width=0.3\linewidth]{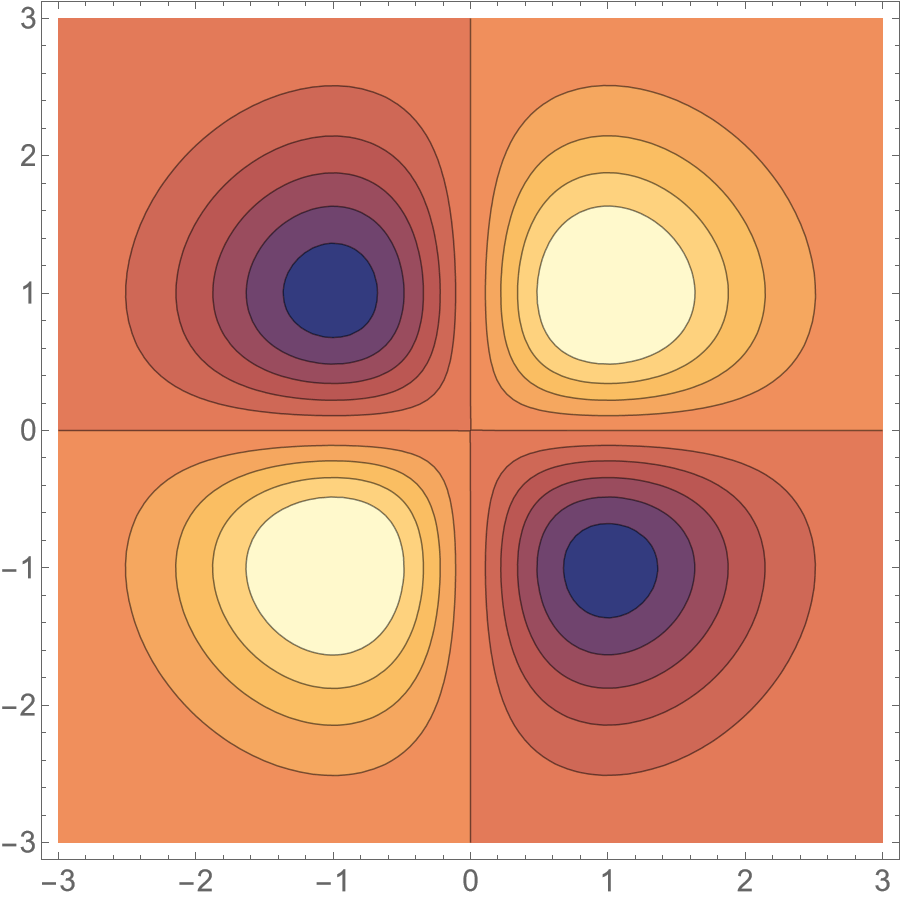}
    \includegraphics[width=0.3\linewidth]{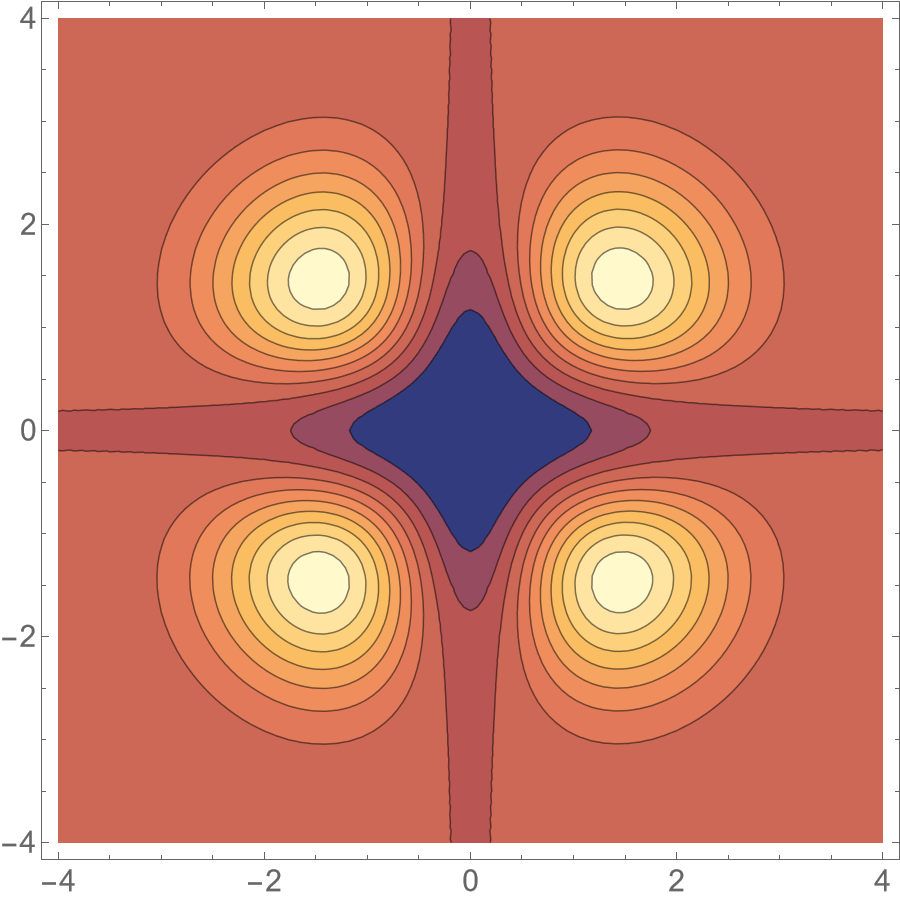}
    \includegraphics[width=0.3\linewidth]{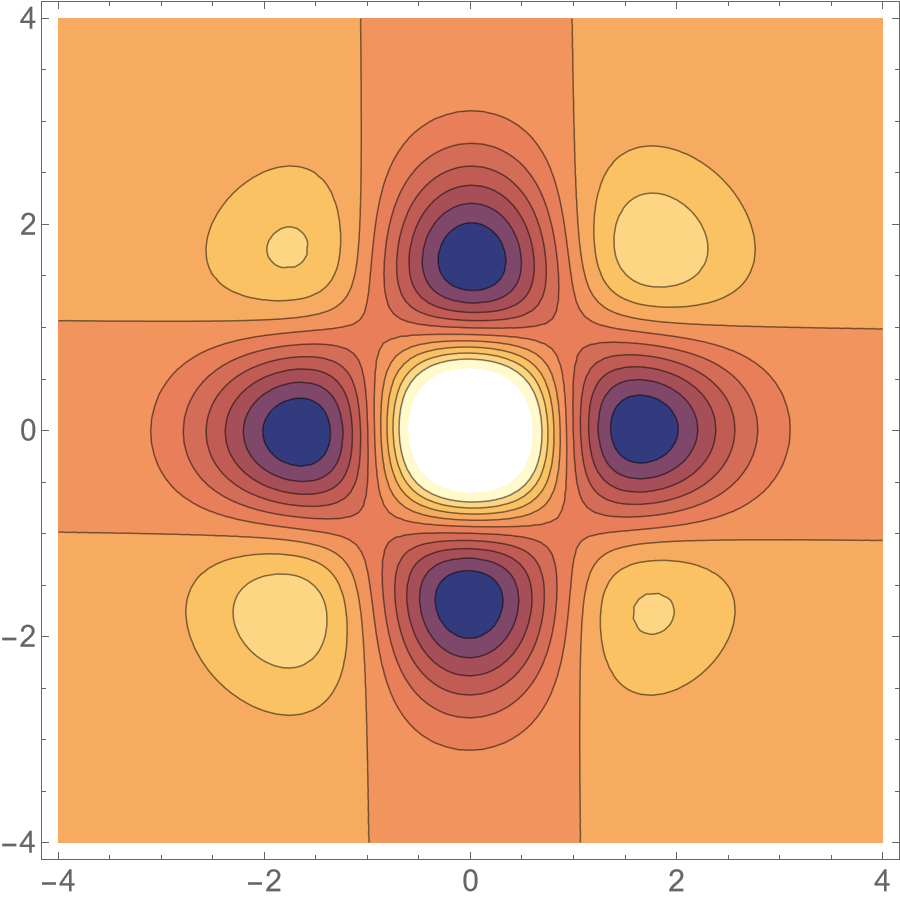}
    \includegraphics[width=0.3\linewidth]{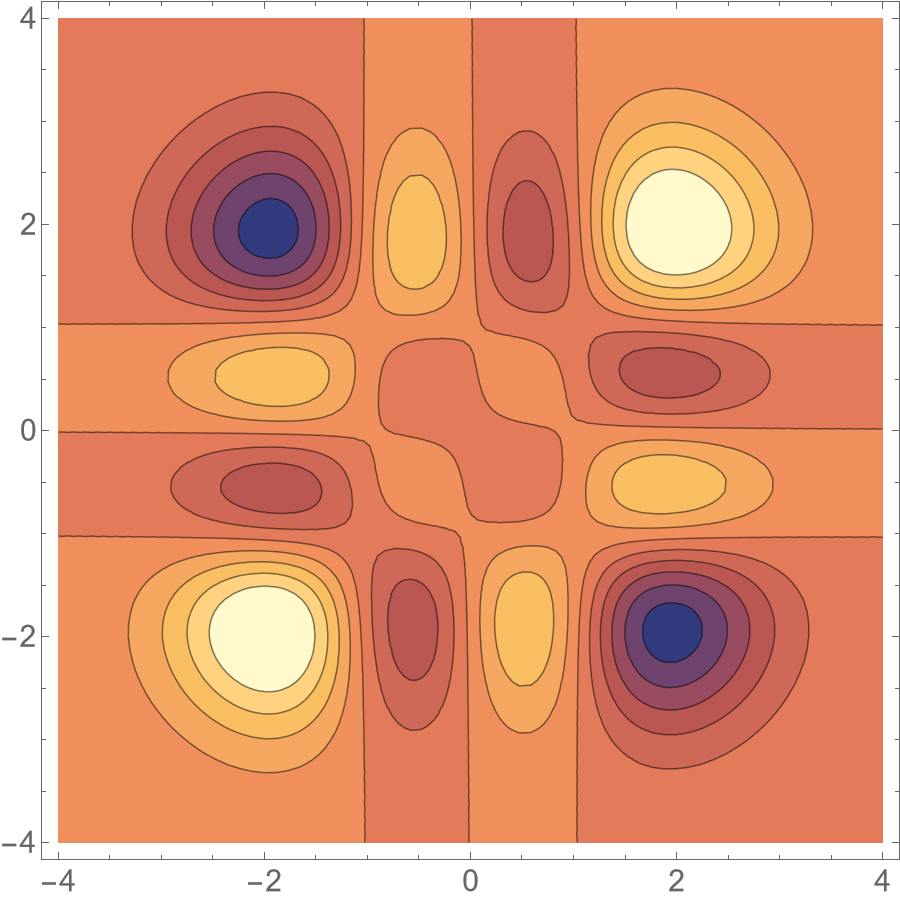}
    \caption{Contour plots of the potential and of the first Krylov wave functions for $g=10^1$.}
    \label{fig:waves g1}
\end{figure}

\section{Modified Toy Model}

\label{ModifedToyModelAppendix}

\begin{figure}
\centering
\begin{minipage}{0.48\textwidth}
		\includegraphics[width=0.95\textwidth]{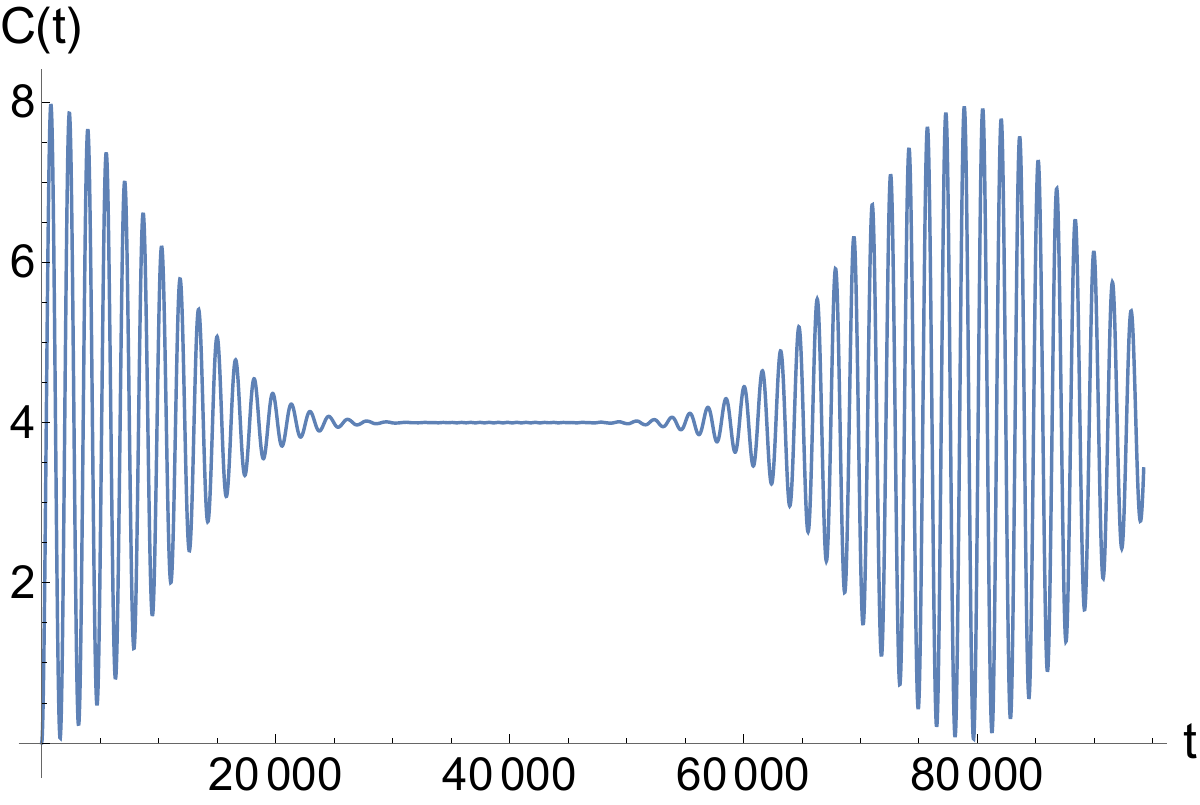}
\end{minipage}
\begin{minipage}{0.48\textwidth}
		\includegraphics[width=0.95\textwidth]{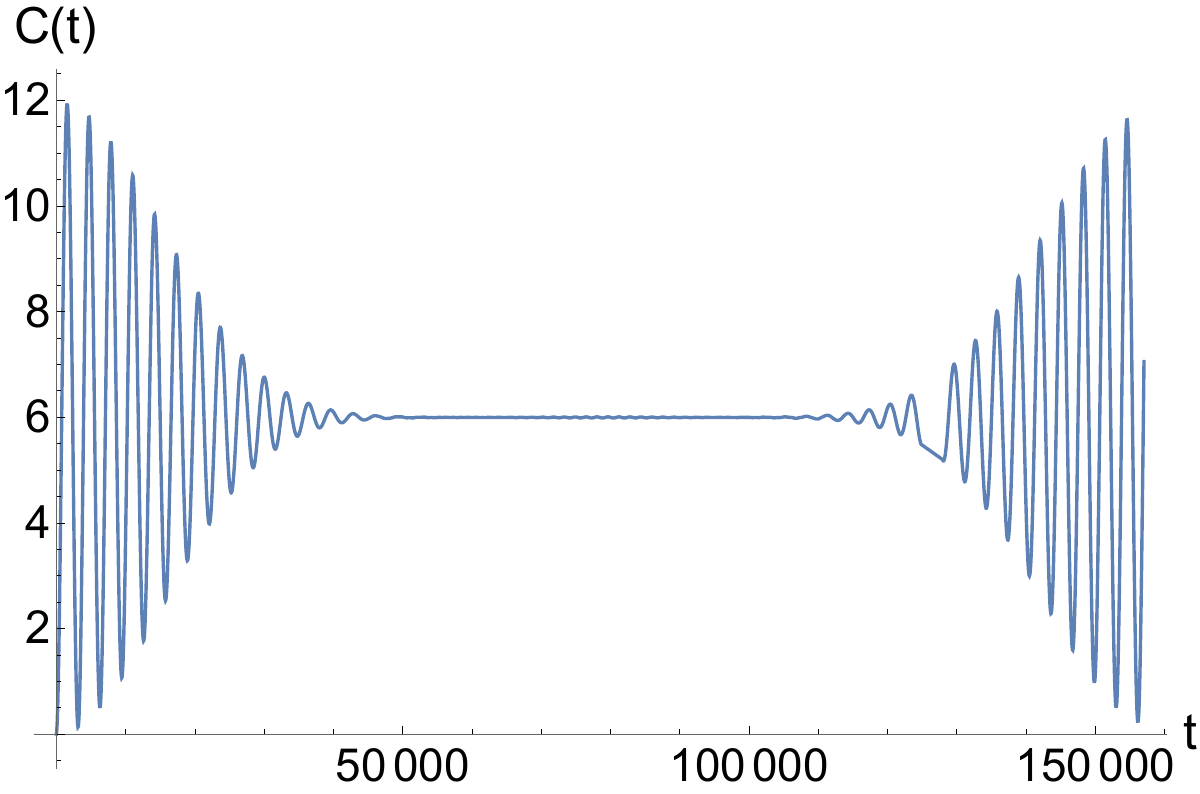}

	\end{minipage}
	\caption{	The complexity of the time-evolved reference state for $q=8, \omega_1=1, g = 0.2, \alpha = \log(1.01)$ (left) and $q=12, g = 0.1, \alpha = \log(1.01)$ (right) .  The complexity is initially oscillatory but saturates at a value of $\frac{q}{2}$ at late times.  At even later times the cycle repeats.  }
		\label{CtForToyModel}
\end{figure}

\begin{figure}
\centering
\begin{minipage}{0.48\textwidth}
		\includegraphics[width=0.95\textwidth]{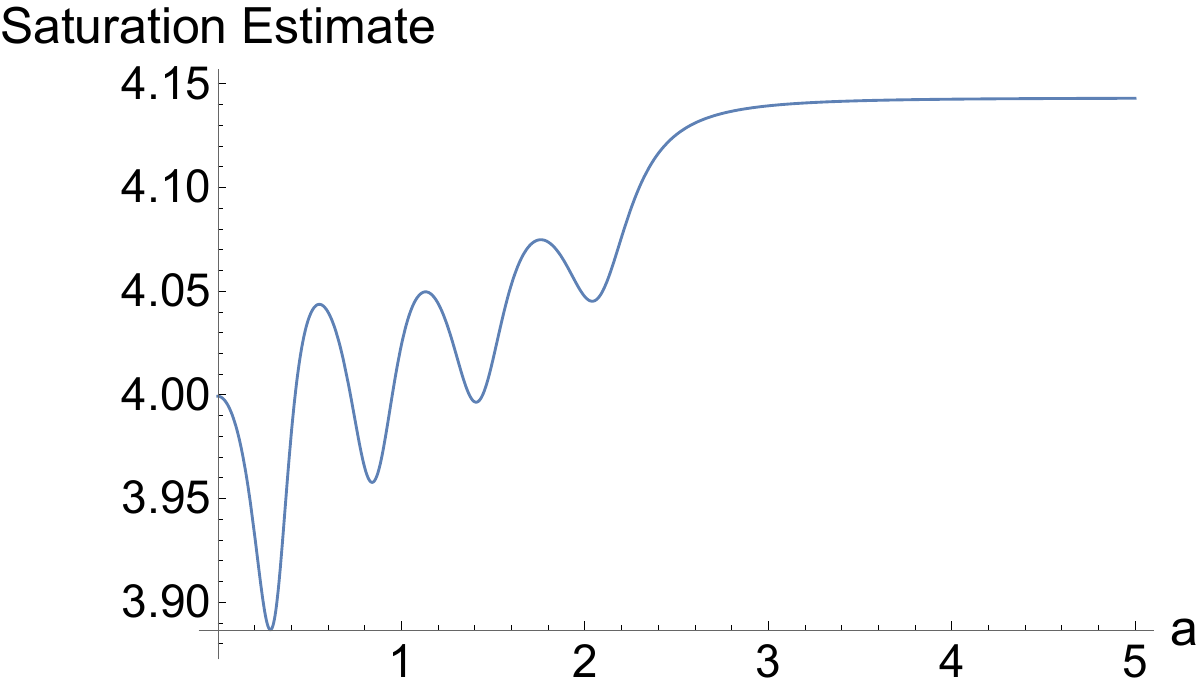}
\end{minipage}
\begin{minipage}{0.48\textwidth}
		\includegraphics[width=0.95\textwidth]{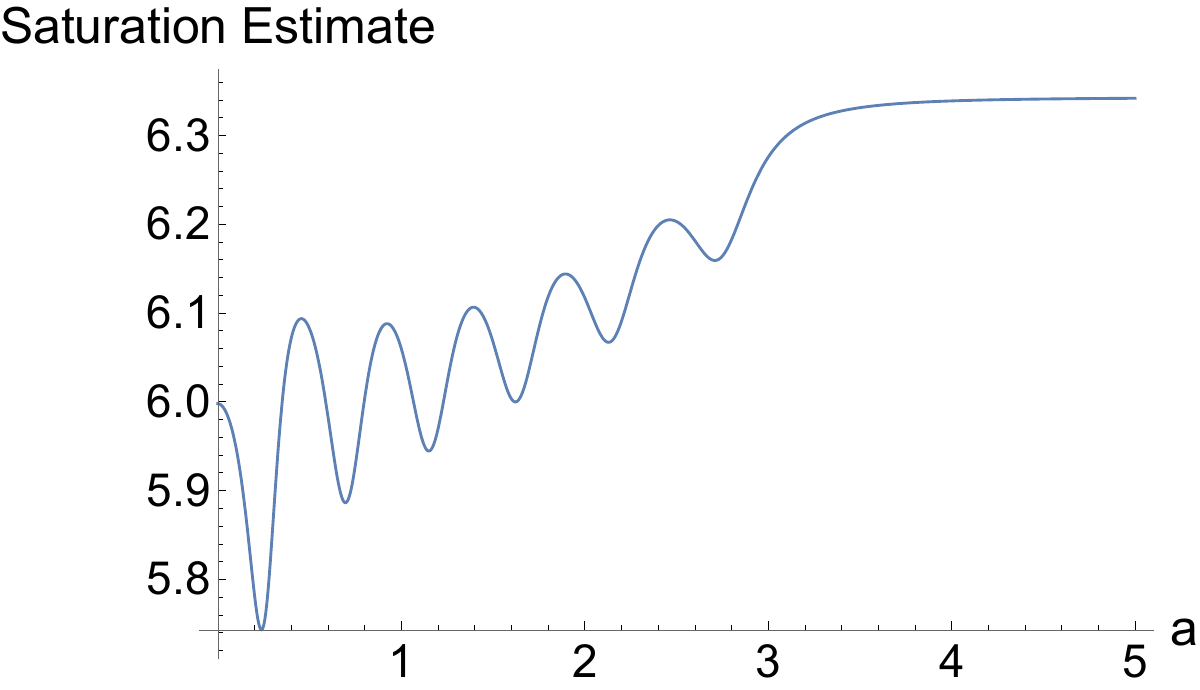}

	\end{minipage}
	\caption{	An estimate for the saturation value obtained by normalising the "instanton sum" formula for $q=8, \omega_1=1, g = 0.2, \alpha = \log(1.01)$ (left) and $q=12, g = 0.1, \alpha = \log(1.01)$ (right).  These agree well with the observed value for all values of $a$.}
		\label{SaturationEstimates}
\end{figure}

In this appendix we present some circumstantial evidence for the saturation formula presented in the main text based on the modifed version of our toy model, eq. (\ref{coupHamexpMod}).  This cannot present a true test, since the simplified version involves terms mixing position and momentum
so that there isn't multiple minima that may support instantons.  Regardless, we will show that blindly using (\ref{instantonApprox}) despite this fact can still provide a great estimate for a saturation value observed for a class of time-evolved states.  As reference state we choose 
\begin{equation}
|K_0\rangle = N (a_{1}^\dag)^q |0, 0\rangle
\end{equation}
and, since the Hamiltonian commutes with $a_1^\dag a_1 + a_2^\dag a_2$, the Krylov subspace is (at most) $q+1$-dimensional.  If we furthermore specialise to $\omega_1 = \omega_2$ then this subspace involves $q+1$ eigenstates of the Hamiltonian with energies
\begin{equation}
E_j = 2 \omega_1 + 2 \omega_1 g e^{\alpha (q - 2 j)   } \ \ \ ; \ \ \ j = 0, 1, 2, \cdots , q
\end{equation}
The spread complexity of the time-evolved reference state 
\begin{eqnarray}
C(t) & = & \sum_n n \langle K_0 |e^{i t H}| K_n\rangle \langle K_n | e^{- i t H} | K_0\rangle    \nonumber \\
& = & \sum_{n} \sum_{j,j'=0}^{q} n \ e^{i t (E_j - E_{j'})} ( \langle K_0| E_j\rangle \langle E_{j'}| K_0\rangle   )    ( \langle K_n| E_{j'}\rangle \langle E_{j}| K_n\rangle   )   \nonumber
\end{eqnarray}
involve complex exponentials of differences of the energies.  At late times we may expect these to oscillate wildly and possibly cancel.  This is not guaranteed, since we have only a finite number of complex phases.  Additionally, since the system is finite-dimensional, the spread complexity must be periodic.  We thus have a time-scale at which we may expect saturation (when the arguments $t \omega_1 g e^{\alpha q}(e^{-2 \alpha j} - e^{-2\alpha j'} )$ are all large) and another after which the system should repeat (when $t E_j = 2 \pi n_j$ for all $j$).  For some choices of parameters we are able to observe complexities that saturate, see Fig. (\ref{CtForToyModel}).  The time-scale for which the complexity saturates increases with $q$.  \\ \\
As mentioned, this model does not support instantons.  In the main text we argued that the saturation value (under assumptions of chaotic dynamics and the existence of instantons) is given by a sum over the instantons paths.  For this to be true we need 
\begin{equation}
\sum_{n}  \sum_{\left\{  x_1    \right\}, \left\{  x_2    \right\} = \textnormal{minima}} \left| K_n^*(  \left\{  x_2    \right\}   )  \psi(\left\{  x_1    \right\} ) e^{-\frac{1}{\hbar} S_E(\textnormal{instanton}, x_1, x_2) }   \right|^2 \approx 1
\end{equation}
As mentioned, the model does not support instantons so that there isn't a discrete set of points that will satisfy the above condition.  In a naive application we could normalise the complexity as (using only a single pair of points for simplicity) as 
\begin{equation}
\frac{ \sum_{n}    n  \left| K_n^*(  \left\{  x_2    \right\}   )  \psi(\left\{  x_1    \right\} ) +  K_n^*(  \left\{  x_1    \right\}   )  \psi(\left\{  x_2    \right\} )  \right|^2    }{ \sum_{n}    \left| K_n^*(  \left\{  x_2    \right\}   )  \psi(\left\{  x_1    \right\} ) +  K_n^*(  \left\{  x_1    \right\}   )  \psi(\left\{  x_2    \right\} )  \right|^2  }
\end{equation}
Based on the symmetries of the function we will take the set of points 
$$ \left\{   x_1 \right\} = \left\{  a, a   \right\} = -\left\{   x_2 \right\}   $$
By now using the naive formula we obtain a good approximation for the saturation value for any value of $a$, see Fig (\ref{SaturationEstimates}).  Note that, since the Krylov space dimension is small enough, we are able to compute these quantities at infinite precision.  

\bibliographystyle{utphys}
\bibliography{KComplexityInstrefs}

\end{document}